\documentclass[11pt]{article}
\usepackage{theorem,amsfonts}
\usepackage{bezier}

{\theoremstyle{break}
\theorembodyfont{\normalfont}

}
\makeatletter
\@addtoreset{equation}{section}
\makeatother

\renewcommand{\theequation}{\thesection.\arabic{equation}}

\font\tengoth=eufm10 \font\sevengoth=eufm7 \font\fivegoth=eufm5
\newfam\gothfam

\textfont\gothfam=\tengoth \scriptfont\gothfam=\sevengoth
  \scriptscriptfont\gothfam=\fivegoth

\def\dibujo #1 by #2 (#3){
  \vbox to #2{
    \hrule width #1 height 0pt depth 0pt
    \vfill
    \special{picture #3}}}

\def\scaledpicture #1 by #2 (#3 scaled #4){{
  \dimen0=#1 \dimen1=#2
  \divide\dimen0 by 1000 \multiply\dimen0 by #4
  \divide\dimen1 by 1000 \multiply\dimen1 by #4
  \dibujo \dimen0 by \dimen1 (#3 scaled #4)}}

\def\be{\begin{equation}}
\def\ee{\end{equation}}
\def\bea{\begin{eqnarray}}
\def\eea{\end{eqnarray}}

\def\R{{\mathbb R}}

\def\1{\'{\i}}

\def\k{\kappa}
\def\dd{{\rm d}}
\def\con{{\rm conf}}
\def\mob{{\rm m\"ob}}
\def\sim{{\rm sim}}
\def\cong{{\rm Conf}}
\def\simg{{\rm Sim}}
\def\cons{C}

\def\semidirprod{\odot} 

\def\aa{a}
\def\bb{b}
\def\xx{x}
\def\yy{y}
\def\te{\phi}

\def\ss{\aa}

\def\ssa{\aa}
\def\ssb{\yy}

\def\rra{r}
\def\rrb{\phi}

\def\vvv{v}
\def\area{{\cal S}}

\def\gudd{{\,\mbox{gd}}}
\def\lan{\Lambda}

\def\Sk{S}           
\def\Ck{C}            
\def\Tk{T}           
\def\Vk{V} 
\def\arcS{{\,\mbox{arcS}}}           
\def\arcC{{\,\mbox{arcC}}}            
\def\arcT{{\,\mbox{arcT}}}           
\def\arcV{{\,\mbox{arcV}}}           

\def\>#1{{\bf #1}}

\def\cnu{+}
\def\cnd{-}

\def\xxtilde{\tilde s}

\def\yyl{\yy^{\wedge}}
\def\xxl{\xx^{\wedge}}

\def\radio{\rho}

\def\kkpp{\eta_+}
\def\kkmm{\eta_-}
\def\kkppmm{\eta_\pm}

\begin{document}

\noindent 
\hfill  \ 
\bigskip

\begin{center}
{\LARGE{\bf{Conformal compactification  and}}}  

{\LARGE{\bf{ cycle-preserving   symmetries of spacetimes }}}

\end{center}
 
\bigskip

\begin{center}
Francisco J. Herranz$^\dagger$
and Mariano Santander$^\ddagger$
\end{center}

\begin{center}
{\it $^\dagger$ Departamento de F\1sica, Escuela
Polit\'ecnica Superior\\
Universidad de Burgos, E-09006 Burgos, Spain}
\end{center}

\begin{center}
{\it $^{\ddagger}$ Departamento de F\1sica Te\'orica,
Facultad de Ciencias\\ Universidad de Valladolid,
E--47011 Valladolid, Spain}
\end{center}

\bigskip

\begin{abstract}
The cycle-preserving symmetries for the nine two-dimensional real
spaces of constant curvature are collectively obtained within a 
Cayley--Klein framework.  This approach affords a  unified and global 
study  of the conformal structure  of the three classical Riemannian
spaces as well as of the six   relativistic and  non-relativistic
spacetimes  (Minkowskian, de Sitter, anti-de Sitter, both Newton--Hooke
and Galilean), and gives rise to  general expressions holding
simultaneously for all of them. Their metric structure and   {\em cycles}
(lines with constant geodesic curvature  that include   geodesics  and  
circles) are explicitly  characterized.  The corresponding cyclic
(M\"obius-like) Lie groups  together with the differential realizations of
their algebras are then deduced; this derivation is new and  much simpler
than the usual ones and applies to any homogeneous space in the
Cayley--Klein family, whether flat or curved and with any signature.
Laplace and  wave-type differential equations with conformal  algebra
symmetry are constructed. Furthermore,  the  conformal groups are realized
as matrix groups acting as globally defined linear transformations in a
four-dimensional `conformal ambient space', which in turn leads to an
explicit description of the  `conformal completion' or compactification of
the nine spaces.  
\end{abstract}

\bigskip

\newpage


\section{Introduction}

The role of conformal groups in physics can hardly be overestimated.
Taking aside conformal invariance in quantum field theory, this role
appears even at a rather basic level.  In special
relativity, spacetime is a flat (zero curvature) pseudo-Riemannian
space $ISO(3,1)/SO(3,1)$ whose kinematical motion group is the Poincar\'e
group $ISO(3,1)$; its `angle'-preserving transformations generate a
group, the (Minkowskian) conformal group
$SO(4,2)$,  which has dimension 15 and contains as a subgroup the
Poincar\'e group. This conformal group is the maximal
invariance group of the vacuum Maxwell equations  
\cite{Cunningham,Bateman,Rosen,Fushchich}, and in general, of a large
number of equations in different areas, such as, for instance, all
equations describing zero-mass systems
\cite{McLennan,McLennanb,Mack,Mickelsson,Jacobsen}.

The group of conformal transformations  of the $N$-dimensional
($N$D)  Euclidean space ${\bf E}^N$ was  firstly  found
by Liouville  (for an explicit construction see \cite{Doub}) and is also
well known. This group can be obtained from two equivalent
approaches. The `conformal' method is to obtain transformations  
$\phi:{\bf E}^N\to {\bf E}^N$ which carry the Euclidean metric $g$  into 
another proportional to it: $g'\equiv \phi^\ast g  = \mu(x) g$,
where   $\mu(x)$ is a  function of the coordinates; these
preserve the angle between any two crossing curves, and generate the
Euclidean conformal group. The `hypersphere' method is to look for (local)
transformations ${\bf E}^N\to {\bf E}^N$ which carry hyperspheres into
hyperspheres, including  hyperplanes as limit cases;
this  approach naturally requires to complete  ${\bf E}^N$ in order to have
these transformations globally defined, and has the advantage of  showing
clearly the structure of the `hyperspherical' group of ${\bf E}^N$ 
\cite{Berger}; it was introduced by Lie and Darboux.   Both constructions
lead to identical results except for the 2D case, where the conformal
angle-preserving group is    infinite-dimensional,  while the
circle-preserving one (so called circular transformations) has dimension
6. In the generic $N$D case, the transformations  obtained through either
method generate, as is well known, a finite-dimensional Lie group
isomorphic to $SO_0(N+1,1)$, that is, the connected component of the
identity of $O(N+1,1)$ which appears when discrete reflections and
inversions are also considered. The corresponding Lie algebra is spanned
by the usual Euclidean generators of translations and rotations together
with some new generators  of dilations and specific conformal
transformations.

The Minkowskian  and Euclidean spaces are two important particular
instances within the family of {\it flat}  symmetric homogeneous spaces
${\bf R}^{p, q} \equiv ISO(p,q)/SO(p,q)$. The conformal group of ${\bf
R}^{p, q}$  is isomorphic to   $SO_0(p+1,q+1)$ (inversions not included)
\cite{Doub,YRY}; its transformations preserve the `angle' between any
two curves and geometrically they carry
${\bf R}^{p, q}$-hyperspheres into themselves. In this context it is
rather natural to inquire which are the conformal groups of the
homogeneous Riemannian and pseudo-Riemannian spaces with {\it non-zero}
constant curvature (for $N=2$ they are the sphere ${\bf S}^2$, the
hyperbolic plane ${\bf H}^2$ and the two de Sitter `spheres'
${\bf AdS}^{1+1}$ and ${\bf dS}^{1+1}$), and also to analyze the conformal
groups of the (contracted) cases with a degenerate metric (for
$N=2$, these are the $(1+1)$D Galilean ${\bf G}^{1+1}$ and the two
Newton--Hooke ${\bf NH}_\pm^{1+1}$ spacetimes). The above spaces together
with the Euclidean ${\bf E}^{2}$ and the Minkowskian ${\bf M}^{1+1}$ spaces
constitute the nine quasi-orthogonal homogeneous spaces or Cayley--Klein
(CK) spaces \cite{Yaglom,CKdos,Tesis,trigo} in two dimensions. 

The CK groups
share many properties, even if the CK family comprises
simultaneously simple as well as non-simple (but still quasi-simple)
orthogonal Lie groups \cite{Ros,Rosb,casimir}. Further to the groups and
algebras themselves, this CK framework allows a  unified and global
approach to the study  of the associated homogeneous spaces (including the
relativistic and non-relativistic spacetimes  of constant curvature). 
General  expressions parametrized through two real coefficients $\k_1,
\k_2$ and holding simultaneously for all nine 2D CK groups/spaces can be
obtained; for each specific case they follow simply by particularizing 
$\k_1, \k_2$. For instance, their corresponding Drinfeld--Jimbo quantum
deformations
\cite{CKdos,Poisson},   superintegrable systems \cite{integrable},
trigonometry \cite{trigo}, etc.\ have already been studied by following
this `universal' approach which displays clearly some family properties
hidden in a case-by-case analysis. The logic behind this approach can be
described as `extended meaning, simpler proofs'.

The aim of this paper is twofold. First, to study in detail the {\em
cycle-preserving} transformations of the nine 2D CK  spaces (the finite
dimensional part of the full set of conformal transformations in 2D),
paying special attention to the six spacetimes. Second,
we propose an explicit description, covering the
nine CK spaces in the same run, of their corresponding inversive or
conformal compactification, where cycle-preserving transformations can be
defined as global transformations. We present a joint
view of the structure of these conformal compactifications together with
their relationships and the way the initial space is embedded into its
compactified. Thus   we recover many known results while adotping a fresh
perspective we provide a clear view of the behaviour of the
compactification when either curvature vanishes or metric degenerates. 

In section 2 we  describe the   nine 2D CK spaces, whose 
metric structure is studied in section 3 by introducing
three sets of geodesic coordinates useful in our further construction.
In section 4 we obtain the equations of  {\em cycles} as lines with
constant geodesic curvature that include, as particular cases, geodesics,
equidistants, horocycles and circles; these complete the results obtained
in \cite{trigo} concerning trigonometry. Next we  give a simple and
apparently new derivation of cycle-preserving transformations in section
5. In this way, the conformal Lie groups  together with the differential
realizations of their corresponding conformal algebras are obtained.
Differential equations with conformal algebra symmetry are constructed in
section 6; these results not only cover the usual  2D Laplace and
$(1+1)$D wave equations, but also encompass their non-zero curvature
versions with the associated symmetry operators. The `conformal
completion' of the 2D CK spaces is developed in section 7 through a
realization of their  conformal groups acting as globally defined linear
transformations in a 4D ambient space; this extends to all the nine CK
spaces the conformal compactification which is  already familiar for the
Euclidean and Minkowskian spaces but which also makes  sense for the
curved and degenerate  spaces.  Some remarks
close the paper.

A large amount of information relative to cycle-preserving conformal
transformations is made available in completely explicit and tabular
form, so that it is easily retrievable. As a byproduct of our
 construction, a clear and global picture of the
conformal  symmetries for $(1+1)$D relativistic and non-relativistic
spacetimes with constant curvature is obtained.  Along the paper we
freely interpret the `geometrical' results by `translating' them into the
kinematical language (as in tables \ref{table3}, \ref{table4},
\ref{table6}, \ref{table8} and \ref{table10}).


\section{The nine two-dimensional Cayley--Klein
spaces}

To begin with we briefly recall the algebraic structure of the
nine 2D CK  spaces (for more
details see \cite{trigo}).  Their motion groups are collectively denoted 
$SO_{\k_1,\k_2}(3)$ where $\k_1$, $\k_2$ are  two
real coefficients. 
 The commutation relations of the CK algebra $so_{\k_1,\k_2}(3)$ in the
basis $\{P_1,P_2,J_{12}\}$  and the Casimir invariant read:
\be
[J_{12},P_1]=P_2 \qquad [J_{12},P_2]=-\k_2 P_1 \qquad 
[P_1,P_2]=\k_1 J_{12}
\label{ba}
\ee
\be
{\cal C}= \k_2 P_1^2 + P_2^2+\k_1 J_{12}^2 .
\label{bbaa}
\ee
Notice that $\k_1, \k_2$ can be reduced to $+1, 0, -1$ by rescaling the
generators.  The {\it plane} (as the space of points) corresponds to the 
2D symmetric homogeneous space
\be
S^2_{[\k_1],\k_2}=SO_{\k_1,\k_2}(3)/SO_{\k_2}(2)
\qquad  SO_{\k_2}(2)=\langle J_{12}\rangle  
\label{bb}
\ee
hence  the generator  $J_{12}$ leaves a point $O$ (the origin) invariant,
thus    acting as the rotation around   $O$, while
$P_1$, $P_2$ generate translations that move $O$ along two basic
directions.

The curvature and metric signature of
these CK spaces are determined by $\k_1,\k_2$:
the space
$S^2_{[\k_1],\k_2}$ has a canonical {\em metric} which at $O$ comes from
the Casimir; as the space is homogeneous the metric is determined once
given at any point. This metric turns out to have {\em constant curvature}
$\k_1$ (written in square brackets in the space notation). At the origin
$O$ the metric matrix is diag$(1,\k_2)$ in the basis $P_1, P_2$ of the
tangent space;  therefore $\k_2$ determines the metric {\em signature}.
 We display the nine 2D CK spaces in table
\ref{table1}; any  vanishing coefficient $\k_i$ can be interpreted as  an
In\"on\"u--Wigner contraction limit and corresponds to either vanishing
curvature ($\k_1\to 0$) or degenerating metric ($\k_2\to 0$).

Spacetimes  with constant curvature  \cite{BLL} appear in
this scheme. If $\{P_1,P_2,J_{12}\}$ are read as generators of
time translations, space translations  and  boosts, respectively, the  
six CK groups with
$\k_2\le 0$  (second and third rows of table \ref{table1}) are the
(kinematical) motion groups of   $(1+1)$D  spacetimes. According to  the
values of $(\k_1,\k_2)$ we  find in table \ref{table1}:

\noindent
$\bullet$ Three `absolute-time' spacetimes for $\k_2=0$:
oscillating Newton--Hooke ${\bf NH}_+^{1+1}$ $(+,0)$, Galilean ${\bf
G}^{1+1}$ $(0,0)$ and expanding
Newton--Hooke   ${\bf NH}_-^{1+1}$ $(-,0)$ (we denote  $ISO(1)\equiv \R$).
 These are non-relativistic spacetimes with a
degenerate Riemannian metric of signature type diag$(+,0)$.

\noindent
$\bullet$ Three  `relative-time' spacetimes for $\k_2<0$: anti-de
Sitter ${\bf AdS}^{1+1}$ $(+,-)$, Minkowskian ${\bf
M}^{1+1}$ $(0,-)$ and de Sitter ${\bf dS}^{1+1}$ $(-,-)$,
with a Lorentzian (indefinite) metric with signature type diag$(+,-)$.

In this kinematical context, the coefficients $\k_i$ are related to the
 universe time radius $\tau$ and  speed of light $c$ by
\be
\k_1=\pm 1/\tau^2\qquad \k_2=-1/c^2 .
\label{bc}
\ee
The curvature $\k_1$ may also be considered as a cosmological constant.
The contraction
$\k_1\to 0$ corresponds to   the flat limit $\tau\to\infty$; likewise the
contraction  
$\k_2\to 0$ is the non-relativistic limit $c\to\infty$.
 The three remaining spaces with $\k_2>0$ have no 
direct kinematical interpretation (unless one passes to imaginary
Euclidean time); they are the three classical Riemannian spaces of
constant curvature with   definite positive  metric  with signature type
diag$(+,+)$.


\section{Metric structure and coordinate systems}


\subsection{Matrix realization of the Cayley--Klein groups}

Let us consider the following 3D real matrix   representation of
$so_{\k_1,\k_2}(3)$:
\be
P_1=-\k_1 e_{01}+e_{10}\qquad
P_2=-\k_1\k_2 e_{02}+e_{20}\qquad
J_{12}=-\k_2 e_{12}+e_{21} 
\label{bd}
\ee
where $e_{ij}$ is a 3D matrix with a single non-zero  entry 1 at row $i$
and column $j$ $(i,j=0,1,2)$. The exponential of these matrices leads to
  one-parametric subgroups of $SO_{\k_1,\k_2}(3)$:
\bea
&&{\rm e}^{\alpha P_1}=\left(\begin{array}{ccc}
\Ck_{\k_1}(\alpha)&-\k_1\Sk_{\k_1}(\alpha)&0 \cr 
\Sk_{\k_1}(\alpha)&\Ck_{\k_1}(\alpha)&0\cr 
0&0&1
\end{array}\right) \qquad
{\rm e}^{\gamma J_{12}}=\left(\begin{array}{ccc}
1&0&0\cr 
0&\Ck_{\k_2}(\gamma)&-\k_2\Sk_{\k_2}(\gamma)\cr
0&\Sk_{\k_2}(\gamma)&\Ck_{\k_2}(\gamma)
\end{array}\right) \nonumber\\[4pt] 
&&{\rm e}^{\beta P_2}=\left(\begin{array}{ccc}
\Ck_{\k_1\k_2}(\beta)&0&-\k_1\k_2\Sk_{\k_1\k_2}(\beta)\cr 
0&1&0\cr 
\Sk_{\k_1\k_2}(\beta)&0&\Ck_{\k_1\k_2}(\beta)
\end{array}\right) 
\label{bf}
\eea
where we have introduced the cosine $\Ck_\k(x)$ and sine
$\Sk_\k(x)$ functions  \cite{CKdos,Tesis,trigo,Poisson}:
\be 
\Ck_{\k}(x) 
=\left\{
\begin{array}{ll}
  \cos {\sqrt{\k}\, x} &\   \k >0 \cr 
  1  &\  \k  =0 \cr 
\cosh {\sqrt{-\k}\, x} &\  \k <0 
\end{array}\right. \qquad 
\Sk_{\k}(x) 
=\left\{
\begin{array}{ll}
    \frac{1}{\sqrt{\k}} \sin {\sqrt{\k}\, x} &\   \k >0 \cr 
  x &\  \k  =0 \cr 
\frac{1}{\sqrt{-\k}} \sinh {\sqrt{-\k}\, x} &\  \k <0 
\end{array}\right.  
\label{bh}
\ee
From them, we define  the `versed sine' (or versine) $\Vk_\k(x)$ and the
tangent
$\Tk_\k(x)$:
\be
\Vk_{\k}(x) =\frac 1\k(1-\Ck_\k(x)) \qquad
\Tk_{\k}(x) =\frac{\Sk_\k(x)}{ \Ck_\k(x)} .
\label{bj}
\ee
These curvature-dependent functions coincide with the  circular
and hyperbolic ones for   $\k=1$ and $\k=-1$, respectively; the
contracted case    $\k=0$  gives rise to the parabolic or Galilean
functions:  $\Ck_{0}(x)=1$,
$\Sk_{0}(x)=x$ and $\Vk_{0}(x)=x^2/2$.  
Identities for the  trigonometric functions (\ref{bh})--(\ref{bj})
will be necessary in the computations carried out throughout the paper;
these relations can be found in the appendix of \cite{trigo} so that we
omit them here. We will also need their derivatives: if we denote  their
corresponding inverse functions  with the prefix `arc-', it can be shown
that \cite{Tesis}:
 \be
\begin{array}{ll}
\displaystyle{ \frac {\dd}{\dd x}\Ck_\k(x)=-\k\Sk_\k(x)  } &\qquad 
\displaystyle{\frac {\dd}{\dd x}\arcC_\k(x) = \frac {-1}{{\k\sqrt{{\frac
1\k}(1-x^2)}}}  }\\
\displaystyle{ \frac {\dd }{\dd x}\Sk_\k(x) =\Ck_\k(x) }&\qquad 
\displaystyle{\frac {\dd }{\dd x}\arcS_\k(x) = \frac {1}{ {\sqrt{1-\k
x^2}}}}\\
\displaystyle{ \frac {\dd }{\dd x}\Tk_\k(x) =\frac
{1}{\Ck^2_\k(x)}}&\qquad 
\displaystyle{\frac {\dd }{\dd x}\arcT_\k(x) =\frac {1}{{1+\k x^2}} }\\
\displaystyle{\frac {\dd }{\dd x}\Vk_\k(x) =\Sk_\k(x) }&\qquad
\displaystyle{\frac {\dd }{\dd x}\arcV_\k(x) =\frac {1}{ {\sqrt{2x-\k
x^2}}}}.
\end{array} 
\label{zd}
\ee
By taking into account the realization (\ref{bf}), the CK
group $SO_{\k_1,\k_2}(3)$ can be seen as a group of linear transformations
in an ambient space $\R^3=(x^0,x^1,x^2)$, acting as
the group of isometries of a bilinear form $\Lambda$ with matrix: 
\be
\Lambda={\mbox{diag}}(1,\k_1,\k_1\k_2 ).
\label{be}
\ee
Therefore, a generic group element $X\in SO_{\k_1,\k_2}(3)$ can be written
as a product of the matrices (\ref{bf})  satisfying 
\be
X^T\, \Lambda\, X=\Lambda 
\label{bk}
\ee
where $X^T$ denotes the transpose matrix of $X$.

The action of $SO_{\k_1,\k_2}(3)$ on $\R^3$ is linear but not
transitive, since it conserves the quadratic form  $(x^0)^2+\k_1 (x^1)^2+
\k_1\k_2 (x^2)^2$ provided by $\Lambda$. In this action the subgroup
$SO_{\k_2}(2)=\langle J_{12}\rangle$ is the isotropy subgroup of the point
$O\equiv (1,0,0)$ which will be taken as the {\em origin} in the space
$S^2_{[\k_1],\k_2}$. The action becomes transitive if we restrict to  the
orbit in $\R^3$ of the point $O$, which is contained in the `sphere'
$\Sigma$:
\be
\Sigma \equiv   (x^0)^2+\k_1 (x^1)^2+  \k_1\k_2 (x^2)^2=1 .
\label{bl}
\ee
This orbit is identified with the CK space, and
$(x^0,x^1,x^2)$ fulfilling  (\ref{bl})  are called {\em Weierstrass
coordinates} in  $S^2_{[\k_1],\k_2}$. This scheme includes
under a common description all the familiar embeddings (the 
vector models) of the 2D sphere, hyperbolic plane, anti-de
Sitter and de Sitter spacetimes in a linear 3D ambient
space with a flat metric of either Euclidean or
Lorentzian type. The Weierstrass coordinates  allow us to
obtain a  differential realization of the  generators as first-order
vector fields in $\R^3$ with $\partial_i = {\partial}/{\partial
x^i}$  (see (\ref{bd})):
\be
P_1=\k_1 x^1\partial_0 - x^0\partial_1 \qquad 
P_2=\k_1\k_2 x^2\partial_0 - x^0\partial_2 \qquad 
J_{12}= \k_2 x^2\partial_1 - x^1\partial_2 .
\label{bbl}
\ee


\subsection{Metric structure}

If both coefficients $\k_i$ are
different from zero,  $SO_{\k_1,\k_2}(3)$ is a simple Lie group, and the
space $S^2_{[\k_1],\k_2}$ is naturally endowed with a non-degenerate
metric $g_0$ coming from the non-singular Killing--Cartan form in the Lie
algebra $so_{\k_1,\k_2}(3)$. At the origin, $g_0$ is given by:
\be
g_0(P_1,P_1)= -2\k_1 \qquad 
g_0(P_2,P_2)= -2\k_1\k_2 \qquad 
g_0(P_1,P_2)= 0 .
\label{bm} 
\ee
 When dealing with spaces associated to
simple Lie algebras it is customary to define the invariant metric by
simply propagating $g_0$ 
to the whole space through the group action (after taking out non-essential
factors). But we want to cover as well the cases with $\k_1=0$ where $g_0$
vanishes identically. To this aim we may take out instead a factor
$-2\k_1$ out of
$g_0$ in the expression (\ref{bm}), and
introduce the space {\em main metric} $g_1$ from the
Killing--Cartan  metric $g_0$ as 
\be
-2g_1:=  g_0/{\k_1} \qquad g_1(P_1,P_1)= 1 \qquad 
g_1(P_2,P_2)=  \k_2 \qquad 
g_1(P_1,P_2)= 0.
\label{bn}
\ee
The metric obtained by propagating $g_1$ to the whole space is invariant,
and is well defined for any value of $\k_1$.

The non-generic situation $\k_2=0$ corresponds to a degenerate
metric and is singular. In this case (second row of
table \ref{table1}), the action of $SO_{\k_1,0}(3)$ on $S^2_{[\k_1],0}$
has an invariant foliation, whose set of leaves can be parametrized
by   $(x^0)^2+\k_1 (x^1)^2=1 \equiv S^1_{[\k_1]}$; each leaf has
codimension 1. The restriction of $g_1$ to each of these leaves
vanishes, but $g_2=\frac 1{\k_2} g_1$ has a non-vanishing and well
defined restriction to each leaf, with  signature type $(+)$;  we will
refer to $g_2$ as the {\em subsidiary metric}.  Summing up, a unified
description of the metric structure for the nine 2D CK spaces  has the
following elements
\cite{Tesis,phasespace}:

\noindent
(a) The space $S^2_{[\k_1],\k_2}$ has a {\em connection} $\nabla$
(the canonical connection of any symmetric homogeneous space
\cite{Nomizu}) invariant under $SO_{\k_1,\k_2}(3)$.

\noindent
(b) The space $S^2_{[\k_1],\k_2}$ has a  {\em hierarchy  of two
metrics} $g_1$ and $g_2=\frac 1{\k_2} g_1$.  Generically both metrics are
compatible with the connection $\nabla$. The action of
$SO_{\k_1,\k_2}(3)$ on $S^2_{[\k_1],\k_2}$ is by isometries  of both
metrics. 

\noindent
(c)  The main metric $g_1$ is
actually a metric in the true sense (yet it may be degenerate) and 
 has constant curvature $\k_1$ and signature
diag$(+, \k_2)$. 

\noindent
(d)   If $\k_2\neq 0$, the connection $\nabla$ is the unique
connection compatible with the metric $g_1$ (thus $\nabla$ is the metric
or Levi--Civita connection), and  the subsidiary metric $g_2$ is  a
true quadratic metric proportional to $g_1$.

\noindent
(e)  If $\k_2=0$ there are more connections compatible with $g_1$, but
anyhow $\nabla$ is singularized as the canonical one in the homogeneous
symmetric space.  The subsidiary metric $g_2$  gives a true metric only in
each leaf of the invariant foliation which for $\k_2=0$ exists  in
$S^2_{[\k_1],0}$ and $g_2$ has signature $(+)$; in this case the
subsidiary metric cannot be overlooked.

In terms of Weierstrass coordinates  in the linear ambient space $\R^3$,
the two metrics in $S^2_{[\k_1],\k_2}$ come from the flat ambient metric
\be
\dd s^2=(\dd x^0)^2+\k_1 (\dd x^1)^2+  \k_1\k_2 (\dd x^2)^2
\label{bo}
\ee
in the form
\be
(\dd s^2)_1=\frac 1{\k_1}\,\dd s^2 \qquad (\dd s^2)_2=\frac 1{\k_2}(\dd
s^2)_1 .
\label{bp}
\ee


\subsection{Geodesic coordinate systems}

\begin{figure}[ht]
\begin{center}
\begin{picture}(170,145)
\put(130,25){\circle*{3}}
\put(50,33){\makebox(0,0){$\te$}}
\put(140,52){\makebox(0,0){$\yy$}}
\put(75,15){\makebox(0,0){$\aa$}}
\put(75,115){\makebox(0,0){$\xx$}}
\put(75,75){\makebox(0,0){$r$}}
\put(25,25){\line(4,3){118}}
\put(120,96){\circle*{3}}
\put(140,15){\makebox(0,0){$Q_1$}}
\put(136,99){\makebox(0,0){$Q$}}
\put(149,80){\makebox(0,0){$l'_1$}}
\put(149,119){\makebox(0,0){$l$}}
\put(25,25){\circle*{3}}
\put(15,15){\makebox(0,0){$O$}}
\put(15,75){\makebox(0,0){$\bb$}}
\put(25,108){\circle*{3}}
\put(15,108){\makebox(0,0){$Q_2$}}
\put(15,130){\makebox(0,0){$l_2$}}
\put(25,10){\vector(0,1){125}}
\put(0,25){\vector(1,0){170}}
\put(168,15){\makebox(0,0){$l_1$}}
\qbezier(39,25)(40,30)(36,33)
\put(35,98){\line(0,1){10}}
\put(25,98){\line(1,0){10}}
\put(120,25){\line(0,1){10}}
\put(120,35){\line(1,0){9}}
\qbezier[50](25,108)(70,110)(140,90)
\qbezier[50](130,25)(125,80)(115,114)
\put(115,124){\makebox(0,0){$l'_2$}}
\end{picture}
\end{center}
\noindent
\\[-45pt]
\caption{The three geodesic coordinate systems $(\aa,\yy)$, $(\xx,\bb)$
and $(r,\te)$ of a point $Q$.}
\label{figure2}
\end{figure}
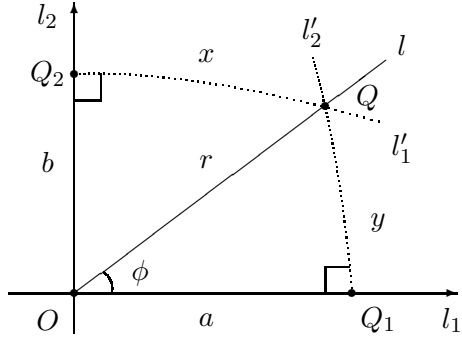

Now we proceed to  introduce three specially relevant coordinate  systems
of geodesic type  in  $S^2_{[\k_1],\k_2}$ (see figure
\ref{figure2}). Let us consider the origin
$O\equiv (1,0,0)$, two (oriented) geodesics $l_1$, $l_2$ which
are orthogonal  through the origin, and a generic point $Q$ with
Weierstrass coordinates $\>x=(x^0,x^1,x^2)$. By taking
into account (\ref{bf}), we have:

\noindent
$\bullet$ If $\>x=\exp(\aa P_1) \exp(\yy P_2)O$, we will call
$(\aa,\yy)$ the {\it geodesic parallel coordinates  of type I}
of $Q$. The point $Q_1$ being the intersection point of  the  geodesic
$l'_2$ orthogonal to $l_1$ through $Q$ with $l_1$ itself, then the first
coordinate  is the (oriented) distance $\aa$ between the origin
$O$ and the point $Q_1$, measured along the geodesic $l_1$. The second
coordinate is the distance $\yy$ between $Q_1$ and $Q$, 
 measured along the geodesic  $l'_2$.

\noindent
$\bullet$ If $\>x=\exp(\bb P_2) \exp(\xx P_1)O$ we will call $(\xx,\bb)$
the {\it geodesic parallel coordinates of type II}   of $Q$. Now we take as 
baseline the geodesic $l_2$  instead of $l_1$. Thus, if $l'_1$ is the
geodesic orthogonal to $l_2$ through $Q$, and $Q_2$ is the  intersection
point of $l'_1$ and $l_2$, then $\xx$ is the distance between
$Q_2$ and $Q$ measured along  $l'_1$, while $\bb$ is the distance between
$O$ and $Q_2$ measured along  $l_2$.

\noindent
$\bullet$ The {\it geodesic polar coordinates} of the point $\>x=\exp(\te
J_{12}) \exp(r P_1)O$ are $(r,\te)$. The coordinate $r$ is the distance 
between $O$ and $Q$ measured along the geodesic $l$ joining both points,
and $\te$ is the angle at $O$ between $l$ and $l_1$.

Weierstrass coordinates $\>x$ of a generic point $Q$ are
displayed in table \ref{table2} in the
three geodesic coordinate systems; of course 
the `sphere' condition (\ref{bl}) is automatically fulfilled. By
introducing these  systems  in the expressions of the metrics 
in Weierstrass coordinates, (\ref{bo}) and (\ref{bp}), we find the
main and subsidiary metrics in these
intrinsic coordinates. From them we may compute the conexion symbols
$\Gamma^i_{jk}$. The area element $\dd\area$ in coordinates say $u^1,
u^2$ is $\sqrt{\det g_1/\k_2}\,\dd u^1 \wedge \dd u^2$. Integration of
$\dd\area$ over a triangle (easily done in polar coordinates) would lead
to the Gauss--Bonnet  theorem for the nine CK spaces, recovering the
equation (4.29) of \cite{trigo}. Both metrics, the non-zero  Christoffel
symbols and the area element are also given in table
\ref{table2}.  When the pair $(\k_1, \k_2)$ is particularized to its nine
essentially different values, relations appearing in table \ref{table2}
provide the explicit description of the nine spaces
$S^2_{[\k_1],\k_2}$. This is performed in   parallel I and polar
coordinates  in tables \ref{table3} and \ref{table4}, respectively.

In ${\bf E}^2$,  geodesic
coordinates reduce to the standard cartesian and polar ones (see for
instance \cite{Klinberger}). In the three flat spaces
$(\k_1=0$), the identities $\xx=\aa\equiv x^1$, $\yy=\bb\equiv
x^2$ hold, so that both parallel I and  II systems are the same but 
this is no longer true for spaces with non-zero
curvature $(\k_1\ne 0)$.

Each of the systems so obtained covers some domain and may have
singularities; both may depend on $\k_1, \k_2$. Polar coordinates 
are always singular at the origin
$r=0$, and in the three relativistic spacetimes with $\k_2=-1/c^2$ they 
only  cover the `time-like' causal cone of $O$ 
limited by the isotropic lines $x^2=\pm c\, x^1$ since
$|x^2/x^1|=c |\tanh (\chi/c)|< c$ no matter of the value of the curvature
$\k_1$. Another `type II' polar coordinate system, with base in the
space-like geodesic $l_2$ instead of the time-like $l_1$, might also be
introduced; this would cover each half of the exterior of the causal cone
in  these cases (in the flat case this part is the Rindler space). Similar
limitations may happen in parallel I coordinates, which do not cover 
fully the de Sitter space. When both  $\k_1,\k_2$ are positive, all three
coordinate systems cover the whole sphere ${\bf S}^2$, and in this case
they are essentially equivalent. 

The non-relativistic (Newtonian) spacetimes with
$c=\infty$ ($\k_2=0$) have foliations with leaves defined by
constant $t$, and the subsidiary (space) metric 
is only relevant within them. In parallel I coordinates we
find a degenerate main temporal  metric $(\dd s^2)_1= \dd t^2$ (providing
`absolute-time'), and an invariant foliation whose leaves are the
`absolute-space' in the moment
$t=t_0$ with a non-degenerate subsidiary spatial metric $(\dd s^2)_2=\dd
y^2$ defined in each leaf. Notice that the `same' metrics are behind the
{\it three} non-relativistic spacetimes with any curvature; they are
distinguished by the canonical connection, which in these cases {\it
cannot} be derived from the metrics.


\section{Cycles}

In any 2D CK  space $S^2_{[\k_1],\k_2}$, {\em cycles} are defined as   
lines  with constant geodesic curvature, including {\em geodesics}
as the particular cases  with zero geodesic curvature. 
From a Lie group  viewpoint, cycles can also be defined as the orbits of
any point by the one-parameter subgroups of the motion group
$SO_{\k_1,\k_2}(3)$. 
When translated to  the vector model in the ambient space $\R^3$, this
gives another possible interpretation of geodesics and cycles: 

\noindent$\bullet$ Geodesics are  intersections of the
`sphere' $\Sigma$ (\ref{bl}) with planes through the origin in $\R^3$.

\noindent$\bullet$ In spaces of {\it non-zero curvature}, cycles 
are the intersections of   $\Sigma$   with
planes in $\R^3$.
The case of zero
curvature, $\k_1=0$, is  special since all plane sections of the
`sphere' $\Sigma$ in $\R^3$ only give  geodesics but not cycles.

In this section we characterize  cycles and  deduce their equations in a
direct way. We will assume  $\k_1\ne 0$, but the final results will also
hold when $\k_1=0$, as a well defined limit $\k_1\to 0$ may be meaningfully
performed. In the case $\k_1\ne 0$ the equation of a generic cycle is a
linear equation in Weierstrass coordinates, where $\alpha, \alpha_0,
\alpha_1,\alpha_2$ are constants:
\be
\alpha_0 x^0+\alpha_1 x^1+\alpha_2 x^2=\alpha .
\label{ca}
\ee
Some properties are better studied in specific coordinate systems in
$S^2_{[\k_1],\k_2}$, and we will freely use the three
geodesic coordinates discussed in the previous section. All properties
discussed below are coordinate-independent, but in order not to burden
the exposition, when using a particular system we will not specify the
domains (each coordinate system may not cover all space) nor will we  
discuss separately their singularities. We start in geodesic parallel I
coordinates.   By introducing $(\aa,\yy)$ of table \ref{table2} in
(\ref{ca}) we obtain the cycle equation in the form:
\be
\alpha_0\Ck_{\k_1}(\aa)+\alpha_1\Sk_{\k_1}(\aa) = \frac
{\alpha}{\Ck_{\k_1\k_2}(\yy)} - \alpha_2 \Tk_{\k_1\k_2}(\yy).
\label{cb}
\ee
This equation can be recast in a simpler way that
involves the  Lambda function $c \to \lan_\k(c)$, fully described in
appendix A. For any quantity $c$ with label $\kappa$, $\lan_\k(c)$ has
the {\it opposite} label $-\k$, and will
be denoted simply as
$c^{\wedge}$. If now we consider the Lambda funcion $\lan_{\k_1\k_2}(y)
\equiv \yyl$ of
$y$, whose label is $-\k_1\k_2$, the  appendix relations
(\ref{wg}), $1/{\Ck_{\k_1\k_2}(\yy)}={\Ck_{-\k_1\k_2}(\yyl)}$ and
$\Tk_{\k_1\k_2}(\yy)=\Sk_{-\k_1\k_2}(\yyl)$, allow us to rewrite
the cycle equation under the (apparently unknown) form:
\be
\alpha_0\Ck_{\k_1}(\aa)+\alpha_1\Sk_{\k_1}(\aa)
= {\alpha}{\Ck_{-\k_1\k_2}(\yyl)} - \alpha_2 \Sk_{-\k_1\k_2}(\yyl).
\label{cbb}
\ee
As is proven in  appendix B, the geodesic curvature $k_g$ of the
cycle (\ref{cb}) (or (\ref{ca})) turns out to be
\be
k_g^2=\k_1^2\alpha^2\frac{\k_2}
{\alpha_2^2+\k_2\alpha_1^2+\k_1\k_2(\alpha_0^2-\alpha^2)}.  
\label{kg}
\ee

 The equations of  cycles in the two remaining  geodesic coordinate
systems can be deduced in a similar way starting  again from
(\ref{ca}). We summarize in table \ref{table5}  the final
results. In each coordinate system we obtain several equivalent expressions for
the rhs of the cycle equations; the last one involves $y$, $x$ or $r$ only
through  $\Tk_{\k_1\k_2}(\yy/2)$, $\Tk_{\k_1}(\xx/2)$ or $\Tk_{\k_1}(r/2)$. In
this last form the cycle equations in polar coordinates can be  rewritten
as a quadratic equation in $\Tk_{\k_1}(r/2)$:
\be
\Tk^2_{\k_1}(r/2) - \frac{1}{\k_1}\,\frac{2}{\alpha+\alpha_0} \left( 
\alpha_1\Ck_{\k_2}(\te)+\alpha_2\Sk_{\k_2}(\te) \right)  \Tk_{\k_1}(r/2) +
\frac{1}{\k_1}\,
\frac{{\alpha-\alpha_0}}{{\alpha+\alpha_0}}=0  .
\ee
This equation clearly displays the universal validity,  for any curved CK
space, of the well known power property of Euclidean circles. Let
$\Tk_{\k_1}(r_1/2),
\Tk_{\k_1}(r_2/2)$ be the two roots of this
equation, where $r_1, r_2$ are the distances to the origin $O$ of
the two intersection points of the cycle with the line
$\te={\rm constant}$ through $O$, then  the product
\be
\Tk_{\k_1}(r_1/2) \Tk_{\k_1}(r_2/2) =   \frac{1}{\k_1}\,
\frac{{\alpha-\alpha_0}}{{\alpha+\alpha_0}} =:  \wp
\label{xyzt}
\ee
is the same for all lines through $O$. As the cycle is arbitrary, this
means that the same property holds for any other point. 
The quantity $\wp$ will be called  {\em power of the point relative to
the cycle}; in the flat limit this reduces to $r_1 r_2 = 4 \wp$.

In what follows we proceed to identify the equations of the three generic
types of cycles:  {\em geodesics} (zero geodesic curvature), {\em
equidistants}  (constant distance to a geodesic) and {\em circles}
(constant distance to a point), that are also displayed in table
\ref{table5}.


\subsection{Geodesics}

If we set $\alpha=0$ in (\ref{cb}) then $k_g=0$ and we obtain
{\em geodesics}. The generic geodesic equations (when $\alpha_2\ne 0$), 
which generalize the usual cartesian equations $y=\beta_0+\beta_1 a$
($\beta_0=-\alpha_0/\alpha_2, \beta_1=-\alpha_1/\alpha_2,\alpha_2\ne 0$)
reducing to them when $\k_1=0$,  are:
\be
\Tk_{\k_1\k_2}(\yy)=
\beta_0\Ck_{\k_1}(\aa)+\beta_1\Sk_{\k_1}(\aa) .
\label{cd}
\ee
The case $\alpha_2= 0$ gives the non-generic geodesics $\aa=\aa_0$. These
expressions  could also have been  deduced by integrating the  usual
geodesic differential equations. In the relativistic spacetimes   with
$\k_2=-1/c^2<0$, the geodesic   (\ref{cd}) can be  either a time-like, 
space-like or isotropic line according to the following conditions
 \be
\begin{array}{ll}
{\mbox {Time-like geodesic:}}&\quad 1+\k_1 \k_2\beta_0^2+ \k_2\beta_1^2>0\\
{\mbox {Space-like geodesic:}}&\quad 1+\k_1 \k_2\beta_0^2+ \k_2\beta_1^2<0\\
{\mbox {Isotropic geodesic:}}&\quad 1+\k_1 \k_2\beta_0^2+ \k_2\beta_1^2=0.
\end{array} 
\label{condition}
\ee

Hereafter, when dealing with geodesics we will distinguish the possible
types;  this will be actually meaningful only whenever $\k_2<0$ and
becomes irrelevant in the Riemannian spaces  with $\k_2>0$, where all
geodesics merge in a single type.


\subsection{Equidistants}

In the relativistic spacetimes, the equidistant
to a given geodesic is of the same type as the geodesic, which can also be 
either  time-like  or  space-like, but its equidistance radius is 
space-like or  time-like, respectively.

The two branches of the {\em equidistant} to a `space-like'
 line given by (\ref{cd})  with 
`time-like' equidistance radius    $d$ is:
\be
\Sk^2_{\k_1}(d)
=\k_2\,\frac{\left\{\Sk_{\k_1\k_2}(\yy) -\Ck_{\k_1\k_2}(\yy)(\beta_0
\Ck_{\k_1}(\aa)+\beta_1 \Sk_{\k_1}(\aa))
\right\}^2}{1+\k_2\beta_1^2+\k_1\k_2\beta_0^2}\qquad \mbox{for}\quad
\k_2\ne 0.
\label{equi}
\ee
These are  particular cycles with equation
(\ref{cb}) for $\alpha_0=-\beta_0 \alpha_2, \alpha_1=-\beta_1 \alpha_2,
\alpha_2\ne 0$ as given for the baseline geodesic, but with 
\be
\alpha=\pm \alpha_2 \Sk_{\k_1}(d)
\left(\beta_1^2+\k_1 \beta_0^2+1/\k_2\right)^{1/2}
\ee
instead of $\alpha=0$, and their   geodesic curvature (\ref{kg}) reads 
\be  
  k_g= | \k_1 \Tk_{\k_1}(d) |  
\label{cee}
\ee 
reducing to $ k_g=0$ (geodesics) in the flat case with $\k_1=0$.


\subsection{Circles}

A `space-like' {\em circle} of center $(a_0, y_0)$  and `time-like' radius
$\radio$   (see appendix B) is: 
\be
\Ck_{\k_1}(\radio)=
\Ck_{\k_1\k_2}(\yy)
\Ck_{\k_1\k_2}(\yy_0)\Ck_{\k_1}(\aa-\aa_0) +
\k_1\k_2\Sk_{\k_1\k_2}(\yy)\Sk_{\k_1\k_2}(\yy_0)
\label{cf}
\ee
and can alternatively be  written in
terms of versed sines  (\ref{bj}): 
\be
\Vk_{\k_1}(\radio)=
\Ck_{\k_1\k_2}(\yy)\Ck_{\k_1\k_2}(\yy_0)\Vk_{\k_1}(\aa-\aa_0)
+\k_2\Vk_{\k_1\k_2}(\yy-\yy_0) .
\label{cg}
\ee
These equations are equivalent to one in the form  (\ref{cb}) for the
following  choice of $\alpha_i$: 
\be
\begin{array}{l}
\alpha_0=\Ck_{\k_1}(a_0)\Ck_{\k_1\k_2}(y_0) \quad
\alpha_1=\k_1\Sk_{\k_1}(a_0)\Ck_{\k_1\k_2}(y_0)\quad 
\alpha_2=\k_1\k_2\Sk_{\k_1\k_2}(y_0) \quad 
\alpha=\Ck_{\k_1}(\radio).
\end{array} 
\label{cce}
\ee
Hence the   geodesic curvature (\ref{kg}) for a circle reads 
\be  
  k_g=  | 1/\Tk_{\k_1}(\radio) | 
\label{ce}
\ee 
reducing to $ k_g=1/\radio$ in the flat case.  The length
$L_\psi(\radio)$ of an arc of circle and the area $\area_\psi(\radio)$ of
a sector of circle, both of radius $\radio$ with central angle $\psi$, 
turn out to be
\be
L_\psi(\radio)=\Sk_{\k_1}(\radio)\psi  \qquad 
\area_\psi(\radio)= \Vk_{\k_1}(\radio)\psi .
\label{ci}
\ee
The circle equations also give
the finite form of the {\em distance} $s\equiv \radio$ between two 
points and can be obtained either by integrating  the main
metric
$(\dd s)_1$ along a geodesic, or directly from  the cosine theorems
of the CK spaces  given in
\cite{trigo}; this is  sketched in appendix B.

To end this section, we analyze the different types of `space-like' cycles
obtained above which arise in the CK spaces; these can be classed 
according to  the geodesic  curvature $k_g$ that ranges in the interval
$k_g\in[0,\infty)$ and the sign of the CK constant $\k_1$:

\noindent
$\bullet$ If $\k_1>0$ (${\bf S}^2$  and ${\bf AdS}^{1+1}$), 
both geodesic curvatures of `space-like' equidistants (\ref{cee}) and
`space-like' circles (\ref{ce}) range in $k_g\in(0, \infty)$, so that such
equidistants (\ref{equi}) are at once circles (\ref{cf}) and conversely;
this is clear on the sphere where parallels are simultaneously
equidistants to the equator and circles with center in the pole. 

\noindent
$\bullet$ If $\k_1=0$ (${\bf E}^2$  and ${\bf M}^{1+1}$), 
equidistants are simply geodesics with $k_g=0$, while circles correspond to 
$k_g\in(0, \infty)$.

\noindent
$\bullet$ If $\k_1<0$  (${\bf H}^2$ and ${\bf dS}^{1+1}$), 
`space-like' equidistants (\ref{equi}) and circles (\ref{cf}) are {\it
different} cycles and a third type, naturally separating them appear; these
are the `space-like' {\em horocycles}, which are the common  limits of
equidistants when the `space-like' base geodesic goes to infinity or of
`space-like' circles when the center goes to infinity. In this case,
equidistants correspond to the values of $k_g\in (0, \sqrt{-\k_1})$ while
circles have $k_g\in (\sqrt{-\k_1}, \infty)$, with
$k_g=0$ for geodesics, $k_g=\sqrt{-\k_1}$ for horocycles and
$k_g=\infty$ for circles of radius $0$.

Equations for `time-like'  equidistants to a `time-like'
line and `time-like' circles (with `space-like' equidistance or radius) can
also be  deduced. For them an analysis similar to the above one can  be
performed.  The classification depends on the sign of
$\k_1\k_2$; for instance, in ${\bf H}^2$ or in ${\bf AdS}^{1+1}$
($\k_1\k_2<0$) there are three types of `time-like'  cycles: equidistants
to a `time-like' geodesic, `time-like' horocycles and `time-like' circles
(`space-like' radius). In ${\bf S}^2$ and in ${\bf dS}^{1+1}$ 
($\k_1\k_2>0$) `time-like' equidistants and  `time-like' circles coincide.

 In table \ref{table6} we explicitly write the equations of geodesics  and 
circles for each of the nine  spaces $S^2_{[\k_1],\k_2}$  in parallel I
coordinates  (a complete study of Galilean lines and cycles can be found in
\cite{Yaglom}). 


\section{Cycle-preserving Lie groups}


\subsection{One-parameter subgroups of geodesic-preserving
transformations}

In this section we give a {\it direct} derivation of the group of
cycle-preserving transformations in $S^2_{[\k_1],\k_2}$. 
We start by
discussing the geodesic-preserving transformations, thus allowing to
recover the CK group $SO_{\k_1,\k_2}(3)$.  Consider the equations of
geodesics in parallel I coordinates given in table \ref{table5}. We look
whether there exists a transformation within the ansatz $(\aa,\yy)\to
(\aa'(\aa),\yy)$ which carries geodesics into geodesics. Noticing that
$\yy$ is assumed not to change, this requirement is equivalent to impose:
\be
\beta'_0\Ck_{\k_1}(\aa'(\aa))+\beta'_1\Sk_{\k_1}(\aa'(\aa))\propto
\beta_0\Ck_{\k_1}(\aa)+\beta_1\Sk_{\k_1}(\aa)  .
\label{da}
\ee
The addition properties of the trigonometric functions
$\Ck_{\k_1}(\aa), \Sk_{\k_1}(\aa)$ \cite{trigo}:  
\be 
  C_\k(x\pm y) = C_\k(x) C_\k(y)\mp \k S_\k(x) S_\k(y) 
\quad
  S_\k(x\pm y)  = S_\k(x) C_\k(y)\pm  S_\k(y) C_\k(x) 
\label{eq:CKSinSumDif}
\ee 
display directly the obvious
solution
$\aa'=\aa+\aa_0$, so that the transformation  
\be
(\aa,\yy)\to (\aa+\aa_0,\yy) 
\label{db}
\ee
is indeed geodesic preserving; it is the  {\em translation}
along the geodesic $l_1$ of $S^2_{[\k_1],\k_2}$ generated by
$P_1$ (see figure \ref{figure2}); in the kinematical language,
this would be a time translation.  Now we look
for a geodesic-preserving transformation described by the ansatz
$(\xx,\bb)\to (\xx,\bb'(\bb))$ in parallel II  coordinates; similar
arguments lead to another geodesic-preserving map:
\be
(\xx,\bb)\to (\xx,\bb+\bb_0)
\label{dc}
\ee
which is the {\em translation} along the geodesic $l_2$
generated by
$P_2$  (a space translation). Finally, we look within the ansatz
$(r,\te)\to (r,\te'(\te))$ in polar coordinates; the condition of
preserving geodesics singles out the transformation:
\be
(r,\te)\to (r,\te+\te_0)
\label{dd}
\ee
which is the {\em rotation} around the origin of
$S^2_{[\k_1],\k_2}$ with generator $J_{12}$ (kinematically a boost). The
fundamental vector fields of (\ref{db})--(\ref{dd}) are given by:
\be
P_1=-\partial_\aa \qquad P_2=-\partial_\bb \qquad 
J_{12}=-\partial_\te .
\label{de}
\ee
By writing them in a single coordinate system (see table \ref{table7}
below) it can be shown that they close a Lie algebra, with commutators
(\ref{ba}): thus we recover the CK algebra
$so_{\k_1,\k_2}(3)$.


\subsection{One-parameter subgroups of cycle-preserving transformations}

Our purpose now is to proceed in a similar way in order to deduce the
{\em cycle-preserving transformations} of the spaces $S^2_{[\k_1],\k_2}$
giving rise to their corresponding conformal groups. Again we start
by assuming $\k_1\neq0$; at the end we will recover the $\k_1\to0$ limit. 

The full CK motion group generated by $\{P_1, P_2, J_{12}\}$ is clearly
cycle-preserving. Are there additional cycle-preserving transformations
which are not geodesic-preserving? A natural idea is to look for them
within three ansatzes complementary to the ones previously explored, 
that is,  $(\aa,\yy)\to (\aa,\yy'(\yy))$, $(\xx,\bb)\to (\xx'(\xx),\bb)$ 
and $(r,\te)\to (r'(r),\te)$, respectively. 

Let us begin with the ansatz $(\aa,\yy)\to (\aa,\yy'(\yy))$ in parallel I
coordinates, and require this to be cycle-preserving; taking 
(\ref{cbb}) into account, and noticing that $\aa$ is
assumed not to change this is equivalent to
enforce
\be
{\alpha'}{\Ck_{-\k_1\k_2}({\yy'}^{\wedge})} - \alpha_2'
\Sk_{-\k_1\k_2}({\yy'}^{\wedge})
\propto  {\alpha}{\Ck_{-\k_1\k_2}(\yyl)} - \alpha_2
\Sk_{-\k_1\k_2}(\yyl)
\label{dff}
\ee
and again the addition  relations (\ref{eq:CKSinSumDif}) display a solution
${\yy'}^{\wedge} = \yyl + \xi$, so that 
\be
(\aa, \yy) \to (\aa,\yy'(\yy))\qquad
\lan_{\k_1\k_2}(\yy')= \lan_{\k_1\k_2}(\yy)+\xi
\label{dj}
\ee
satisfies (\ref{dff}) and is
cycle-preserving. This is a one-parametric subgroup of transformations
with canonical parameter
$\xi$; the fundamental vector field follows using (\ref{wh}): 
\be
L_2=-\partial_{\yyl} =-\Ck_{\k_1\k_2}(\yy)\,\partial_\yy .
\label{dk}
\ee
Transformations generated by $L_2$  behave in the neighbourhood of
the origin $O$ as  translations; we call 
 $L_2$ the generator of   {\it $\lan$--translations} along
the  geodesic $l_2$   of $S^2_{[\k_1],\k_2}$. On the variable
$\yyl=\lan_{\k_1\k_2}(\yy)$, $L_2$ generates actually ordinary 
translations. If one now takes into account that the inverse function of
$\lan_\k$ is $\lan_{-\k}$ (\ref{wf}), then (\ref{dj}) can be rewritten as
\be
(\aa, \yy) \to (\aa,\yy'(\yy))\qquad
\yy'=
\lan_{-\k_1\k_2}\bigl(\lan_{\k_1\k_2}(\yy)+\xi) .
\label{djj}
\ee  

Likewise, the transformation 
\be
(\xx,\bb)\to (\xx'(\xx),\bb) \qquad
\xx'=\lan_{-\k_1}\bigl(\lan_{\k_1}(\xx)+\zeta\bigr)
\label{dl}
\ee
with canonical parameter $\zeta$ is also cycle-preserving, and 
its  fundamental vector field   is the generator of
{\em $\lan$--translations} along the 
geodesic $l_1$ of $S^2_{[\k_1],\k_2}$:
\be
L_1=-\partial_{\xxl} =-\Ck_{\k_1}(\xx)\,\partial_\xx .
\label{do}
\ee

There remains to search for a cycle preserving transformation
within $(r,\te)\to (r'(r),\te)$. In this case it is better to use the
cycle equation in the second alternative form given in table
\ref{table5}, introducing the power $\wp$ of the origin relative to the
 cycle (\ref{xyzt}), and taking into account that $\te$ is assumed not to
change. Thus it suffices to enforce:
\be
\frac 1{\wp'}\,\Tk_{\k_1}(r'/2)+\frac 1{\Tk_{\k_1}(r'/2)}\propto 
\frac 1{\wp}\,\Tk_{\k_1}(r/2)+\frac 1{\Tk_{\k_1}(r/2)} .
\label{dp}
\ee
Hence, the one-parametric family of transformations with canonical
parameter $\lambda$:
\be
\Tk_{\k_1}(r'/2)={\rm e}^\lambda \,\Tk_{\k_1}(r/2) 
\label{dq}
\ee
are cycle-preserving as they  verify (\ref{dp}) with  ${\wp'} = {{\rm
e}^{2\lambda}}  {\wp}$; they can be interpreted as homoteties around the
origin with scale factor ${\rm e}^\lambda $.  The fundamental vector
field is obtained with the aid of (\ref{zd}):
\be
D=-\Sk_{\k_1}(r)\,\partial_r.
\label{ds}
\ee
We remark the natural
appearance of a one-parameter family of `dilations' in spaces like the
sphere and the hyperbolic plane, with non-zero curvature, in spite of a
vague but widespread belief against this possibility.

When $\k_1= 0$, the approach discussed in this paragraph only leads  to
the standard {\em dilation}  $r'={\rm e}^\lambda r$ as a new
transformation beyond the motions. The generators
$L_1=-\Ck_{\k_1}(\xx)\,\partial_\xx$,
$L_2=-\Ck_{\k_1\k_2}(\yy)\,\partial_\yy$, are still meaningful in the
limit $\k_1=0$ where they  reduce to
$P_1, P_2$ and do not provide any new transformation. Nevertheless, the
flat conformal groups can still be obtained as some suitable limit, to be
discussed below.


\subsection{Involutive discrete cycle-preserving transformations}

The equations (\ref{dff}) or (\ref{dp}) allow  other
discrete solutions not belonging to one-parametric families. These
solutions correspond to {\it inversions in cycles} and are involutive.
There are three families, each matching perfectly with the three earlier
types $D,L_1, L_2$. We start with the {\em inversions in circles},
associated to $D$. The discrete transformation:
\be
\Tk_{\k_1}(r'/2) \cdot \Tk_{\k_1}(r/2) =  \wp_0
\label{dt}
\ee
is also a solution of (\ref{dp}) with $\wp'=\wp=\wp_0$.   The circle with
center at the origin and radius $\radio_0$ such that
 $\Tk^2_{\k_1}(\radio_0/2)=-\wp_0$ is invariant under this transformation,
which therefore corresponds to the  inversion  in that circle; notice
that, in this case,  if $r>0$, then $r'<0$ and conversely, so that
$\wp_0<0$. For any value of $\k_1$, the product of two inversions in two
concentric circles in the family (\ref{dt}) with constants  $\wp_0,
\wp_0'$ is a dilation (\ref{dq}) with the same center and scale factor
${\rm e}^\lambda = \wp_0'/\wp_0$. All this is well known for the $\k_1=0$
case, where (\ref{dt})  reduces to the standard flat inversion
in circles $r'=4 \wp_0/r$.

In the generic $\k_1\neq0$ case, a fully analogous  situation happens for
the two additional one-parameter subgroups generated by $L_1, L_2$. Both
$\Lambda$-translations (\ref{djj}) and (\ref{dl}) can be expressed as a
product of two involutive discrete transformations, namely {\it
inversions in   equidistants}. In the $\k_1=0$ limit, however, these new
inversions reduce to the ordinary reflections in geodesics (recall that
$L_1, L_2$ also reduce in this limit to
$P_1, P_2$). 

Specifically, the rhs of the cycle equations in parallel II
coordinates can be recast in the third form given in table 5. If the
condition of carrying cycles into cycles is enforced, this gives an
equation fully analogous to (\ref{dp}), where the rhs is replaced by:
\be
\frac{1}{\k_1\wp_2} 
\left(
\frac{1-\sqrt{\k_1}\,\Tk_{\k_1}(\xx/2)}{1+\sqrt{\k_1}\,\Tk_{\k_1}(\xx/2)}   
\right) +
\left(
\frac{1-\sqrt{\k_1}\,\Tk_{\k_1}(\xx/2)}{1+\sqrt{\k_1}\,\Tk_{\k_1}(\xx/2)}   
 \right)^{-1}\quad 
\wp_2=\frac{1}{\k_1}\,\frac{
\alpha\sqrt{\k_1} -\alpha_1}{ \alpha\sqrt{\k_1}+\alpha_1 }  
\ee
and the lhs is formally similar with $\alpha', \alpha_1', \xx', \wp_2'$. In
this form the equation allows us to read directly a discrete and involutive
cycle-preserving transformation similar to (\ref{dt}):
\be
\left(\frac{1-\sqrt{\k_1}\,\Tk_{\k_1}(\xx'/2)}
{1+\sqrt{\k_1}\,\Tk_{\k_1}(\xx'/2)} \right)  
\cdot
\left(\frac{1-\sqrt{\k_1}\,\Tk_{\k_1}(\xx/2)}
{1+\sqrt{\k_1}\,\Tk_{\k_1}(\xx/2)}\right)
=\k_1\wp_2. 
\label{equiline}
\ee
This transformation keeps invariant one branch of a certain equidistant
(with parameters $\alpha_0=\beta_0=0, \alpha_1\ne 0, \alpha_2=\beta_2=0,
\alpha=\alpha_1\Sk_{\k_1}(d)$) to the line
$l_2$, and the `time-like' equidistance $d$ coincides,  up to the sign,
with the $x$ coordinate which is constant along the equidistant. The
quantity $\wp_2$ may well  be called {\em power of the  baseline relative
to that cycle}.  Again the product of two such
inversions in two coaxial equidistants (with the same baseline) is a
$\Lambda$-translation along the line $l_1$, in total analogy with the
decomposition of a dilation as the product of two inversions in concentric
circles. Everything is similar for the third family corresponding to the
cycle equations in parallel I coordinates, described by similar
expressions with the replacements $\alpha\rightarrow \alpha$,
$\alpha_1\rightarrow \alpha_2$, $\yy\rightarrow \xx$ and $\k_1\rightarrow
\k_1\k_2$, so that $\wp_2\to\wp_1=\frac{1}{\k_1\k_2}\,\frac{
\alpha\sqrt{\k_1\k_2}-\alpha_2 }{\alpha\sqrt{\k_1\k_2} + \alpha_2}$. 

Thus these discrete inversions are even more basic than the
one-parameter transformations generated by $D, L_1, L_2$, and they
display a behaviour which is completely symmetric in the generic curved
case; this symmetry  disappears in
the flat limit.


\subsection{Conformal algebras}

Summing up, for $\k_1\neq0$  we have found six one-parametric subgroups
 of cycle-preserving transformations in  $S^2_{[\k_1],\k_2}$ with
generators $\{P_i,J_{12},L_i,D\}$ $(i=1,2)$. These   close a Lie
algebra.  This fact can be checked by
explicit computation once all the generators are written in the same
coordinate system as is shown in table \ref{table7}; their Lie
brackets are:
\be
\begin{array}{lll}
 [J_{12} ,P_1]=P_2 &\qquad [J_{12},P_2]=-\k_2 P_1 &\qquad 
[P_1,P_2]=\k_1 J_{12} \cr
 [J_{12},L_1]=L_2 &\qquad [J_{12},L_2]=-\k_2 L_1 &\qquad 
[L_1,L_2]=-\k_1 J_{12} \cr
 [D,P_i]=L_i &\qquad 
[D,L_i]=P_i  &\qquad  [D,J_{12}]=0 \cr
[P_1,L_1]=\k_1 D &\qquad [P_2,L_2]=\k_1\k_2 D &\qquad  \cr
[P_1,L_2]=0 &\qquad [P_2,L_1]=0 .&\qquad   
\end{array} 
\label{odv}
\ee
This Lie algebra will be denoted $\con_{\k_1,\k_2}$ 
($\hbox{\mob}_{\k_1,\k_2}$ would be adequate as well) and has two Casimir
invariants: 
\be
\begin{array}{l}
 {\cal C}_1=-\k_1 J_{12}^2+\k_1\k_2 D^2
+\k_2 (L^2_1-P_1^2)+ L_2^2-P_2^2\cr
 {\cal C}_2=\k_1J_{12}D+L_1P_2 -P_1L_2.
\end{array} 
\label{odvv}
\ee
The basis $\{P_i,J_{12},L_i,D\}$ shows a clear conformal {\em duality} or
algebraic symmetric role between the  generators of translations and 
$\lan$--translations: the map defined by
\be
P_i\leftrightarrow L_i\qquad  J_{12}\leftrightarrow
J_{12}\qquad D\leftrightarrow D
\ee
interchanges the set of conformal algebras
$\con_{\k_1,\k_2}\leftrightarrow \con_{-\k_1,\k_2}$. 

Now we discuss the limit $\k_1\to0$. In this case both
generators $L_i$ coincide with the  translation generators  $P_i$, but  as
far as $\k_1\neq0$, we may take  other two generators
\be
G_i=\frac 1{\k_1}(L_i-P_i) \qquad i=1,2  
\label{du}
\ee
generating the so called {\it specific conformal transformations}. The 
 $G_i$ are always  defined, but when $\k_1=0$ they continue to be
independent of the four remaining generators $P_i, J_{12}, D$. The
differential realization of generators $\{P_i, J_{12}, G_i, D\}$
for any value of $\k_1$ are given in table \ref{table7}. When
$\k_1=0$ and in parallel I coordinates  they reproduce the known
vector fields for the conformal generators of flat spaces  (${\bf E}^2$ or
${\bf M}^{1+1}$ according to the sign of $\k_2$):
\be
\begin{array}{l}
P_1 = -\partial_a  \qquad P_2 = -\partial_y \qquad
J_{12}= \k_2 y\, \partial_a -a\,\partial_y  \qquad
 D   = -a\,\partial_a - y\,\partial_y  \\[2pt]
G_1  = \frac 12 ({a^2-\k_2 y^2})\partial_a + ay\, \partial_y \qquad 
G_2 =\k_2 ay\, \partial_a -\frac 12 ({a^2-\k_2 y^2})\,\partial_y. 
\end{array} 
\label{limit}
\ee

In the basis $\{P_i,J_{12},G_i,D\}$  the commutation rules and Casimirs of
$\con_{\k_1,\k_2}$ read
\be
\begin{array}{lll}
 [J_{12} ,P_1]=P_2 &\qquad [J_{12},P_2]=-\k_2 P_1 &\qquad 
[P_1,P_2]=\k_1 J_{12} \cr
 [J_{12},G_1]=G_2 &\qquad [J_{12},G_2]=-\k_2 G_1 &\qquad 
[G_1,G_2]=0 \cr
 [D,P_i]=P_i+\k_1 G_i &\qquad 
[D,G_i]=-G_i  &\qquad  [D,J_{12}]=0 \cr
[P_1,G_1]=D &\qquad [P_2,G_2]=\k_2 D &\qquad  \cr
[P_1,G_2]=-J_{12} &\qquad [P_2,G_1]=J_{12}  &\qquad   
\end{array} 
\label{dv}
\ee
\be
\begin{array}{l}
 {\cal C}_1=-J_{12}^2+\k_2 D^2
+\k_2 (P_1G_1+G_1P_1)+P_2G_2+G_2P_2+\k_1\k_2 G_1^2+\k_1 G_2^2\cr
 {\cal C}_2=J_{12}D+G_1P_2 -P_1G_2.
\end{array} 
\label{dvv}
\ee
This is the basis we will use in the next sections.  Remarkable 
subalgebras of  $\con_{\k_1,\k_2}$ are:

\noindent
$\bullet$   $\{P_1,P_2,J_{12}\}$ span a CK algebra
$so_{\k_1,\k_2}(3)$ (see table \ref{table1}), but the set
$\{P_1,P_2,J_{12}; D\}$ only close a Lie algebra (isometries plus
dilations) if $\k_1=0$.

\noindent
$\bullet$   $\{G_1,G_2,J_{12}\}$ span a {\it flat} CK algebra
$so_{0,\k_2}(3)$, no matter of the value of $\k_1$. The corresponding CK
group has  a semidirect product structure:
$SO_{0,\k_2}(3)=ISO_{\k_2}(3)=T_2\semidirprod SO_{\k_2}(2)$ where
$T_2=\langle G_1,G_2\rangle$ and $SO_{\k_2}(2)=\langle J_{12}\rangle$.

\noindent
$\bullet$   $\{G_1,G_2,J_{12},D\}$ span a Lie algebra 
$\sim_{0,\k_2}$ of a similitude group of a  2D flat
space with the following structure: $\simg_{0,\k_2}=
T_2 \semidirprod \left( SO_{\k_2}(2)\otimes SO(1,1) \right)$, where
$SO(1,1)=\langle D\rangle$. Thus this is either the Euclidean
$(\k_2>0)$, Galilean  $(\k_2=0)$, or Poincar\'e $(\k_2<0)$ similitude group.

Another interesting basis for $\con_{\k_1,\k_2}$ is 
  $\{R_i,J_{12},G_i,D\}$ where
\be
R_i=P_i+\frac 12  {\k_1}G_i=\frac 12 (P_i+L_i) \qquad i=1,2.
\label{dw}
\ee
The commutation relations and Casimir operators are now given by
\be
\begin{array}{lll}
 [J_{12},R_1]=R_2 &\qquad [J_{12},R_2]=-\k_2 R_1 &\qquad 
[R_1,R_2]=0 \cr
 [J_{12},G_1]=G_2 &\qquad [J_{12},G_2]=-\k_2 G_1 &\qquad 
[G_1,G_2]=0 \cr
 [D,R_i]=R_i  &\qquad 
[D,G_i]=-G_i  &\qquad  [D,J_{12}]=0 \cr
[R_1,G_1]=D &\qquad [R_2,G_2]=\k_2 D &\qquad  \cr
[R_1,G_2]=-J_{12} &\qquad [R_2,G_1]=J_{12} &\qquad   
\end{array} 
\label{dx}
\ee
\be
\begin{array}{l}
 {\cal C}_1=-J_{12}^2+\k_2 D^2
+\k_2 (R_1G_1+G_1R_1)+R_2G_2+G_2R_2\cr
 {\cal C}_2=J_{12}D+G_1R_2 -R_1G_2.
\end{array} 
\label{dxx}
\ee
Therefore, the coefficient $\k_1$ disappears from the commutators.
This means that  {\em all\/}
spaces in the family  $S^2_{[\k_1],\k_2}$ with the {\em same $\k_2$}
have isomorphic conformal algebras, no matter of the value of $\k_1$.
These are either:

\noindent
$\bullet$ ${so}(3,1)$ ($(2+1)$D de Sitter algebra) as the conformal
algebra of the  three 2D homogeneous  Riemannian spaces  with  $\k_2>0$.

\noindent
$\bullet$ ${iso}(2,1)$ ($(2+1)$D Poincar\'e) for
 the $(1+1)$D non-relativistic  spacetimes with $\k_2=0$.

\noindent
$\bullet$ ${so}(2,2)$ ($(2+1)$D anti-de Sitter) for
the $(1+1)$D relativistic  spacetimes with $\k_2<0$.

The known symmetry between $P_i$ and $G_i$ which exists in the flat case
{\it does not} survive  in the curved case.  Note also that the vector
fields of $\con_{\k_1,\k_2}$ displayed  in table \ref{table7}  are 
zero-realizations  for both Casimirs. All conformal generators  so far
determined actually satisfy the conformal Killing equations for the
metrics $g_1$ and $g_2$; this is checked in  appendix C.


\section{Conformal symmetries of Laplace/wave-type equations}

As a byproduct of the conformal vector fields deduced in the previous
section, we now proceed to obtain differential equations with conformal
algebra symmetry.

Let us consider a 2D space with coordinates $(u^1,u^2)$, a differential
operator $E=E(u^1,u^2,\partial_1,\partial_2)$ acting  on functions
$\Phi(u^1,u^2)$ defined on the space ($\partial_i\equiv
\partial/\partial u^i$), and consider the differential equation:
\be
E\Phi(u^1,u^2)=0  . 
\label{xxa}
\ee
 An operator ${\cal O}$ is a symmetry of   (\ref{xxa}) if
${\cal O}$ transforms solutions into  solutions:
\be
 E\, {\cal O} = {\cal Q} \, E   \qquad \mbox{or}\qquad
[E,{\cal O}]={\cal Q}'\, E  
\label{xxb}
\ee
 where ${\cal Q}$ is another operator and ${\cal Q}'={\cal Q}-{\cal O}$.

We now focus attention in the differential  equation obtained by
taking as $E$   the  Casimir $\cal C$ of the CK algebra
$so_{\k_1,\k_2}(3)$ (\ref{bbaa}) in the space $S^2_{[\k_1],\k_2}$:
${\cal C}
\Phi=0$. In the three geodesic coordinate systems, such an equation turns
out to be
\bea
&&\left(  
\frac{\k_2}{\Ck_{\k_1\k_2}^2(\yy)}\,\partial_\aa^2+\partial_\yy^2-\k_1\k_2 
\Tk_{\k_1\k_2}(\yy)\,\partial_\yy
\right)\Phi(\aa,\yy)=0\nonumber\\[4pt]
&&\left(  
\k_2\partial_\xx^2-\k_1 \k_2\Tk_{\k_1}(\xx)\,\partial_\xx
+\frac{1}{\Ck_{\k_1}^2(\xx)}\,\partial_\bb^2
\right)\Phi(\xx,\bb)=0\label{xxc}\\[4pt]
&&\left(\k_2\,\partial_r^2+
\frac{\k_2} { \Tk_{\k_1}(r) }\partial_r+
\frac{1}{\Sk_{\k_1}^2(r)}
\partial_\te^2\right)  \Phi(r,\te)=0 .
\nonumber
\eea

The conformal
algebra $\con_{\k_1,\k_2}$  is a symmetry algebra of these equations,
and by using table \ref{table7} the
generators $\{P_i,J_{12},G_i,D\}$ are shown to fulfil
\be
\begin{array}{l}
[{\cal C},X]=0\qquad X\in\{P_1,P_2,J_{12}\}\\[2pt]
[{\cal C},D]=- 2 \Ck_{\k_1}(\aa)\Ck_{\k_1\k_2}(\yy) \, {\cal C}
=- 2 \Ck_{\k_1}(\xx)\Ck_{\k_1\k_2}(\bb) \, {\cal C}
=-2 \Ck_{\k_1}(r) \, {\cal C}
\equiv -2 x^0\, {\cal C}\\[2pt]
[{\cal C},G_1]=2\Sk_{\k_1}(\aa)\Ck_{\k_1\k_2}(\yy) \, {\cal C}
=2\Sk_{\k_1}(\xx)\, {\cal C}
=2\Sk_{\k_1}(r)\Ck_{\k_2}(\te) \, {\cal C}
 \equiv  2 x^1\, {\cal C}\\[2pt]
[{\cal C},G_2]=2\k_2\Sk_{\k_1\k_2}(\yy) \, {\cal C}
=2\k_2\Ck_{\k_1}(\xx)\Sk_{\k_1\k_2}(\bb) \, {\cal C} 
=2\k_2\Sk_{\k_1}(r)\Sk_{\k_2}(\te) \, {\cal C}
\equiv  2\k_2 x^2\, {\cal C}.
\end{array} 
\label{xxd}
\ee

The operator $\cal C$ leading to the equations (\ref{xxc}) is written for
each specific CK space in table   \ref{table8} in geodesic parallel I and
polar coordinates; in the six spacetimes we introduce the notation used in
tables
\ref{table3} and
\ref{table4}.  By taking into account table 
\ref{table8} we see:

\noindent
$\bullet$  The usual 2D Laplace equation appears in ${\bf E}^2$;
the corresponding non-zero curvature Laplace--Beltrami  versions arise in
the sphere and hyperbolic plane. All of them share the same symmetry
algebra $so(3,1)$.

\noindent
$\bullet$  In the non-relativistic spacetimes we find an equation which
does not involve  time and indeed reduces to a 1D `Laplace'
equation. This agrees with the known absence of a true Galilean
invariant wave equation and is the main reason  precluding a further
development of non-relativistic electromagnetic theories, where only two
separate electric and magnetic essentially static limits  are allowed 
\cite{Levy,LeBellac}; see also \cite{negroa,negrob} for a
recent study of  Newtonian electromagnetisms.

\noindent
$\bullet$ The proper $(1+1)$D wave equation is
associated to ${\bf M}^{1+1}$ \cite{Barut}; its 
curvature versions correspond to anti-de Sitter and de Sitter
electromagnetisms in both ${\bf AdS}^{1+1}$ and ${\bf dS}^{1+1}$. These
three wave-type equations have $so(2,2)$ as their conformal symmetry
algebra.

It is worth noting that this
collective  treatment enables a clear view of the contraction limits
between  these  Laplace/wave-type equations together with their associated
conformal symmetry algebras as shown in table   \ref{table8}.


\section{Conformal completion of the Cayley--Klein spaces}

Up to now, we have only identified the conformal group as a {\it local}
group of transformations. The question of whether or not this group can
be realized globally escapes the infinitesimal approach provided by Lie
algebras. The need to suitably complete $S^2_{[\k_1],\k_2}$ with
`additional' points in order to have conformal transformations defined as
a global {\it conformal group} $\cong_{\k_1,\k_2}$ arises, 
because the vector fields given in table
\ref{table7} are, in general, not complete. 
An   example is provided by the hyperbolic space
 ${\bf H}^2\equiv S^2_{[-],+}$ and the finite one-parameter transformations
generated by the dilation $D$ (\ref{dq}):
\be
\tanh(r'/2)={\rm e}^\lambda \tanh(r/2) 
\label{fe}
\ee
where ${\rm e}^\lambda$ is a `similarity factor'. A dilation with
center $O$ and factor ${\rm e}^\lambda>1$  transforms the interior of a
disk of center $O$ and radius 
$\radio_0=2\,\mbox{argtanh}({\rm e}^{-\lambda})$ into the {\it entire}
hyperbolic space, because $\radio_0'=\infty$. What about the images of
points {\it outside} this disk? If one wants to have a global group, the 
images  of these points outside the disk should be added to the `ordinary'
points. 

The same conclusion follows from a direct analysis of the
discrete inversions, expressed in terms of the functions
$\Tk_{\k_1}(r/2),
\Tk_{\k_1}(x/2),
\Tk_{\k_1\k_2}(y/2)$, so that the problem of adding new
points is related with the properties of the function $c\to
\Tk_{\k}(c/2)$. This leads us to a study of the Lambda  function (see
appendix A) and shows that  a suitable extension of this function (mapping
in a one-to-one way the {\it complete} elliptic line into {\it two copies}
of the hyperbolic line glued by their final points) is required and
suffices to deal with this problem algebraically.

We will however not follow here the details of this path, and to
accomplish the conformal  completion of the spaces
$S^2_{[\k_1],\k_2}$ we sketch in this section an alternative 
approach which leads to the same results. This procedure  
extends to any value of curvature and signature the familiar
linealization of the conformal flat Euclidean or Minkowskian groups. 

 From the group theoretical viewpoint, the conformal completion of the CK
space $S^2_{[\k_1],\k_2}$ is defined as a  homogeneous space 
$\cong_{\k_1,\k_2}/ \simg_{0,\k_2}$, 
where $\cong_{\k_1,\k_2}$ is a matrix Lie group defined below and
$\simg_{0,\k_2}$ is the subgroup of $\cong_{\k_1,\k_2}$ generated by
$\{J_{12}, G_1, G_2, D\}$, isomorphic to the similitude group of the flat
CK space $S^2_{[0],\k_2}$. These groups are determined through the Lie
algebras $\con_{\k_1,\k_2}$ and $\sim_{0,\k_2}$ by means of a 4D matrix
realization which allows us  to {\it define} 
$\cong_{\k_1,\k_2}$  as a group of {\it linear} transformations in a 
4D {\it conformal ambient space}. Therefore,  while
the action of the CK group
$SO_{\k_1,\k_2}(3)$ on the 2D  spaces
$S^2_{[\k_1],\k_2}$ can be linearized in an ambient space $\R^3$ with one
extra dimension, the conformal group 
$\cong_{\k_1,\k_2}$ is realized as a matrix group acting  as 
globally defined linear transformations in a  `conformal ambient space'
$\R^4$, with  {\em two} extra dimensions. To establish these ideas we first
study the completion of the 1D spaces  where $\k_1$
cannot be interpreted as the gaussian curvature, as the three 1D spaces
are flat; there are however differences better described after the 
values of $\k_1$.


\subsection{One-dimensional conformal spaces}

The 1D CK spaces $S^1_{[\k_1]}$ are the circle ${\bf S}^1\equiv
S^1_{[+]}$, the Euclidean line ${\bf E}^1 \equiv S^1_{[0]}\equiv \R$ and
the hyperbolic line ${\bf H}^1\equiv S^1_{[-]}$ (which topologically can
be identified with the open segment $(-1,+1)$); the  corresponding CK group
$SO_{\k_1}(2)=\langle P_1\rangle$ is isomorphic, in this order, to 
$SO(2)$, $ISO(1)\equiv\R$, and $SO(1,1)$.  The  conformal algebra 
$\con_{\k_1}$  has three generators $\{P_1,G_1,D\}$  verifying the
following commutation relations:
\be
[D,P_1]=P_1+\k_1G_1 \qquad [D,G_1]=-G_1 \qquad 
[P_1,G_1]=D  .
\label{ga}
\ee
The differential realization in terms of a
single   coordinate
$\aa$ is given by:
\be
P_1=-\partial_\aa \qquad G_1=\Vk_{\k_1}(\aa)\,\partial_\aa 
\qquad D=-\Sk_{\k_1}(\aa)\,\partial_a .
\label{gb}
\ee

 In order to  describe the conformal completion of the three spaces
$S^1_{[\k_1]}$ we consider a $3\times 3$ real matrix representation of 
$\con_{\k_1}$ obtained through the
isomorphism $\con_{\k_1}\simeq so(2,1)$ which can
directly be   seen in terms of the alternative generators $\{R_1, G_1,
D\}$:
\be
[D,R_1]=R_1 \qquad [D,G_1]=-G_1 \qquad 
[R_1,G_1]=D .
\label{gaa}
\ee
Starting from the  realization of
$so(2,1)$ in a pseudo-orthogonal basis
$\{\Omega_{01},\Omega_{02},\Omega_{12}\}$ given by  (\ref{bd}) with the
two labels equal to $-1$:
\be
\Omega_{01}=  e_{01}+e_{10}\qquad
\Omega_{02}=-  e_{02}+e_{20}\qquad
\Omega_{12}=  e_{12}+e_{21} 
\label{ggb}
\ee
we find the matrices for the  generators $\{R_1, G_1, D\}$:
\be
R_1=\frac{1}{2\ell}(\Omega_{02}-\Omega_{12})\qquad
G_1=-\ell (\Omega_{02}+\Omega_{12})\qquad
D=\Omega_{01}  
\label{ggc}
\ee
 where $\k_1$ does not appear and $\ell$ is a non-zero constant with
dimensions of length, which is required by dimensional reasons ($R_1$ and
$G_1$ must be dimensionally inverse of each other after (\ref{gaa}); see
\cite{AFF}). Changing back to the basis
$\{P_1, G_1, D\}$ by means of (\ref{dw}), we obtain (remark that the
product $\k_1\ell^2$ is dimensionless):
\bea
&& P_1=
\frac 1{2\ell}\left(\begin{array}{ccc}
0&0&-1-\k_1 \ell^2\cr 0&0&-1+\k_1 \ell^2\cr 1+\k_1 \ell^2&-1+\k_1 \ell^2&0
\end{array}\right) 
\nonumber\\[2pt]
&& G_1= \ell \left(\begin{array}{ccc}
 0&0&1\cr 0&0&-1\cr -1&-1&0\end{array}\right) \qquad 
D=\left(\begin{array}{ccc}
  0&1&0\cr 1&0&0\cr 0&0&0\end{array}\right).
\label{gd}
\eea
Exponentiation gives rise to the following one-parametric
subgroups of $\cong_{\k_1}$:
 \bea
&&\!\!\!\!\!\!\!\!\!\!\!\!\!\!\! 
{\rm e}^{{\mu P_1}}=\left(\begin{array}{ccc}
1-\frac 1{4\ell^2}(1+\k_1\ell^2)^2\Vk_{\k_1}( \mu )&
\frac 1{4\ell^2} (1-\k_1^2\ell^4)\Vk_{\k_1}( \mu )&
-\frac 1{2\ell}(1+\k_1\ell^2)\Sk_{\k_1}(\mu ) \\[4pt] 
-\frac 1{4\ell^2} (1-\k_1^2\ell^4)\Vk_{\k_1}( \mu )&
1+\frac 1{4\ell^2}(1-\k_1\ell^2)^2\Vk_{\k_1}(\mu)&
-\frac 1{2\ell}(1-\k_1\ell^2)\Sk_{\k_1}( \mu )\\[4pt]
\frac 1{2\ell}(1+\k_1\ell^2)\Sk_{\k_1}( \mu)&
-\frac 1{2\ell}(1-\k_1\ell^2)\Sk_{\k_1}( \mu )&
\Ck_{\k_1}(\mu )\end{array}\right) \nonumber\\[4pt]
&&\!\!\!\!\!\!\!\!\!\!\!\!\!\!\! 
{\rm e}^{{\nu G_1}} =\left(\begin{array}{ccc}
1-\frac 12 \nu^2 \ell^2& -\frac 12 \nu^2 \ell^2&\nu\ell\\[2pt]  
\frac 12 \nu^2\ell^2&1+\frac 12 \nu^2 \ell^2&-\nu \ell\\[2pt] 
-\nu \ell&-\nu \ell&1\end{array}\right) \qquad  
{\rm e}^{{\xi D}}=\left(\begin{array}{ccc}
\cosh\xi&\sinh\xi&0\cr 
\sinh\xi&\cosh\xi&0\cr
0&0&1\end{array}\right)  .
\label{ge}
\eea

In this representation the conformal group $\cong_{\k_1}\simeq SO(2,1)$
acts  in a `conformal  linear ambient space' $\R^3=(s^\cnu,s^\cnd,s^1)$ and
one may easily check that this action is as a  group of isometries of a
bilinear form $\Upsilon={\mbox{diag}}(1,-1,1)$, that is,  any element
$X\in  \cong_{\k_1}$ fulfils
\be
X^T\, \Upsilon\, X=\Upsilon .
\label{ggee}
\ee
Therefore the action of $\cong_{\k_1}$ preserves the quadratic
form  $(s^\cnu)^2-(s^\cnd)^2+(s^1)^2$.   The one-parameter subgroup
$\exp({\nu G_1})$ turns out to be the isotropy subgroup  of the point
${\cal O}=(1,-1,0)$, while
$\exp({\xi D})$ transforms this point into 
   $({\rm e}^{-\xi},-{\rm e}^{-\xi},0)$. Thus the  subgroup
$\simg_{0}$,  generated by $\{G_1, D\}$,  coincides with the isotropy
subgroup of the {\it ray} of ${\cal O}$. The orbit of
${\cal O}$ under $\cong_{\k_1}$ is henceforth contained in the {\em cone}
$\Gamma_0$ given by
\be
\Gamma_0\equiv   (s^\cnu)^2-(s^\cnd)^2+(s^1)^2=0.
\label{gf}
\ee
In this way we find an explicit model of the conformal space  as
contained in the {\it projective cone} $\Gamma_0/\!\propto$ obtained from
$\Gamma_0$ modulo the proportionality relation in the ambient space; more
precisely, as the orbit of the ray of ${\cal O}$ in $\Gamma_0/\!\propto$,
and $\cong_{\k_1}$ acts transitively on this orbit. As a sideline remark,
two possible notions of ray can be used here, allowing proportionality
with any non-zero factor or with positive factors; they lead to two
related  conformal spaces, the latter being a twofold covering of the
former. We shall stick to the relation given as proportionality with a
non-zero factor. 

Since the cone has no direction with  $s^\cnd=0$, the projective
cone, that is, the set of rays in $\Gamma_0$ can directly be  identified
with the section $s^\cnd=-1$ of $\Gamma_0$. Each point in the cone
$(s^\cnu,s^\cnd,s^1)$ is represented by a point
$(\xxtilde^\cnu,\xxtilde^1)$ in this section according to
$(\xxtilde^\cnu=-s^\cnu/s^\cnd, \xxtilde^1=-s^\cnu/s^1)$, and fulfilling
\be
(\xxtilde^\cnu)^2+(\xxtilde^1)^2=1 .
\ee
Consequently, the 1D
completed conformal space  
\be
\cons^1_{[\k_1]}\equiv \cong_{\k_1}/\simg_0=SO(2,1)/(T_1\semidirprod 
SO(1,1))
\qquad \simg_0=\langle G_1,D\rangle
\label{gge}
\ee
can be identified with a {\em circle} 
${\bf S}^1$:  
\be
\cons^1_{[\k_1]}\equiv \Gamma_0/\!\propto\, \equiv
(\xxtilde^\cnu)^2+(\xxtilde^1)^2=1
\longleftrightarrow  {\bf S}^1 .
\label{gh}
\ee
Note that $\cons^1_{[\k_1]}$ is indeed independent of $\k_1$, but we
maintain reference to $\k_1$ to remind the original space. The natural
parametric description of $\cons^1_{[\k_1]}\equiv{\bf S}^1$  is in terms
of a single coordinate $A\in [-\pi, \pi]$ with $\pi$ and $-\pi$ identified:
\be
\xxtilde^\cnu=\cos A    \qquad 
\xxtilde^1=\sin A .
\label{para}
\ee


\subsubsection{Conformal embedding $S^1_{[\k_1]} \longmapsto
\cons^1_{[\k_1]}$ }

The complete CK space $S^1_{[\k_1]}$ is naturally embedded in
$\cons^1_{[\k_1]}$; to describe the embedding the natural procedure is to
choose ${\cal O}$ as an {\em origin} point,  to identify  $S^1_{[\k_1]}$ to
the orbit of ${\cal O}$ in ${\bf S}^1$ under the subgroup $\exp({\mu
P_1})$ that  moves ${\cal O}$ in the cone $\Gamma_0$, and after to
consider the projective identification (\ref{gh}) in the conformal space
${\bf S}^1$. The action of $\exp({\mu P_1})$ on
${\cal O}$  provides a parametrization of the orbit of this point in the
conformal ambient space
$\R^3$ in terms of a single coordinate 
$\ss$: 
\be
\left(\begin{array}{c}
1\cr -1\cr 0\end{array}\right)
 \in {\bf S}^1
\to
\left(\begin{array}{c}
s^\cnu\cr s^\cnd\cr s^1\end{array}\right)
={\rm e}^{\ss P_1}
\left(\begin{array}{c}
1\cr -1\cr 0\end{array}\right)
=\left(\begin{array}{c}
1-\frac 1{2\ell^2} (1+\k_1\ell^2)\Vk_{\k_1}( \ss )\\[3pt] 
-1-\frac 1{2\ell^2} (1-\k_1\ell^2)\Vk_{\k_1}( \ss )\\[3pt] 
\Sk_{\k_1}( \ss )/\ell\end{array}\right) \in \Gamma_0.
\label{gg}
\ee
Within the identification (\ref{gh})  we should put the image of
$(1,-1,0)$ back to the circle determined by the intersection of the
cone with the plane 
$s^\cnd=-1$. Hence the explicit expression for the parametrization
$(\xxtilde^\cnu,\xxtilde^1)$ of the conformal space  is:
\be
\xxtilde^\cnu=
\frac{\ell^2-\Tk^2_{\k_1}(\ss/2)}{\ell^2+\Tk^2_{\k_1}(\ss/2)} 
\qquad 
\xxtilde^1=\frac{2\ell\,\Tk_{\k_1}(\ss/2)}{\ell^2+\Tk^2_{\k_1}(\ss/2)}.
\label{gj}
\ee
By comparing with (\ref{para})
we find the universal
description of the embedding as: 
\be
a \in S^1_{[\k_1]} \longmapsto A \in \cons^1_{[\k_1]}    \qquad\qquad
\frac{1}{\ell^2} \Tk^2_{\k_1}(\ss/2) = \tan^2(A/2)  .
\label{ggjj}
\ee

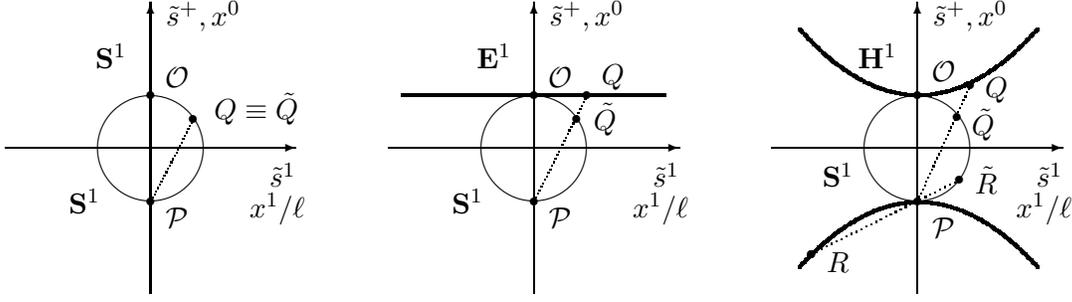
\begin{figure}[ht]
\begin{center}
\begin{picture}(400,130)
\put(0,65){\vector(1,0){110}}
\put(145,65){\vector(1,0){110}}
\put(290,65){\vector(1,0){110}}
\put(55,10){\vector(0,1){110}}
\put(200,10){\vector(0,1){110}}
\put(345,10){\vector(0,1){110}}
\put(55,65){\circle{50}}
\put(200,65){\circle{50}}
\put(345,65){\circle{50}}
\put(105,55){\makebox(0,0){${\tilde s}^1$}}
\put(103,42){\makebox(0,0){${x}^1/\ell$}}
\put(250,55){\makebox(0,0){${\tilde s}^1$}}
\put(248,42){\makebox(0,0){${x}^1/\ell$}}
\put(395,55){\makebox(0,0){${\tilde s}^1$}}
\put(393,42){\makebox(0,0){${x}^1/\ell$}}
\put(75,115){\makebox(0,0){${\tilde s}^+,x^0$}}
\put(220,115){\makebox(0,0){${\tilde s}^+,x^0$}}
\put(365,115){\makebox(0,0){${\tilde s}^+,x^0$}}
\put(30,45){\makebox(0,0){${\bf S}^1$}}
\put(175,45){\makebox(0,0){${\bf S}^1$}}
\put(315,55){\makebox(0,0){${\bf S}^1$}}
\put(40,100){\makebox(0,0){${\bf S}^1$}}
\put(185,100){\makebox(0,0){${\bf E}^1$}}
\put(330,100){\makebox(0,0){${\bf H}^1$}}
\put(55,85){\circle*{3}}
\put(200,85){\circle*{3}}
\put(345,85){\circle*{3}}
\put(55,45){\circle*{3}}
\put(200,45){\circle*{3}}
\put(345,45){\circle*{3}}
\put(65,92){\makebox(0,0){${\cal O}$}}
\put(210,92){\makebox(0,0){${\cal O}$}}
\put(355,94){\makebox(0,0){${\cal O}$}}
\put(65,39){\makebox(0,0){${\cal P}$}}
\put(210,39){\makebox(0,0){${\cal P}$}}
\put(355,35){\makebox(0,0){${\cal P}$}}
\put(71,76){\circle*{3}}
\put(95,80){\makebox(0,0){$Q\equiv {\tilde Q}$}}
\qbezier[25](55,45)(63,60)(71,76)
\put(220,85){\circle*{3}}
\put(216,76){\circle*{3}}
\put(230,92){\makebox(0,0){$Q$}}
\put(227,76){\makebox(0,0){$\tilde Q$}}
\qbezier[30](200,45)(210,65)(220,85)
\qbezier[30](345,45)(355,67)(365,89)
\put(365,89){\circle*{3}}
\put(375,87){\makebox(0,0){$Q$}}
\put(305,25){\circle*{3}}
\qbezier[30](305,25)(333,39)(361,53)
\put(305,25){\circle*{3}}
\put(315,22){\makebox(0,0){$R$}}
\put(361,53){\circle*{3}}
\put(371,53){\makebox(0,0){$\tilde R$}}
\put(360,77){\circle*{3}}
\put(370,74){\makebox(0,0){$\tilde Q$}}
\linethickness{1pt}
 \put(150,85){\line(1,0){100}}
\qbezier(300,110)(345,60)(390,110)
\qbezier(300,20)(345,69)(390,20)
\end{picture}
\end{center}
\noindent
\\[-45pt]
\caption{2D visualization of the three conformal embeddings $S^1_{[\k_1]}
\longmapsto \cons^1_{[\k_1]}\equiv {\bf S}^1$ with 
coordinates $(\xxtilde^\cnu \leftrightarrow x^0, \xxtilde^1
\leftrightarrow x^1/\ell)$. The point  $Q=(x^0,x^1/\ell)$ belongs to 
the CK space  $S^1_{[\k_1\ell^2]}$, while ${\tilde
Q}=(\xxtilde^\cnu,\xxtilde^1)$ is  its stereographic projection with pole 
${\cal P}=(-1,0)$ in the conformal compactification ${\bf S}^1$; the point
$O\equiv {\cal O}=(1,0)$ is the origin in the spaces. The point $\tilde R$
is obtained through  stereographic projection of
$R$ that belongs to another copy of $\>H^1$; hence $\tilde R$ is a new
point   to be added in  the conformal completion $\bf S^1$.}
\label{figure3}
\end{figure}

This embedding, obtained through a group theoretical argument, has a very
simple geometrical description as a {\em stereographic projection}. The
initial CK space $S^1_{[\k_1]}$ lives in an ambient space $\R^2=(x^0,
x^1/\ell)$, with equation $(x^0)^2 + \k_1 \ell^2 (x^1/\ell)^2$ =1, which
can be considered as the CK space $S^1_{[\k_1\ell^2]}$. The conformal
compactification $\cons^1_{[\k_1]}$ is identified with the circle ${\bf
S}^1$ (\ref{gh}) in an ambient space  $\R^2=(\xxtilde^\cnu, \xxtilde^1)$.
Identify both ambient spaces as $(\xxtilde^\cnu \leftrightarrow x^0,
\xxtilde^1 \leftrightarrow x^1/\ell)$; in this common space, the embedding
of $S^1_{[\k_1]}$ into $\cons^1_{[\k_1]}$ is simply the stereographic
projection  with pole ${\cal P}=(-1,0)$ of
$S^1_{[\k_1\ell^2]}$ into ${\bf S}^1$, that is,
\be
(x^0+1,x^1/\ell)=\mu (\xxtilde^\cnu+1, \xxtilde^1)
\label{proj}
\ee
where $\mu$ is a real factor  which   can easily be derived  and
reads: 
\be
\mu=\frac{2}{(1+\xxtilde^\cnu) +
\k_1\ell^2(1-\xxtilde^\cnu)}=\frac{x^0+1}{2}+\frac{1-x^0}{2\k_1\ell^2}.
\ee
Hence the stereographic projection equations turn out to be
\be
\begin{array}{ll}
\displaystyle
\xxtilde^\cnu = \frac{\ell^2 - \frac{1}{\k_1}\frac{1-x^0}{1+x^0}}{\ell^2 +
\frac{1}{\k_1}\frac{1-x^0}{1+x^0}} \\[15pt]
\displaystyle
\xxtilde^1 = \frac{2\ell\frac{1}{1+x^0}}{\ell^2 +
\frac{1}{\k_1}\frac{1-x^0}{1+x^0}} \, x^1
\end{array}  \qquad
\begin{array}{ll}
\displaystyle
x^0=\frac{(1+\xxtilde^\cnu) - \k_1\ell^2(1-\xxtilde^\cnu)
}{(1+\xxtilde^\cnu) +
\k_1\ell^2(1-\xxtilde^\cnu)} \\[15pt]
\displaystyle
x^1=\frac{2\ell}{(1+\xxtilde^\cnu) +
\k_1\ell^2(1-\xxtilde^\cnu)} \, \xxtilde^1 .
\end{array}
\label{est1}
\ee
By substituting $x^0=\Ck_{\k_1}(\ss)$ and $x^1=\Sk_{\k_1}(\ss)$ in
(\ref{est1}) we recover (\ref{gj}). The three particular projections are
represented in figure \ref{figure3}.

The embedding has a non-canonical character, as it depends on
the choice of $\ell$ (or on the arbitrary choice $s^\cnd=-1$ in
(\ref{gf})). Only the {\it sign} of
$\k_1$ matters, as for any fixed $\k_1$, a suitable choice
of  $\ell$ can reduce the dimensionless product $\k_1 \ell^2$ to
either $1, 0, -1$. This assumed,
parametrization (\ref{gj}) encompasses the three 1D conformal spaces:
$$
\begin{array}{llll}
\k_1\ell^2=1& \mbox{Elliptic}\ \cons^1_{[+]}\equiv  {\bf S}^1:& \quad 
\xxtilde^\cnu =\cos(\ss/\ell)&\qquad 
\xxtilde^1 = \sin (\ss/\ell) \\[4pt]
\k_1\ell^2=0& \mbox{Euclidean}\ \cons^1_{[0]}\equiv  {\bf S}^1:&\quad 
\displaystyle{\xxtilde^\cnu =\frac {4- (\ss/\ell)^2}{4+(\ss/\ell)^2}}
&\qquad
\displaystyle{\xxtilde^1 = \frac {4(\ss/\ell)}{4+(\ss/\ell)^2} }\\[10pt]
\k_1\ell^2=-1& \mbox{Hyperbolic}\ \cons^1_{[-]}\equiv  {\bf S}^1:& \quad 
\displaystyle{ \xxtilde^\cnu = \frac 1{\cosh (\ss/\ell)}} &\qquad
\xxtilde^1 = \tanh (\ss/\ell) 
\end{array}
$$ 
where we have used:
\be
 \Ck_{\k_1}(\ss) = \Ck_{\k_1\ell^2}(\ss/\ell)\qquad
\Sk_{\k_1}(\ss) =\ell \,\Sk_{\k_1\ell^2}(\ss/\ell)\qquad
\Tk_{\k_1}(\ss) =\ell \,\Tk_{\k_1\ell^2}(\ss/\ell) .
\ee

While the conformal completion of either the elliptic, Euclidean or
hyperbolic line is the same circle ${\bf S}^1$, there are
some differences between the three cases. For  
$\k_1>0$ the image of the whole elliptic space $S^1_{[+]}$ fills in
completely the conformal space   $\cons^1_{[+]} \equiv {\bf S}^1$. When 
$\k_1<0$ the image of the whole hyperbolic line   $S^1_{[-]}\equiv {\bf
H}^1$   does only fill a proper subinterval ${\cal{I}}_\ell =
(-2\arctan(1/\ell\sqrt{-\k_1}),
2\arctan(1/\ell\sqrt{-\k_1}))\subset\cons^1_{[-]} \equiv {\bf S}^1$; this
is clear from stereographic projection (see figure \ref{figure3}). Due to
the arbitrariness of $\ell$, the position of the  images in the conformal
embedding of the two final points of ${\bf H}^1$ is conventional  (e.g.\
for  $\k_1\ell^2=-1$, ${\cal{I}}_\ell =(-\frac \pi 2,\frac \pi 2)$), but
the fact that the completed conformal hyperbolic space $\cons^1_{[-]}$
contains an infinity of {\em new} points not already present in
${\bf H}^1$ is {\it not} conventional. These new points, filling in a
subinterval complementary to $\cal{I}_\ell$ and centered in $A=\pi$, can
be interpreted as another copy of a  hyperbolic line, which is glued to
the original one by their final points which are identified. In the
limit $\k_1\to 0$, this new copy collapses to a single point since 
 ${\cal{I}}_\ell \to(- \pi  ,  \pi  )$. Hence  only the point $A=\pi\equiv
-\pi$ of $\cons^1_{[0]} \equiv {\bf S}^1$ does not appear as image of a
proper point of the Euclidean line $S^1_{[0]}\equiv {\bf E}^1$ under the
embedding  (\ref{ggjj}); such a point coincides with  the stereographic
projection   pole ${\cal P}$.  Therefore one has to add no points to ${\bf
S}^1$, a single point to ${\bf E}^1$  and an infinity of points to ${\bf
H}^1$ (another copy of ${\bf H}^1$ indeed) in order to obtain their
conformal completion  $\cons^1_{[\k_1]}
\equiv {\bf S}^1$. 
 

\subsection{Two-dimensional conformal spaces}

We now follow the same steps described in the previous subsection for the
2D case. The generators $\{P_i, J_{12},G_i,D\}$ of the conformal algebra 
$\con_{\k_1,\k_2}$ of the 2D spaces  $S^2_{[\k_1],\k_2}$ can be regarded
as generators of linear transformations in a  `conformal ambient space' 
$\R^4=(s^\cnu,s^\cnd,s^1,s^2)$ by means of the following $4\times 4$
matrix representation of (\ref{dv}):
\bea
&&
P_1=\frac 1{2\ell }
\left(\begin{array}{cccc}
0&0&-1-\k_1 \ell^2 &0 \cr 0&0&-1+\k_1 \ell^2 &0\cr 
1+\k_1 \ell^2&-1+\k_1 \ell^2&0&0\cr
0&0&0&0  \end{array}\right) \qquad
J_{12}= \left(\begin{array}{cccc}
0&0&0&0\cr 0&0&0&0\cr 
0&0&0&-\k_2\cr
0&0&1&0  \end{array}\right) \nonumber\\[4pt]
&&
P_2 =\frac 1{2\ell }\left(\begin{array}{cccc}
0&0&0&-\k_2(1+\k_1 \ell^2)\cr 0&0&0&-\k_2(1-\k_1 \ell^2)\cr
0&0&0&0\cr
1+\k_1 \ell^2&-1+\k_1 \ell^2&0&0 \end{array}\right) \qquad
D= \left(\begin{array}{cccc}
0&1&0&0\cr 1&0&0&0\cr 
0&0&0&0\cr
0&0&0&0 \end{array}\right)  \nonumber\\[4pt]
 && G_1 =  \ell \left(\begin{array}{cccc}
0&0&1&0\cr 0&0&-1&0\cr 
-1&-1&0&0\cr
0&0&0&0  \end{array}\right) \qquad
 G_2= \ell \left(\begin{array}{cccc}
0&0&0&\k_2 \cr 0&0&0&-\k_2\cr 
0&0&0&0\cr
-1&-1&0&0  \end{array}\right) 
\label{gl}
\eea
which follows \cite{granada} as before from the identification of the
conformal algebras to  either $so(3,1)$, $iso(2,1)$ or $so(2,2)$
according to the sign of $\k_2$. The corresponding one-parametric
subgroups of   $\cong_{\k_1,\k_2}$  turn out to be:
\bea
&&\!\!\!\!\!\!\!\!\!\!\!\!
{\rm e}^{\mu_1 P_1} =\left(\begin{array}{cccc}
1-\frac {(1+\k_1\ell^2 )^2}{4\ell^2}\Vk_{\k_1 }(  {\mu_1}  )&
\frac {(1-\k_1^2 \ell^4 )}{4\ell^2}\Vk_{\k_1 }(  {\mu_1}  )&
- \frac {(1+\k_1\ell^2 )}{2\ell}\Sk_{\k_1 }( {\mu_1}  )&0 \\[4pt] 
-\frac {(1-\k_1^2 \ell^4 )}{4\ell^2}\Vk_{\k_1 }(  {\mu_1}  )&
 1+\frac {(1-\k_1 \ell^2)^2}{4\ell^2}\Vk_{\k_1 }(  {\mu_1}  )&
- \frac {(1-\k_1 \ell^2)}{2\ell}\Sk_{\k_1 }(  {\mu_1}  )&0 \\[4pt] 
\frac {(1+\k_1 \ell^2)}{2\ell}\Sk_{\k_1 }(  {\mu_1}  )&
-\frac {(1-\k_1 \ell^2)}{2\ell}\Sk_{\k_1 }( {\mu_1}  )&
\Ck_{\k_1 }( {\mu_1}  )&0 \\[4pt] 
0&0&0&1  \end{array}\right) \nonumber\\[4pt] 
&&\!\!\!\!\!\!\!\!\!\!\!\!
{\rm e}^{\mu_2 P_2} =\left(\begin{array}{cccc}
1-\frac {\k_2(1+\k_1 \ell^2)^2}{4\ell^2}\Vk_{\k_1\k_2 }(  {\mu_2}  )&
\frac {\k_2(1-\k_1^2 \ell^4 )}{4\ell^2}\Vk_{\k_1\k_2 }(  {\mu_2}  )&0&
 -\frac {\k_2(1+\k_1\ell^2 )}{2\ell}\Sk_{\k_1\k_2 }(  {\mu_2}  )  \\[4pt] 
-\frac {\k_2(1-\k_1^2 \ell^4  )}{4\ell^2}\Vk_{\k_1\k_2 }(  {\mu_2}  )&
 1+\frac {\k_2(1-\k_1\ell^2 )^2}{4\ell^2}\Vk_{\k_1\k_2 }(  {\mu_2}  )&0&
- \frac {\k_2(1-\k_1\ell^2 )}{2\ell}\Sk_{\k_1\k_2 }(  {\mu_2}  ) \\[4pt] 
0&0&1&0\\[2pt] 
\frac {(1+\k_1 \ell^2)}{2\ell}\Sk_{\k_1\k_2 }( {\mu_2}  )&
-\frac {(1-\k_1\ell^2 )}{2\ell}\Sk_{\k_1\k_2 }(  {\mu_2}  )&0&
\Ck_{\k_1\k_2 }(  {\mu_2}  ) \end{array}\right) \nonumber\\[4pt] 
&&\!\!\!\!\!\!\!\!\!\!\!\!
{\rm e}^{\nu_1 G_1}  =\left(\begin{array}{cccc}
1-\frac 12 \nu_1^2 \ell^2& -\frac 12 \nu_1^2\ell^2 &\nu_1\ell &0\\[4pt] 
\frac 12 \nu_1^2\ell^2 &1+\frac 12 \nu_1^2 \ell^2&-\nu_1 \ell&0\\[2pt] 
-\nu_1\ell&-\nu_1\ell&1&0\cr 0&0&0&1 \end{array}\right) 
\quad
{\rm e}^{\psi J_{12}}= \left(\begin{array}{cccc}
1&0&0&0 \cr0&1&0&0\cr 
0&0&\Ck_{\k_2}(\psi)&-\k_2\Sk_{\k_2}(\psi)\cr
0&0&\Sk_{\k_2}(\psi)&\Ck_{\k_2}(\psi)  \end{array}\right)
\nonumber\\[4pt] 
&&\!\!\!\!\!\!\!\!\!\!\!\!
{\rm e}^{ \nu_2 G_2}   =\left(\begin{array}{cccc}
1-\frac 12 \k_2\nu_2^2 \ell^2&-\frac 12
\k_2\nu_2^2 \ell^2&0&\k_2\nu_2 \ell \\[4pt] 
\frac 12 \k_2\nu_2^2 \ell^2&1+\frac 12\k_2
\nu_2^2 \ell^2&0&-\k_2\nu_2\ell\\[2pt]  
0&0&1&0\cr
-\nu_2\ell &-\nu_2 \ell&0&1  \end{array}\right) 
\quad
{\rm e}^{\xi D}= \left(\begin{array}{cccc}
\cosh\xi&\sinh\xi&0&0\cr \sinh\xi&\cosh\xi&0&0\cr
0&0&1&0\cr
0&0&0&1  \end{array}\right) .\cr
&& \label{gm}
\eea

The conformal group $\cong_{\k_1,\k_2}$ acts as the isometry group of
the bilinear form 
\be
\Upsilon={\mbox{diag}}(1,-1,1,\k_2)
\ee
 verifying the relation (\ref{ggee}). Thus $\cong_{\k_1,\k_2}$ preserves
the quadratic form  $(s^\cnu)^2-(s^\cnd)^2+(s^1)^2+\k_2(s^2)^2$. We  
consider  the cone $\Gamma_0$ defined by:
\be
\Gamma_0\equiv (s^\cnu)^2-(s^\cnd)^2+(s^1)^2+\k_2(s^2)^2=0.
\label{gn}
\ee
The subgroup    $\langle J_{12},G_1,G_2\rangle$ leaves invariant the
{\em origin} point  ${\cal O}=(1,-1,0,0)\in \Gamma_0$, while the dilation
subgroup transforms ${\cal O}$ into $({\rm e}^{-\xi},-{\rm e}^{-\xi},0,0)$.
 Therefore  $\simg_{0,\k_2}=\langle J_{12},G_1,G_2,D\rangle$ is the
isotropy subgroup of the {\it ray} of ${\cal O}$; this  is isomorphic to
the similitude  group of a flat space as commented in section 5.4.  
 The two remaining subgroups generated by $P_1$ and $P_2$ move the ray of
${\cal O}$.

The 2D completed conformal
 space $\cons^2_{[\k_1],\k_2}$ is defined as the homogeneous space
\bea
&&\cons^2_{[\k_1],\k_2}\equiv \cong_{\k_1,\k_2}/\simg_{0,\k_2}
\qquad \simg_{0,\k_2 }=T_2\semidirprod \left(SO_{\k_2}(2)\otimes SO(1,1)
\right)\nonumber\\
 && T_2=\langle  G_1,G_2\rangle
\qquad  SO_{\k_2}(2)=\langle  J_{12}\rangle
\qquad SO(1,1)=\langle  D\rangle
\label{go}
\eea
and can again be identified with the orbit of the ray of ${\cal O}$ under
the linear action of $\cong_{\k_1,\k_2}$  in  $\R^4$. Now the situation is
a bit more complicated and more interesting than in the previous 1D case,
because when $\k_2<0$ the cone will contain directions with  $s^\cnd=0$,
and the direct identification with the section $s^\cnd=-1$ of $\Gamma_0$
cannot be longer made.  Consider first the rays for which $s^\cnd\neq 0$;
each such a ray determines a unique and well defined point in the section
of $\Gamma_0$ by $s^\cnd=-1$. The natural coordinates
$(\xxtilde^\cnu,\xxtilde^1,\xxtilde^2)$ on this section are defined by
$\xxtilde^i=-s^i/s^\cnd$ and verify 
\be
(\xxtilde^\cnu)^2+(\xxtilde^1)^2+\k_2(\xxtilde^2)^2 =1
\longleftrightarrow  S^2_{[+],\k_2}  
\label{gpa}
\ee 
displaying the possible identification of the set of these rays with the
orbit of ${\cal O}$ in a CK `sphere' (\ref{bl}), hence to a CK space
$S^2_{[+],\k_2}$. If $\k_2>0$ the cone cannot contain directions with
$s^\cnd=0$. In this way we find that

\noindent
$\bullet$ The  three 2D  {\em Riemannian spaces}  with 
$\k_2>0$ and {\it any} $\k_1$ have as conformal completion the CK space
$S^2_{[+],+}$, that is, the ordinary {\em sphere} ${\bf S}^2$:
$$
\cons^2_{[\k_1],+}\equiv
\cong_{\k_1,+}/\simg_{0,+}\equiv {SO}(3,1)/ (T_2\semidirprod (SO(2)\otimes 
SO(1,1)))\equiv S^2_{[+],+}\equiv {\bf S}^2.
$$ 

The situation is a bit more complicated when $\k_2\leq0$, as  in these
cases the cone $\Gamma_0$ will always contain rays with $s^\cnd=0$.
Points in the conformal ambient space with $s^\cnd=0$ will have a ray
with no proper intersection with the former section, yet they will
appear as points at infinity in (\ref{gpa}). Further, the fact that we
are dealing with rays imply that these points should be identified
through the ordinary antipodal identification in $\R^4$, which when
$s^\cnd=0$ translates into antipodal identification of these points at
infinity in the section 
$(\xxtilde^\cnu,\xxtilde^\cnd=-1,\xxtilde^1,\xxtilde^2)$ of the ambient
space.  Thus we obtain the result:
$$
\cons^2_{[\k_1],\k_2} = \widetilde{S}^2_{[+],\k_2} := S^2_{[+],\k_2}  \cup
\{ \hbox{points at infinity in }  S^2_{[+],\k_2} \hbox{with antipodal
identification}\}.
$$
Of course, this description  is still valid when $\k_2>0$, but then there
are no points at infinity in $S^2_{[+],\k_2}$, so that
$\widetilde{S}^2_{[+],+}= S^2_{[+],+}$. The space $\cons^2_{[\k_1],\k_2}$
so obtained is always compact (hence the name of conformal
compactification). We now consider the two remaining possibilities
$\k_2=0$ and $\k_2<0$ (see table
\ref{table1}):

\noindent
$\bullet$ The three $(1+1)$D {\em non-relativistic spacetimes} with 
$\k_2=0$ and {\it any} $\k_1$   have as their
conformal compactification 
$$
\cons^2_{[\k_1],0}\equiv  \cong_{\k_1,0}/\simg_{0,0}\equiv {ISO}(2,1)/ 
(T_2\semidirprod ( ISO(1)\otimes SO(1,1)))\equiv
\widetilde{S}^2_{[+],0}\equiv \widetilde{{\bf NH}}_+^{1+1} 
$$
obtained from the {\em oscillating NH spacetime}
through antipodal identification of points at infinity in ${\bf
NH}_+^{1+1}$. 

\noindent
$\bullet$ The three $(1+1)$D {\em relativistic spacetimes} with 
$\k_2<0$ and {\it any} $\k_1$  have as their conformal
compactification 
$$
\cons^2_{[\k_1],-}\equiv
\cong_{\k_1,-}/\simg_{0,-}\equiv  {SO}(2,2)/ (T_2 \semidirprod(SO(1,1)
\otimes SO(1,1)))\equiv
\widetilde{S}^2_{[+],-}\equiv \widetilde{{\bf AdS}}^{1+1} 
$$
obtained from the {\em anti-de Sitter spacetime}  
through antipodal identification of points at infinity in ${\bf
AdS}^{1+1}$.


\subsubsection{Conformal embedding $S^2_{[\k_1],\k_2} \longmapsto
\cons^2_{[\k_1],\k_2}$}

Consider a point $Q$ in $S^2_{[\k_1],\k_2} $ with coordinates 
$(\aa, \yy)$, $(\xx, \bb)$, $(r, \te)$, and Weierstrass coordinates $(x^0,
x^1, x^2)$. Its image under the conformal embedding  is the point
$\tilde Q$ obtained from  ${\cal O}=(1, -1, 0, 0)\in\Gamma_0$ under the
pairs of subgroups in (\ref{gm}) naturally associated to the three types
of  coordinates, and the {\em same} canonical parameters. By following
the pattern of section 3.3 we obtain  parametrizations for $\tilde
Q=(s^\cnu, s^\cnd, s^1, s^2)\in \Gamma_0$ (as in (\ref{gg})) in either 
parallel I, parallel II or polar coordinates of the initial space
$S^2_{[\k_1],\k_2}$. Next the embedding into ${\tilde{S}^2_{[+],\k_2}}$ is
described explicitly by
$(\xxtilde^\cnu,\xxtilde^1,\xxtilde^2)$ with $\xxtilde^i=-s^i/s^\cnd$. The
final results are displayed in table \ref{table9};
notice the  simple expressions   that arise  in polar
coordinates. 

As in the 1D case, this algebraic construction allows a very neat
geometrical interpretation as a {\em stereographic projection}. Proceed as
in the 1D case, and identify the linear ambient space $\R^3=(x^0,
x^1/\ell, x^2/\ell)$ where the initial CK space $S^2_{[\k_1],\k_2}$ can be
realized as
$S^2_{[\k_1\ell^2],\k_2}$ to the  space $\R^3=(\xxtilde^\cnu,
\xxtilde^1,\xxtilde^2)$ where  ${S}^2_{[+],\k_2}$ lives. 
In this common space $(\xxtilde^\cnu
 \leftrightarrow x^0, \xxtilde^1 \leftrightarrow x^1/\ell, \xxtilde^2
\leftrightarrow x^2/\ell)$ the
embedding of $S^2_{[\k_1],\k_2}$ into $\cons^2_{[\k_1],\k_2}$ is simply a
stereographic projection of $S^2_{[\k_1\ell^2],\k_2}$ into
$S^2_{[+],\k_2}$ with pole ${\cal P}=(-1,0,0)$. Its corresponding
equations  are again given by (\ref{proj})--(\ref{est1}) together with 
similar expressions for 
$\xxtilde^2 \leftrightarrow x^2$ as for 
$\xxtilde^1 \leftrightarrow x^1$. 

Such equations provide a description of the conformal embedding which
makes obvious the identification of antipodal points at infinity in
$S^2_{[+],\k_2}$ because these are mapped by stereographical projection
into the same image. Further, this description in ambient Weierstrass
coordinates can be translated to any choice of intrinsic coordinate
systems in either the initial space $S^2_{[\k_1],\k_2}$   or in the
conformal compactified one $\cons^2_{[\k_1],\k_2}
\equiv \widetilde{S}^2_{[+],\k_2}$. For the former, if in the 2D version of
(\ref{est1}) we introduce the parametrizations given in table
\ref{table2}, we recover the results displayed in table \ref{table9}. For
the latter, considered as CK spaces  in their natural  coordinates,
denoted here in capital letters (parallel I $(A, Y)$, parallel II  $(X,
B)$, polar  $(R, \Phi)$; see table
\ref{table2}), the embedding into the ambient space $\R^3=(\xxtilde^\cnu,
\xxtilde^1,\xxtilde^2)$ is: 
\be
\begin{array}{l}
\xxtilde^\cnu = \cos A\  \Ck_{\k_2 }(Y) = \cos X\  \Ck_{\k_2 }(B) = \cos
R  \\
\xxtilde^1 = \sin A\   \Ck_{\k_2 }(Y) = \sin X = \sin R\  \Ck_{\k_2
}(\Phi)\\
\xxtilde^2 = \Sk_{\k_2 }(Y) = \cos X\  \Sk_{\k_2 }(B) =
 \sin R\  \Sk_{\k_2 }(\Phi) . 
\end{array}
\ee  
Thus we can find the description of the embedding 
$S^2_{[\k_1],\k_2} \longmapsto \cons^2_{[\k_1],\k_2}$ in any coordinates
required that are also written in table \ref{table9}. As an example, when
parallel I type coordinates are used in both
$S^2_{[\k_1],\k_2}$  and  ${\tilde S}^2_{[+],\k_2}$, the explicit  
embeddings of the  nine CK spaces are shown in table \ref{table10}; note
that whenever   $\k_1\ell^2=\{1,0,-1\}$ the expressions in
 tables \ref{table9} and \ref{table10} are substantially  simplified.

By taking into account the results given in  table \ref{table10},
the {\em conformal completion} for each CK space can clearly be 
 visualized   in the section $s^\cnd=-1$ of the conformal ambient space
 $\R^4=(\xxtilde^+,\xxtilde^-=-1,\xxtilde^1,\xxtilde^2)$. 
 The  {\em conformal embedding} can be  visualized
 through  a stereographic projection with pole
${\cal P}=(-1,0,0)$ in the common space obtained by 
identifying the initial ambient  CK space $(x^0, x^1/\ell, x^2/\ell)$
to the section $(\xxtilde^+,\xxtilde^1,\xxtilde^2)$
of the conformal ambient space, as depicted for the
1D case in figure \ref{figure3}.
 In particular, the nine CK spaces are embedded according to the sign of
$\k_2$, into its three completions 
$\widetilde{{\bf S}}^{2} \equiv {\bf S}^{2},
\widetilde{{\bf NH}}_+^{1+1}, \widetilde{{\bf AdS}}^{1+1}$ as follows.


\subsubsection{Riemannian spaces with $\k_2>0$: $S^2_{[\k_1],+} \longmapsto
\cons^2_{[\k_1],+}\equiv {\bf S}^2$}

The embedding of the {\em sphere} $S^2_{[+],+}$ covers the full sphere
${\bf S}^2$ once, so that the sphere coincides with its conformal
compactification. The embedding of the {\em Euclidean} plane ${\bf E}^2$
covers the full sphere ${\bf S}^2$ minus a single point, with conformal
ambient coordinates $(-1,-1,0,0)$, that is, the pole ${\cal P}$ of the
stereographic projection. This is the usual point at infinity required to
globally define inversions and transforms ${\bf E}^2$ into the Riemann
sphere. The embedding of the {\em Lobachewski} plane
${\bf H}^2$ only covers one half of the sphere ($\xxtilde^+>0$ when
$\ell$ is chosen such that $\k_1\ell^2=-1$), and points at infinity in
${\bf H}^2$ appear as ordinary points in the equator ($\xxtilde^+=0$); the
new points added in the compactification make up another half of
the sphere  ($\xxtilde^+<0$), which may be considered as another copy of
${\bf H}^2$ glued to the proper embedding of the hyperbolic plane in ${\bf
S}^2$ by the points in the equator
$\xxtilde^+=0$.


\subsubsection{Non-relativistic spacetimes with $\k_2=0$: $S^2_{[\k_1],0}
\longmapsto
\cons^2_{[\k_1],0}\equiv \widetilde{{\bf NH}}_+^{1+1}$}

 The transition from ${\bf NH}_+^{1+1}$ to $\widetilde{{\bf NH}}_+^{1+1}$
requires antipodal identification of the two  circles
$(\xxtilde^+)^2+(\xxtilde^1)^2=1$ at the infinity $\xxtilde^2=\pm
\infty$. Thus the new space is compact. Once this is assumed, the
description of the compactification of the three non-relativistic
spacetimes is straightforward. For {\em oscillating NH} spacetime ${\bf
NH}_+^{1+1}$, one has simply to embed identically
${\bf NH}_+^{1+1}$  into itself. For the {\em Galilei} spacetime  
${\bf G}^{1+1}$, its embedding into ${{\bf NH}}_+^{1+1}$ covers this
space minus the  line  $\xxtilde^+=-1,\xxtilde^-=-1,\xxtilde^1=0$, any
$\xxtilde^2$; the (pole) point  ${\cal P}=(-1,-1,0,0)$ on this line plays
the role of point at infinity in ${\bf G}^{1+1}$ and other  points on
this line correspond to the `instantaneous 1D space' of the point at
infinity. The embedding of the {\em expanding NH} spacetime ${\bf
NH}_-^{1+1}$ into ${{\bf NH}}_+^{1+1}$ covers one `upper'  half of the
cylinder ($\xxtilde^+>0$), so that one has to add as new points the
remaining  `lower' half of the cylinder ($\xxtilde^+<0$) 
 glued to the other proper half   by the two lines in the plane
$\xxtilde^+=0$. Further in the three cases one has   to  consider the
antipodal identification of points at infinity  in 
${\bf NH}_+^{1+1}$ giving rise finally to $\widetilde{{\bf NH}}_+^{1+1}$
as the conformally compactified space.


\subsubsection{Relativistic spacetimes with $\k_2=-1/c^2<0$:
$S^2_{[\k_1],-} \longmapsto
\cons^2_{[\k_1],-}\equiv \widetilde{{\bf AdS}}^{1+1}$}

The transition from 
${\bf AdS}^{1+1}$ to $\widetilde{{\bf AdS}}^{1+1}$ follows by antipodal
identification of the  two circles at the spatial infinity in the
{\em anti-de Sitter}
spacetime: $(\xxtilde^+)^2+(\xxtilde^1)^2-(\xxtilde^2)^2/c^2=1$; hence
$(\xxtilde^+)^2+(\xxtilde^1)^2=\infty, \xxtilde^2=\pm\infty$ and each
circle $\xxtilde^2=\infty, \xxtilde^2=-\infty$ is antipodal to the other.
The (compact) time-like lines (label $\k_1>0$) embed homeomorphically,
while the space-like lines, whose label is $\k_1\k_2<0$, hence hyperbolic
and originally not compact, are glued by their points at infinity with
their antipodals lines in ${\bf AdS}^{1+1}$, following the antipodal
identification of their points at infinity; hence they become also
compact. Topologically the space so obtained is ${\bf S}^1 \times {\bf
S}^1$. One ${\bf S}^1$ corresponds to the originally compact time-like
line $l_1$. The other comes from the 1D compactification of the originally
hyperbolic, hence non-compact space-like line $l_2$, which is obtained by
glueing    two copies of a hyperbolic line ($l_2$ and its antipodal) by
their points at infinity,  as discussed in the section 7.1.1.

For {\em Minkowskian} spacetime ${\bf M}^{1+1}$, the
completion  requires to add  two lines with
equations  $\xxtilde^+=-1,\xxtilde^-=-1,\xxtilde^2=\pm c\,\xxtilde^1$ 
crossing through the intersection (pole) point $\cal P$. This is rather
well known: the new point $\cal P$ corresponds to the point `at infinity'
in ${\bf M}^{1+1}$ and the two new lines are the light-cone of the point
at infinity. 

Finally, the {\em de Sitter} spacetime 
${\bf dS}^{1+1}$, with hyperbolic non-compact time-like lines ($\k_1<0$)
and compact space-like ones ($\k_1\k_2>0$) is the only CK space where
parallel I coordinates do not cover completely the space, so the
expressions   in table 10 do not provide a  complete description. In this
case it is better to use  directly stereographic projection, which maps
the whole de Sitter spacetime into ${\bf AdS}^{1+1}$ in a one-to-one way
compatible with antipodal identification in ${\bf AdS}^{1+1}$.  This could
have been foreseen, as essentially ${\bf dS}^{1+1}$ and
${\bf AdS}^{1+1}$ are the same space, with   an interchange time-like
$\leftrightarrow$ space-like, so we can expect
${\bf dS}^{1+1}$ and ${\bf AdS}^{1+1}$ to have essentially the same
compactification. If we are interested only in the double wedge in ${\bf
 dS}^{1+1}$ covered by parallel I coordinates (with focal points at the
poles of the initial time-like line $l_1$), the embedding of this region
is determined by $(\xxtilde^1)^2< 1$, that is, 
$0<(\xxtilde^+)^2-(\xxtilde^2)^2/c^2\le 1$,  which is limited by four
lines with equations  $\xxtilde^-=-1,\xxtilde^1=\pm 1,\xxtilde^2=\pm
c\,\xxtilde^+$; these four lines are the two pairs of isotropic lines
through the two poles of the initial line $l_1$. However, the other
regions limited by these lines in the conformal compactification are not
new points, but the images by the embedding of the region of ${\bf
dS}^{1+1}$ not covered by the parallel I coordinates.


\section{Concluding remarks}

The present paper affords an approach to
the conformal algebras, groups and spaces comprehensive enough to give a
global  understanding of those aspects of conformal invariance related with
either the space curvature or the metric signature in the initial space.
In particular, it produces a very explicit description of the {\it
conformal completion} of the homogeneous spacetimes where the fact that
the conformal completion of a  homogeneous space of constant curvature
$\k_1$ does not depend on the curvature can be clearly seen.   This
result  may be foreseen from the  well known fact that every 2D metric,
constant curvature or not, is conformally flat ---and hence {\em a
fortiori} the conformal compactification of a curved space will coincide
with that of its corresponding flat one---, but we have attempted here to
make an {\em ab-initio} approach to cycle-preserving transformations
because this provides a lot of additional details, and further can also
be  extended to higher dimensions, where situation is rather different.  A
topic has not deliberately been   touched upon: the use of the three
one-step Cayley--Dickson extensions of $\R$ (complex, dual or double
numbers \cite{Yaglom}), in terms of which circular transformations of the
nine CK spaces can be represented as fractional linear (M\"obius-like)
transformations; this is because this possibility only exists for two
dimensions.

In both Euclidean or Minkowski spaces the transition
from the motion to the conformal group  can be looked at in two stages: the
motion group can be extended first by a one-parameter dilation subgroup,
obtaining a similitude group, and then by the specific conformal
transformations, closing the whole conformal group. It is a widespread
belief that dilation-like transformations do not exist in spaces with
non-zero curvature (sphere,  anti-de Sitter, \dots),  but the explicit
results obtained here show this is not so. What is actually different for
non-zero curvature is that the intermediate stage provided by similitudes
does not exist, and once a single dilation is added to the motion group,
the full conformal group is obtained.  Another trait follows from the
subalgebra structure. For $\k_1=0$ there is a complete symmetry between
translation generators $P_i$ and
 specific conformal generators  $G_i$, and usually these are introduced as
conjugate to translations by an inversion in the origin. This symmetry {\it
does not} extend to the non-zero curvature case, and while the 
$P_i$ {\em do not} commute among themselves, the
$G_i$ do always commute. There is however another
{\it conformal duality} between translations and $\Lambda$-translations,
with respective generators $P_i$, $L_i$, but this duality is invisible in
the conformal algebra of  flat spaces, where $\Lambda$-translations
coincide with translations, leaving the specific conformal transformations
as a kind of vestigial residue of the difference between $P_i$ and $L_i$. 

From a more general viewpoint, a study with $\k_1, \k_2$ as {\it
parameters} displays many illuminating clues which are clear in this
generic approach but cannot be seen in a single  flat case. In previous
papers \cite{CKdos,trigo,Poisson,integrable}  we have established the
suitability of this type of approach and have developed many of the 
geometrical and algebraic aspects involved. And further, as this approach
has a built-in scheme of contractions, its separate study is made
completely redundant. 

Some problems in the conventional Minkowskian quantum
field theories can be seen in a new light if dealt with
in the non-zero curvature case  (anti-de Sitter or de Sitter) taking
afterwards a flat limit; in this sense to study dependence on the
curvature is a natural inquiry. The structure of conformal
compactification of Minkowskian spacetime appears as a particular instance
within our parametric approach; it allows a clear understanding and
visualization of how the embedding of the initial spacetime  changes when
its curvature  vanishes or when the metric degenerates. The first issue is
of relevance in view of the current interest in AdS-CFT correspondence as
a conjecture relating local QFT on $\bf AdS\rm^{1+(d-1)}$ to a conformal
QFT on the compactified Minkowski spacetime
$Comp\bf M\rm^{1+(d-2)}$ (see \cite{Reh} and references therein). In this
context an explicit description of the geometry behind these spaces in a
way as general as possible should be helpful. In another context, the
degeneration of a Lorentzian-type metric produces Newtonian theories, and
`non-relativistic  electromagnetic theories', Maxwell--Le
Bellac--L\'evy-Leblond equations \cite{Levy,LeBellac}, as non-relativistic
limits of Maxwell equations also fit inside this parametrized scheme.
Furthermore, the results here obtained constitute a starting point for the
development of anti-de Sitter and de Sitter  electromagnetic theories. 

Finally, the techniques and results we have presented
here can be generalized straightforwardly to higher dimensions, where the
differential realizations of conformal groups can be obtained in a
completely explicit and closed form for any curvature and signature. This
will be summarized in a forthcoming paper.

\noindent
\\[-45pt]


\section*{Acknowledgments}

\noindent
\\[-25pt]
This work was partially supported  by the Ministerio de Ciencia y
Tecnolog\1a, Spain (Projects  PB98-0370 and   BFM2000-1055) and by Junta
de Castilla y Le\'on, Spain (Project  CO2/399).

 
\section*{Appendix A. The Lambda function}
\setcounter{equation}{0}
\renewcommand{\theequation}{A.\arabic{equation}}

Let us recall the definitions for the Gudermannian function $\gudd(x)$ and
the  function $\lambda(x)$ \cite{Dwight} which often appear  in
hyperbolic geometry:
\bea
&&\gudd(x) =\frac \pi 2-2\arctan({\rm e}^{-x})=2\arctan({\rm e}^{x})- \frac
\pi 2 
\label{wa}\\
&&\lambda(x) =-i \frac \pi 2-2\arg\!\tanh(-i {\rm e}^{-ix})
=2 i \arctan({\rm e}^{-i x})- i\frac \pi 2  .
\label{wb}
\eea
They are related by $\lambda(x)=i \gudd(-i x)$ and they are inverse of
each other, $\gudd(\lambda(x))=x, \lambda(\gudd(x))=x$. Alternatively,
these functions may be defined by the functional relations: 
\be
\tanh\left(\frac{\lambda(x)}{2} \right) = \tan\left(\frac{x}{2}\right) 
\qquad  
\tanh\left(\frac{x}{2} \right) = \tan\left(\frac{\gudd(x)}{2}\right)  
\label{wbb}
\ee
showing that if $x\in(-\infty, \infty)$, then $\gudd(x)\in(-\pi/2,
\pi/2)$. If $\gudd(x)$ is considered as a point in the circle ${\bf
S}^1 \equiv (-\pi, \pi]$, the image of $\mathbb R$ by the map $\gudd(x)$
only fills half the circle. Alternatively, the map $\lambda(x)$ is only
defined in half the circle $x\in(-\pi/2,
\pi/2)$, and the image is the whole line $\mathbb R$.

Within the parametrized CK approach  it is natural to consider
the {\it Lambda function} 
$\lan_\k(x)$ defined as:
\be
\lan_\k(x)=\int_0^x\frac 1{\Ck_\k(t)}\, \dd t .\label{wc}
\ee
 By taking into account the derivative of  $\arcT_\k(x)$ (\ref{zd}) we
find:  
\be
\int_0^x\frac 1{\Ck_\k(t)}\, \dd t=\left[-2\arcT_{-\k}\left(\frac
1{\sqrt{-\k}}
\, {\rm e}^{-\sqrt{-\k}t}\right)\right]_{t=0}^{t=x} 
\ee
and since
\be
\arcT_{-\k}(1/\sqrt{-\k})=
\frac 1{\sqrt{-\k}}\,\frac \pi 4 
\label{wwd}
\ee
we obtain
\be
\lan_\k(x)=\frac 1{\sqrt{-\k}}\,\frac \pi 2
-2\arcT_{-\k}\left(\frac 1{\sqrt{-\k}}
\,{\rm e}^{-\sqrt{-\k}x}\right)=
\left\{\begin{array}{cl}
 \frac{1}{\sqrt{\k}} \,\lambda( {\sqrt{\k}\, x}) &  \k >0 \cr
  x   & \k =0 \cr 
\frac{1}{\sqrt{-\k}} \, \gudd ({\sqrt{-\k}\, x})  & \k <0 
\end{array}\right. .
\label{wd}
\ee
Therefore both the lambda function  (\ref{wb}) and the gudermannian 
(\ref{wa})   are the two particular elliptic $\k>0$
or hyperbolic $\k<0$ instances of a single CK labelled `Lambda function':
$\lan_{+1}(x)\equiv\lambda(x)$ and  $\lan_{-1}(x)\equiv\gudd(x)$.  The
analogous of the functional definition (\ref{wbb}), which can be obtained 
starting from (\ref{wd}), reads:
\be
\Tk_{-\k}\left(\frac{\lan_\k(x)}2\right)=\Tk_\k\left(\frac x2\right)  
\qquad
\Tk_{-\k}\left(\frac
x2\right)=\Tk_{\k}\left(\frac{\lan_{-\k}(x)}2\right)
\label{wwe}
\ee
whence
\be
\lan_{-\k}(\lan_\k(x))=x  .
\label{wf}
\ee
This property extends the known fact that   $\lambda(x)$ and $\gudd(x)$
are inverse each other. 

From this viewpoint, if $x$ is a quantity with
label $\k$, i.e.\ a point in the
1D CK space $S^1_{[\k]}$, then   $\lan_\k(x)$ (also denoted $\xxl$) is a
quantity with label $-\k$, i.e.\ a point in  $S^1_{[-\k]}$, and
the Lambda function should be seen as providing a canonical identification
between $S^1_{[\k]}$ and $S^1_{[-\k]}$, that is, between quantities  with
elliptic and hyperbolic labels.  In particular,
from  (\ref{wwe}) it can be found that 
\bea
&& \Ck_{-\k}(\lan_\k(x)) =\frac 1{\Ck_\k(x)} \quad
\Sk_{-\k}(\lan_\k(x))=\Tk_\k(x) \quad
\Tk_{-\k}(\lan_\k(x))= \Sk_\k(x) \label{wg}\\
 && \Ck_{-\k}^2(\lan_\k(x))-\k \Sk_{-\k}^2(\lan_\k(x))=1 
\label{wgg}
\eea
and thus  $\lan_\k(x)$ has label $-\k$. The derivative of $\lan_\k(x)$ is:
\be
\frac{\dd\lan_\k(x)}{\dd x}=\frac 1{\Ck_\k(x)}=\Ck_{-\k}(\lan_\k(x)).
\label{wh}
\ee

As commented before, for {\it positive} $\k$ the Lambda function is
{\it not}  defined in the whole  CK space $S^1_{[\k]}$. However this
function can be extended in a natural way to a function, still denoted
under the same name, which is defined in the whole $S^1_{[\k]}$. In this
way the image of the elliptic  1D space $S^1_{[\k]}$ (with $\k>0$) by
$\lan_\k$ should be identified with {\it two} copies of the hyperbolic 
  1D space $S^1_{[-\k]}$ which are glued by their final points. The
transcription of the geometrical approach done in the paper to an algebraic
approach involves the use of this extension of the Lambda function. 

 The Lambda function has a clear geometrical interpretation in the 
anti-de Sitter, Minkowskian and de Sitter spacetimes with curvature
$\k_1$ and $\k_2=-1/c^2$: 
if  $t$ (label $\k_1$) is a time-like
length, and we build an orthogonal triangle with an isotropic
hypothenuse (angle $\equiv$ rapidity equals infinity), then the space
length of the other side is
$c \lan_{\k_1}(t)$ \cite{Tesis}. In particular, in Minkowskian spacetime,
$\k_1=0$,  this reduces to the well known $ct$.

 
\section*{Appendix B. Geodesic curvature and finite distance}
\setcounter{equation}{0}
\renewcommand{\theequation}{B.\arabic{equation}}

In a 2D space, the geodesic curvature of a curve can be defined as
 $k_g= |\frac{\dd \beta}{\dd s} |$, where $\beta(s)$ is the
angle between any vector in parallel transport along the line and the
tangent vector to the curve at the point with canonical parameter $s$ 
\cite{Doub}.  In terms of the metric of a 2D space  
$\dd s^2=\sum_{i,j=1}^2\,g_{ij}\dd u^i \dd u^j$ and conexion symbols
$\Gamma_{jk}^i$, the  geodesic curvature for a line 
$(u^1(s),u^2(s))$ is known to be
\be
k^2_g= \sum_{i,j=1}^2\,g_{ij} \vvv^i \vvv^j
\qquad
\vvv^i=\frac{\dd^2 u^i}{\dd s^2}+\sum_{j,k=1}^2\Gamma_{jk}^i
\frac{\dd u^j}{\dd s}\,\frac{\dd u^k}{\dd s}
\qquad i =1,2 .
\label{ya}
\ee

We proceed to deduce the  general expression of $k_g$ for a cycle in
parallel I coordinates $(u^1,u^2)=(\aa,\yy)$. By taking into account the
cycle equation (\ref{cb}) and the main metric $(\dd s)^2_1$ written in
table \ref{table2}, and using the shorthand
$$
f(\aa,\yy)=\frac{-\alpha_0 \k_1 \Sk_{\k_1}(\aa) +\alpha_1
\Ck_{\k_1}(\aa)}{\alpha \k_1\k_2
\Sk_{\k_1\k_2}(\yy)-\alpha_2}\,\Ck_{\k_1\k_2}(\yy)
$$
we find for the first derivatives
\bea
\frac{\dd \aa}{\dd s}=\frac{1}{\Ck_{\k_1\k_2}(\yy)}
\left(1+\k_2 f^2(\aa,\yy) \right)^{-1/2}
\qquad
\frac{\dd \yy}{\dd s}=f(\aa,\yy)
\left(1+\k_2 f^2(\aa,\yy) \right)^{-1/2}  
\nonumber
\eea
(implying $
\frac{\dd y}{\dd \aa}= f(\aa,\yy) \Ck_{\k_1\k_2}(\yy)$) and for the second
derivatives
\bea
&&\!\!\!\!\!\!\!\!\!
\frac{\dd^2 \aa}{\dd s^2}=\k_1\k_2 
\frac{f(\aa,\yy)}{\Ck^2_{\k_1\k_2}(\yy)}
\left( \Sk_{\k_1\k_2}(\yy) + \frac{\alpha-\alpha_2
\Sk_{\k_1\k_2}(\yy)}{\alpha \k_1\k_2 \Sk_{\k_1\k_2}(\yy) -\alpha_2}\right)
\left(1+\k_2 f^2(\aa,\yy) \right)^{-1}\nonumber\\[2pt]
&&\!\!\!\!\!\!\!\!\!
\frac{\dd^2 \yy}{\dd s^2}=-
\frac{\k_1}{\Ck_{\k_1\k_2}(\yy)}
\left( \frac{\alpha-\alpha_2
\Sk_{\k_1\k_2}(\yy)}{\alpha \k_1\k_2 \Sk_{\k_1\k_2}(\yy) -\alpha_2}\right)
\left(1+\k_2 f^2(\aa,\yy) \right)^{-1} .
\nonumber
\eea
Therefore, in this case the two components $\vvv^i$ (\ref{ya}) turn out to
be
\bea
&&\!\!\!\!\!\!\!\!\!\!\!\!\!\!\!\!\!\!\!\!\!\!\!\!
\vvv^1=\frac{\dd^2 \aa}{\dd s^2}-2\k_1\k_2 \Tk_{\k_1\k_2}(\yy)
\frac{\dd \aa}{\dd s}\,\frac{\dd \yy}{\dd s}
= \frac{\alpha\k_1\k_2 f(\aa,\yy)}{\alpha \k_1\k_2 \Sk_{\k_1\k_2}(\yy)
-\alpha_2}\left(1+\k_2 f^2(\aa,\yy) \right)^{-1}
\nonumber\\[2pt]
&&\!\!\!\!\!\!\!\!\!\!\!\!\!\!\!\!\!\!\!\!\!\!\!\!
\vvv^2=\frac{\dd^2 \yy}{\dd s^2}+\k_1\Sk_{\k_1\k_2}(\yy) 
\Ck_{\k_1\k_2}(\yy)
\left(\frac{\dd \aa}{\dd s}\right)^2=
-\frac{\alpha \k_1 \Ck_{\k_1\k_2}(\yy)}{\alpha \k_1\k_2 \Sk_{\k_1\k_2}(\yy)
-\alpha_2}  \left(1+\k_2 f^2(\aa,\yy) \right)^{-1}
\nonumber
 \eea
so $k_g$ is finally obtained as the length of the vector $v^i$ in
the main metric:
 \be
k_g^2=\Ck^2_{\k_1\k_2}(\yy)(\vvv^1)^2+\k_2(\vvv^2)^2=
\frac{\k_1^2\k_2\alpha^2}
{\alpha_2^2+\k_2\alpha_1^2+\k_1\k_2\alpha_0^2-\k_1\k_2\alpha^2}  
\label{ye}
\ee
where we have used  the cycle equation and the identity
\be
\k_1\left (\alpha_0\Ck_{\k_1}(\aa)+\alpha_1 \Sk_{\k_1}(\aa)\right)^2
+\left(-\alpha_0 \k_1 \Sk_{\k_1}(\aa) +\alpha_1
\Ck_{\k_1}(\aa)\right)^2=\alpha_1^2+\k_1\alpha_0^2 .
\label{yf}
\ee
If $\alpha=0$, then $\vvv^i=0$   reproduces the
differential equations of geodesics   with   $k_g=0$.

On the other hand, the finite form of the (`time-like') distances
associated to the main metric $(\dd s)_1$ between two points (given  in
table \ref{table5}) can be obtained by integrating $(\dd s)_1$ along the
geodesic through the points. Working again with parallel I coordinates and
considering the corresponding geodesic equation in table
\ref{table5} and the identity (\ref{yf}), one finds:
$$
\dd s= \frac{  (1+\k_1\k_2\beta_0^2+\k_2
\beta_1^2)^{1/2}  }{1+\k_1\k_2 \left
(\beta_0\Ck_{\k_1}(\aa)+\beta_1 \Sk_{\k_1}(\aa)\right)^2} 
\,\dd\aa .
$$
Hence the distance between two points with parallel
coordinates $(\aa_0,\yy_0)$ and 
$(\aa_1,\yy_1)$ 
reads
$$
s=\left[ \arcT_{\k_1}\left( \frac{\k_2\beta_0\beta_1
+(1+\k_2\beta_1^2)
\Tk_{\k_1}(\aa)}{(1+\k_1\k_2\beta_0^2+\k_2
\beta_1^2)^{1/2}}
\right)\right]_{\aa=\aa_0}^{\aa=\aa_1} .
$$
Since both points must lie on the same geodesic, 
$\Tk_{\k_1\k_2}(\yy_i)=\beta_0\Ck_{\k_1}(\aa_i)+\beta_1
\Sk_{\k_1}(\aa_i)$ $(i=0,1)$, we may obtain the constants $\beta_0,
\beta_1$ in terms of the end points as:
$$
\beta_0=\frac{\Sk_{\k_1}(\aa_1)\Tk_{\k_1\k_2}(\yy_0)
-\Sk_{\k_1}(\aa_0)\Tk_{\k_1\k_2}(\yy_1)}{\Sk_{\k_1}(\aa_1-\aa_0)}
\quad
\beta_1=\frac{\Ck_{\k_1}(\aa_0)\Tk_{\k_1\k_2}(\yy_1)
-\Ck_{\k_1}(\aa_1)\Tk_{\k_1\k_2}(\yy_0)}{\Sk_{\k_1}(\aa_1-\aa_0)}.
$$
Therefore the expression for the tangent of   the distance is given by
$$
T_{\k_1}(s)=\frac{\Sk_{\k_1}(\aa_1-\aa_0)  
(1+\k_1\k_2\beta_0^2+\k_2 \beta_1^2)^{1/2} }
{ \Ck_{\k_1}(\aa_1-\aa_0) +\k_1\k_2\Tk_{\k_1\k_2}(\yy_0)
\Tk_{\k_1\k_2}(\yy_1) }   
$$
and the expression for the cosine of the distance given in table
\ref{table5} is directly recovered 
\be
C^2_{\k_1}(s)=\frac{1}{1+\k_1 T^2_{\k_1}(s)}
=\left(\Ck_{\k_1\k_2}(\yy_0)\Ck_{\k_1\k_2}(\yy_1)\Ck_{\k_1}(\aa_1-\aa_0)
+\k_1\k_2 \Sk_{\k_1\k_2}(\yy_0)\Sk_{\k_1\k_2}(\yy_1)\right)^2 .
\label{yg}
\ee

An alternative trigonometric procedure to find the finite (`time-like')
distance between  two points in polar coordinates $(r_0,\phi_0)$ and
$(r_1,\phi_1)$ is to consider the cosine theorem for a side $c$ in a CK
triangle as is given in the equations (4.4) of \cite{trigo}:
$$
\Ck_{\k_1}(c)=\Ck_{\k_1}(a)\Ck_{\k_1}(b)+\k_1\Sk_{\k_1}(a)\Sk_{\k_1}(b)
\Ck_{\k_2}(C) .
$$
Taking the vertex $C$ as the origin, then the sides $a$ and $b$ are
the coordinates $r_0$ and $r_1$, the angle $C$ is the difference
$\phi_1-\phi_0$ and the side $c$ is the distance $s$ between both points:
\be
\Ck_{\k_1}(s)=\Ck_{\k_1}(r_0)\Ck_{\k_1}(r_1)+\k_1\Sk_{\k_1}(r_0)
\Sk_{\k_1}(r_1)\Ck_{\k_2}(\phi_1-\phi_0)  .
\ee
This is the expression given in table \ref{table5} and it is equivalent to
(\ref{yg}).


\section*{Appendix C. Conformal transformations of the metrics}
\setcounter{equation}{0}
\renewcommand{\theequation}{C.\arabic{equation}}

To obtain the vector fields $X$ corresponding to the
conformal generators of a given metric $g$, the usual method is
to enforce the proportionality by a positive factor
$\mu(x)$ between 
$g$ and the Lie derivative of
$g$; this leads to the {\it conformal Killing
equations} of the metric (for the flat spaces
$ISO(p,q)/SO(p,q)$ this approach can be found in \cite{Doub}).
We study here such transformations for the  metrics $g_1$ and $g_2$ in the
spaces $S^2_{[\k_1],\k_2}$. Let $g$ be the metric of a 2D space with
coordinates $(u^1,u^2)$:
$\dd s^2=\sum_{i,j=1}^2\,g_{ij}\dd u^i \dd u^j$.
The Lie derivative  $L_X g$ of $g$, when $X$ is a generator of a
Lie group $G$  acting on the space is \cite{Warner}: 
\be
L_X g=\sum_{i,j,k=1}^2
\left(g_{ik}\partial_jX^k+g_{kj}\partial_iX^k+X^k\partial_k g_{ij}
\right)\dd u^i \dd u^j  
\label{eb}
\ee
where  $X^k$ are the components of the generator $X$ written as a vector
field: $X=\sum_k X^k\partial_k$. 

Let us consider first the main metric $g_1$ of the spaces  
$S^2_{[\k_1],\k_2}$ in parallel I coordinates as given in table
\ref{table2}. The non-zero components of $g_1$ are
$g_{\aa\aa}=\Ck^2_{\k_1\k_2}(\yy)$ and
$g_{\yy\yy}=\k_2$, so that for  a generic vector field    $X\equiv
X^\aa\partial_\aa+X^\yy\partial_\yy$ the Lie derivative (\ref{eb})  
reads: 
\bea
&&L_Xg_1 =2\left(\Ck^2_{\k_1\k_2}(\yy)\,\partial_\aa X^\aa -  \k_1\k_2
X^\yy \Ck_{\k_1\k_2}(\yy)\Sk_{\k_1\k_2}(\yy)\right)\dd\aa^2\nonumber\\[2pt]
&&\qquad\qquad +2\k_2\left( \partial_\yy X^\yy\right)\dd\yy^2
+2\left( \Ck^2_{\k_1\k_2}(\yy)\,\partial_\yy X^\aa+\k_2\,\partial_\aa X^\yy
 \right)\dd\aa\,\dd\yy .
\label{ee}
\eea
If  $X$ is the vector field of the generators $\{P_1 ,  P_2,J_{12}\}$
given in table \ref{table7}, the expression (\ref{ee}) vanishes, in plain
concordance with their role as isometries of  $S^2_{[\k_1],\k_2}$. 
Let us see what happen with the three remaining generators of  
$\con_{\k_1,\k_2}$. The components of the dilation generator
 $D$ are
\be
D^\aa=-\frac{ \Sk_{\k_1}(\aa)}{ \Ck_{\k_1\k_2}(\yy)} \qquad
D^\yy=- { \Ck_{\k_1}(\aa)}{ \Sk_{\k_1\k_2}(\yy)} .
\label{ef}
\ee
By taking into account (\ref{zd}) we obtain
\bea
&&\partial_\aa D^\aa =-  \frac{ \Ck_{\k_1}(\aa)}{ \Ck_{\k_1\k_2}(\yy)} 
\qquad
\partial_\yy D^\aa=-\k_1\k_2 \frac{ \Sk_{\k_1}(\aa)\Sk_{\k_1\k_2}(\yy)}{
\Ck^2_{\k_1\k_2}(\yy)} \cr
&&\partial_\aa D^\yy =\k_1{ \Sk_{\k_1}(\aa)}{ \Sk_{\k_1\k_2}(\yy)} 
\qquad
\partial_\yy D^\yy=-{ \Ck_{\k_1}(\aa)}{ \Ck_{\k_1\k_2}(\yy)} 
\label{eg}
\eea
and by substituting this result in (\ref{ee}) we finally have
\be
L_D g_1=-2{ \Ck_{\k_1}(\aa)}{ \Ck_{\k_1\k_2}(\yy)}\left(
\Ck^2_{\k_1\k_2}(\yy)\dd\aa^2+\k_2\dd\yy^2 \right)=
-2{ \Ck_{\k_1}(\aa)}{ \Ck_{\k_1\k_2}(\yy)}\, g_1 .
\label{eh}
\ee
Therefore, the Lie derivative of the main metric along the dilation
generator  is proportional to $g_1$ by a {\it conformal factor}:
\be
\mu_D(a,y)=-2{ \Ck_{\k_1}(\aa)}{ \Ck_{\k_1\k_2}(\yy)}.
\label{ei}
\ee
 Likewise, we  find the conformal factors associated to the generators
$G_i$:
\be
\mu_{G_1}(\aa,\yy)= 2{ \Sk_{\k_1}(\aa)}{ \Ck_{\k_1\k_2}(\yy)} \qquad
 \mu_{G_2}(\aa,\yy)= 2\k_2 { \Sk_{\k_1\k_2}(\yy)}.
\label{ej}
\ee

For the the subsidiary metric, we recall that   $g_2$ is only
relevant whenever  $\k_2=0$, and in this case it is a metric defined only
in each foliation leaf  $\aa=\rm{constant}$, with expression 
$(\dd s^2)_{2}=
\dd\yy^2$. It is straightforward to check that
the results obtained starting from
$g_2$ fully coincide with those obtained for $g_1$ by means of the
contraction $\k_2\to 0$: $\mu_{P_1} =\mu_{P_2}=\mu_{J_{12}}= \mu_{G_2}=0$,
$\mu_D=-2{\Ck_{\k_1}(\aa)}$ and $\mu_{G_1}= 2{ \Sk_{\k_1}(\aa)}$; this
could also have been derived from the fact that $g_2$ is generically
proportional to $g_1$.

Hence we conclude  that both metrics  verify
\be
L_Xg_i=\mu_X(u^1,u^2)\, g_i 
 \qquad X\in\{P_i,J_{12},G_i,D\} \quad i=1,2 
\ee
where the conformal factors  $\mu_X(u^1,u^2)$ are   given in terms of
the three intrinsic geodesic and Weierstrass coordinates by
\bea
&&\mu_{P_1} =\mu_{P_2}=\mu_{J_{12}}=0 \cr
&&\mu_D =-2{ \Ck_{\k_1}(\aa)}{ \Ck_{\k_1\k_2}(\yy)}=
-2{ \Ck_{\k_1}(\xx)}{ \Ck_{\k_1\k_2}(\bb)}=
-2{ \Ck_{\k_1}(r)} \equiv -2 x^0\cr
&&\mu_{G_1} =2{ \Sk_{\k_1}(\aa)}{ \Ck_{\k_1\k_2}(\yy)}=
2{ \Sk_{\k_1}(\xx)} =
2{ \Sk_{\k_1}(r)} \Ck_{\k_2}(\te) \equiv 2 x^1\cr
&&\mu_{G_2} =2\k_2 { \Sk_{\k_1\k_2}(\yy)}=
2\k_2{ \Ck_{\k_1}(x)}\Sk_{\k_1\k_2}(\bb) =
2\k_2{ \Sk_{\k_1}(r)} \Sk_{\k_2}(\te) \equiv 2\k_2 x^2.
\label{el}
\eea
When  $\k_1=0$ we recover the known results concerning  ${\bf E}^2$ and
${\bf M}^{1+1}$ \cite{Doub}:
\bea
&&\mu_{P_1} =\mu_{P_2}=\mu_{J_{12}}=0  \qquad
\mu_{G_1}   =2\aa=2\xx=2 r\Ck_{\k_2}(\te) \cr
&&\mu_D =-2  \qquad
\mu_{G_2} =2\k_2\yy=2\k_2\bb=2\k_2 r\Sk_{\k_2}(\te). 
\eea
Notice that the conformal factors (\ref{el}) coincide with the factors
multiplying $\cal C$ in (\ref{xxd}); this could be expected since the
metrics in the CK space are related with the second-order Casimir $\cal C$
through the Killing--Cartan  form.


\newpage


\parskip=1ex
\oddsidemargin= 0cm 
\evensidemargin= 0cm
\parindent=1.5em
\textheight=23.0cm
\textwidth=16cm
\topmargin=-1.0cm


\begin{table}[ht]
{\footnotesize
 \noindent
\caption{{The nine two-dimensional CK spaces
$S^2_{[\k_1],\k_2}=SO_{\k_1,\k_2}(3)/SO_{\k_2}(2)$.}}
\label{table1}
\smallskip
\noindent\hfill
\begin{tabular}{lll}
\hline
\\[-8pt]
Elliptic:\quad ${\bf S}^2$&Euclidean:\quad ${\bf E}^2$&Hyperbolic:\quad
${\bf H}^2$\\
$S^2_{[+],+}=SO(3)/SO(2)$&
$S^2_{[0],+}=ISO(2)/SO(2)$&
$S^2_{[-],+}=SO(2,1)/SO(2)$\\[12pt]
Oscillating NH:\quad ${\bf NH}_+^{1+1}$&Galilean:\quad ${\bf
G}^{1+1}$&Expanding NH:\quad ${\bf NH}_-^{1+1}$\\ 
(Co-Euclidean)&
&(Co-Minkowskian)\\
$S^2_{[+],0}=ISO(2)/ISO(1)$&
$S^2_{[0],0}=IISO(1)/ISO(1)$&
$S^2_{[-],0}=ISO(1,1)/ISO(1)$\\[12pt]
Anti-de Sitter:\quad ${\bf AdS}^{1+1}$&Minkowskian:\quad ${\bf
M}^{1+1}$& De Sitter:\quad ${\bf dS}^{1+1}$\\ 
(Co-Hyperbolic)& &(Doubly Hyperbolic)\\
$S^2_{[+],-}=SO(2,1)/SO(1,1)$&
$S^2_{[0],-}=ISO(1,1)/SO(1,1)$&
$S^2_{[-],-}=SO(2,1)/SO(1,1)$\\[6pt]
\hline
\end{tabular}\hfill}
\end{table}

\begin{table}[ht]
{\footnotesize
 \noindent
\caption{{Parametrized expressions for Weierstrass coordinates, metric,
canonical  connection and area element for
$S^2_{[\k_1],\k_2}$ given in the three systems of geodesic coordinates. The
subsidiary metric $(\dd s^2)_2$ is only relevant whenever $\k_2=0$.}}
\label{table2}
\medskip
\noindent\hfill
\begin{tabular}{lll}
\hline
\\[-6pt]
Parallel I $(\aa,\yy)$&Parallel II $(\xx,\bb)$&Polar $(r,\te)$\\[6pt]
$ x^0 =\Ck_{\k_1}(\aa)\Ck_{\k_1\k_2}(\yy)$ &
$x^0 =\Ck_{\k_1}(\xx)\Ck_{\k_1\k_2}(\bb)$&$x^0 =\Ck_{\k_1}(r)$\\
$  x^1 =\Sk_{\k_1}(\aa)\Ck_{\k_1\k_2}(\yy)$& 
$x^1 =\Sk_{\k_1}(\xx)$&$x^1=\Sk_{\k_1}(r)\Ck_{\k_2}(\te)$\\
$x^2 =\Sk_{\k_1\k_2}(\yy)$& 
$x^2 =\Ck_{\k_1}(\xx)\Sk_{\k_1\k_2}(\bb)$&
$ x^2 =\Sk_{\k_1}(r)\Sk_{\k_2}(\te)$\\[6pt]
$(\dd s^2)_1=\Ck_{\k_1\k_2}^2 (\yy) \dd \aa^2+\k_2\,  \dd\yy^2$&
$(\dd s^2)_1=\dd\xx^2+\k_2\Ck_{\k_1}^2(\xx) \dd\bb^2$&
$(\dd s^2)_1=\dd r^2+\k_2\Sk_{\k_1}^2 (r)\dd \te^2$\\[3pt]
$(\dd s^2)_2= \dd\yy^2$\ \ for\ \ $\aa=\aa_0$&
$(\dd s^2)_2=\Ck_{\k_1}^2(\xx) \dd\bb^2$\ \ for\ \
$\xx=\xx_0$&
$(\dd s^2)_2 =\Sk_{\k_1}^2(r) \dd\te^2$\ \ for\ \
$r=r_0$\\[6pt]
$\Gamma_{\aa\aa}^\yy =\k_1\Sk_{\k_1\k_2}(\yy)\Ck_{\k_1\k_2}(\yy)$& 
$\Gamma_{\bb\bb}^\xx =\k_1\k_2\Sk_{\k_1}(\xx)\Ck_{\k_1}(\xx)$& 
$\Gamma_{\te\te}^r =-\k_2\Sk_{\k_1}(r)\Ck_{\k_1}(r)$\\[3pt]
 $\Gamma_{\aa\yy}^\aa =-\k_1\k_2\Tk_{\k_1\k_2}(\yy)$& 
$\Gamma_{\bb\xx}^\bb =-\k_1\Tk_{\k_1}(\xx)$& 
$\Gamma_{\te r}^\te =  1/{\Tk_{\k_1}(r)}$\\[6pt]
$\dd\area=\Ck_{\k_1\k_2}  (\yy)\, \dd \aa\wedge \dd\yy$&
$\dd\area=\Ck_{\k_1}(\xx)    \,\dd \xx\wedge\dd\bb$&
$\dd\area=\Sk_{\k_1}(r)   \,\dd r\wedge\dd\te$\\[4pt]
\hline
\end{tabular}\hfill}
\end{table}

\begin{table}[t]
{\footnotesize
 \noindent
\caption{{Particularized expressions for Weierstrass coordinates, metric,
canonical  connection and area element for the nine spaces
$S^2_{[\k_1],\k_2}$ in geodesic parallel I coordinates $(a,y)$. In the
three Riemannian spaces (upper line)
$\k_1, \k_2$ are normalized to $\k_1\in\{1, 0, -1\}$
and $\k_2=1$. In the six spacetimes,
$\k_1=\pm 1/\tau^2$, $\k_2=-1/c^2$,  $a\equiv t$ is the time coordinate
and $y$ is the space one.}}
\label{table3}
\medskip
\noindent\hfill
\begin{tabular}{lll}
\hline
\\[-6pt]
${\bf S}^2=S^2_{[+],+}$&${\bf
E}^2=S^2_{[0],+}$&
${\bf H}^2=S^2_{[-],+}$\\[6pt]
$x^0 =\cos \aa \,\cos \yy$  &$x^0  =1$ &$x^0 =\cosh \aa\,\cosh \yy$\\
$x^1 =\sin \aa \, \cos \yy$  &$ x^1 =\aa$ &$   x^1=\sinh \aa\, \cosh
\yy$\\ 
$x^2 =\sin \yy $  & $x^2 =\yy $&  $ x^2 =\sinh \yy$\\[4pt]
$(\dd s^2)_1=\cos^2 \yy \  \dd \aa^2+ \dd\yy^2$&
$(\dd s^2)_1= \dd \aa^2+ \dd\yy^2$&
$(\dd s^2)_1=\cosh^2 \yy \ \dd \aa^2+ \dd\yy^2$\\[4pt] 
$ \Gamma_{\aa\aa}^\yy =\sin \yy\,\cos \yy $& 
 $ \Gamma_{\aa\aa}^\yy =0$ &   
$  \Gamma_{\aa\aa}^\yy =-\sinh \yy\,\cosh \yy$\\[2pt]
$\Gamma_{\aa\yy}^\aa =-\tan \yy$ &  
$\Gamma_{\aa\yy}^\aa =0 $&  
 $ \Gamma_{\aa\yy}^\aa =\tanh \yy$\\[4pt]
$\dd\area=\cos \yy \,\dd \aa\wedge\dd\yy$&
$\dd\area=  \dd \aa\wedge\dd\yy$&
$\dd\area=\cosh \yy\,  \dd \aa\wedge\dd\yy$\\[12pt]
 ${\bf NH}_+^{1+1}=S^2_{[+1/\tau^2],0}$&${\bf
G}^{1+1}=S^2_{[0],0}$&
${\bf NH}_-^{1+1}=S^2_{[-1/\tau^2],0}$\\[6pt] 
$x^0=\cos (t/\tau)$  &$x^0  =1$ &$x^0=\cosh (t/\tau)$\\
$x^1=\tau \sin (t/\tau)$  &$ x^1 =t$ &$   x^1=\tau \sinh (t/\tau)$\\ 
$x^2 =  y $  & $x^2 =y $&  $ x^2 =  y$\\[4pt]
$(\dd s^2)_1=  \dd t^2 $&
$(\dd s^2)_1=  \dd t^2 $&
$(\dd s^2)_1=  \dd t^2 $\\ 
$(\dd s^2)_2=  \dd \yy^2$\qquad
$t=t_0$&
$(\dd s^2)_2=  \dd \yy^2 $\quad
$t=t_0$&
$(\dd s^2)_2=  \dd \yy^2 $\qquad
$t=t_0$\\[4pt] 
$\Gamma_{tt}^\yy= \frac{1}{\tau^2}\, \yy \qquad  \Gamma_{t\yy}^t =0 $
& $\Gamma_{tt}^\yy=0\qquad  \Gamma_{t\yy}^t =0  $&  
 $ \Gamma_{tt}^\yy=-  \frac{1}{\tau^2}\, \yy \quad  \Gamma_{t\yy}^t
=0$\\[4pt]
$\dd\area= \dd t\wedge\dd\yy$&
$\dd\area=  \dd t\wedge\dd\yy$&
$\dd\area= \dd t\wedge\dd\yy$\\[12pt]
 ${\bf AdS}^{1+1}=S^2_{[+1/\tau^2],-1/c^2}$&
${\bf M}^{1+1}=S^2_{[0],-1/c^2}$&${\bf
dS}^{1+1}=S^2_{[-1/\tau^2],-1/c^2}$\\ [6pt]
$x^0 =\cos (t/\tau)\cosh (\yy/c\tau)$  &$x^0  =1$ &$x^0 =\cosh
(t/\tau) \cos (\yy/c\tau)$\\
$x^1 =\tau\sin (t/\tau)\cosh (\yy/c\tau)$  &$ x^1 =t$ &$   x^1=\tau\sinh
(t/\tau)\cos (\yy/c\tau)$\\ 
$x^2 =c\tau\sinh (\yy/c\tau)$&$x^2 =y $&  $ x^2 =c\tau\sin
(\yy/c\tau)$\\[4pt]
$(\dd s^2)_1=\cosh^2 (\yy/c\tau)   \dd t^2- \frac 1{c^2}\,\dd\yy^2$&
$(\dd s^2)_1= \dd t^2- \frac 1{c^2}\,\dd\yy^2$&
$(\dd s^2)_1=\cos^2 (\yy/c\tau)    \dd t^2- \frac 1{c^2}\,\dd\yy^2$\\[4pt] 
$ \Gamma_{tt}^\yy =\frac{c}{\tau}\sinh (\yy/c\tau)\cosh (\yy/c\tau) $& 
 $ \Gamma_{tt}^\yy =0$ &   
$  \Gamma_{tt}^\yy =-\frac{c}{\tau}\sin (\yy/c\tau)\cos
(\yy/c\tau)$\\[2pt]
$\Gamma_{t\yy}^t =\frac 1{c\tau}\tanh (\yy/c\tau)$ &  
$\Gamma_{t\yy}^t =0 $&  
 $ \Gamma_{t\yy}^t =-\frac 1{c\tau}\tan (\yy/c\tau)$\\[4pt]
$\dd\area=\cosh(\yy/c\tau) \,\dd t\wedge\dd\yy$&
$\dd\area=  \dd t\wedge\dd\yy$&
$\dd\area=\cos(\yy/c\tau)\,  \dd t\wedge\dd\yy$\\[6pt]
\hline
\end{tabular}\hfill}
\end{table}

\begin{table}[t]
{\footnotesize
 \noindent
\caption{{Same information as in table \ref{table3}, with the same
conventions, for the nine spaces $S^2_{[\k_1],\k_2}$ in geodesic polar
coordinates $(r,\te)$. In the six spacetimes, $r$  is a time-like length
(a proper time), while $\te\equiv\chi$ is a rapidity.}}
\label{table4}
\medskip
\noindent\hfill
\begin{tabular}{lll}
\hline
\\[-6pt]
${\bf S}^2=S^2_{[+],+}$&${\bf
E}^2=S^2_{[0],+}$&
${\bf H}^2=S^2_{[-],+}$\\[6pt]
$x^0 =\cos r $  &$x^0  =1$ &$x^0 =\cosh r$\\
$x^1 =\sin r \,\cos \te $  &$ x^1 =r\cos \te $ &$   x^1=\sinh r\,\cos
\te$\\ 
$x^2 =\sin r \,\sin \te$  & $x^2 =r \sin \te$&  $ x^2 =\sinh r \,\sin
\te$\\[4pt]
$(\dd s^2)_1=\dd r^2+ \sin^2 r \   \dd\te^2$&
$(\dd s^2)_1= \dd r^2+   r^2\,   \dd\te^2$&
$(\dd s^2)_1=\dd r^2+ \sinh^2 r \   \dd\te^2$\\[4pt] 
$\Gamma_{\te\te}^r =-\sin r\,\cos r$ &  
$\Gamma_{\te\te}^r =-r $&  
 $ \Gamma_{\te\te}^r =-\sinh r\,\cosh r $\\[2pt]
$ \Gamma_{\te r}^\te =1/{\tan r} $& 
 $ \Gamma_{\te r}^\te  =1/r $ &   
$ \Gamma_{\te r}^\te =1/{\tanh r} $\\[4pt]
$\dd\area=\sin r\,   \dd r\wedge\dd\te$&
$\dd\area=  r\,   \dd r\wedge\dd\te$&
$\dd\area=\sinh r\,   \dd r\wedge\dd\te$\\[12pt]
${\bf NH}_+^{1+1}=S^2_{[+1/\tau^2],0}$&${\bf
G}^{1+1}=S^2_{[0],0}$&
${\bf NH}_-^{1+1}=S^2_{[-1/\tau^2],0}$\\[6pt] 
$x^0=\cos (r/\tau)$  &$x^0  =1$ &$x^0=\cosh (r/\tau)$\\
$x^1=\tau \sin (r/\tau)$  &$ x^1 =r$ &$   x^1=\tau \sinh (r/\tau)$\\ 
$x^2 = \tau \sin (r/\tau) \chi$  & $x^2 = r \chi$&  $ x^2 = \tau \sinh
(r/\tau)\chi$\\[4pt]
$(\dd s^2)_1=  \dd r^2 $&
$(\dd s^2)_1=  \dd r^2 $&
$(\dd s^2)_1= \dd r^2 $\\ 
$(\dd s^2)_2=\tau^2 \sin^2(r/\tau)    \dd\chi^2$\quad 
$r=r_0$&
$(\dd s^2)_2=   r^2\,   \dd\chi^2$\quad
$r=r_0$&
$(\dd s^2)_2= \tau^2 \sinh^2 (r/\tau)  \dd\chi^2$\quad 
$r=r_0$\\[4pt] 
$\displaystyle{ \Gamma_{\chi\chi}^r=0\quad \Gamma_{\chi r}^\chi=\frac
1 {\tau{\tan (r/\tau)}} }$  &  $\Gamma_{\chi\chi}^r=0\quad\Gamma_{\chi
r}^\chi=1/r$&  
$\displaystyle{  \Gamma_{\chi\chi}^r=0 \quad \Gamma_{\chi r}^\chi=\frac
1{\tau{\tanh (r/\tau)}} }$ \\[8pt]
$\dd\area=\tau\sin(r/\tau)\,   \dd r\wedge\dd\chi$&
$\dd\area=  r\,   \dd r\wedge\dd\chi$&
$\dd\area=\tau\sinh (r/\tau)\,   \dd r\wedge\dd\chi$\\[12pt]
 ${\bf AdS}^{1+1}=S^2_{[+1/\tau^2],-1/c^2}$&
${\bf M}^{1+1}=S^2_{[0],-1/c^2}$&${\bf
dS}^{1+1}=S^2_{[-1/\tau^2],-1/c^2}$\\ [6pt]
$x^0 =\cos (r/\tau)$  &$x^0  =1$ &$x^0 =\cosh(r/\tau)$\\
$x^1 =\tau\sin (r/\tau)\cosh (\chi/c)$  &$ x^1 =r \cosh (\chi/c)$ &$  
x^1=\tau\sinh (r/\tau)\cosh (\chi/c)$\\ 
$x^2 =c\tau\sin (r/\tau)\sinh (\chi/c)$&$x^2 =c\, r\sinh (\chi/c) $&  $ x^2
=c\tau\sinh (r/\tau) \sinh (\chi/c)$\\[4pt]
$(\dd s^2)_1=\dd r^2- \frac {\tau^2}{c^2}\sin^2 (r/\tau)\dd\chi^2$&
$(\dd s^2)_1= \dd r^2- \frac {1}{c^2}r^2\,\dd\chi^2$&
$(\dd s^2)_1=\dd r^2- \frac {\tau^2}{c^2}\sinh^2 (r/\tau)\dd\chi^2$\\[4pt] 
$\Gamma_{\chi\chi}^r =\frac {\tau}{c^2}\, \sin (r/\tau) \cos (r/\tau)$
&  $\Gamma_{\chi\chi}^r =\frac {1}{c^2}\, r $&  
 $ \Gamma_{\chi\chi}^r =\frac {\tau}{c^2}\, \sinh (r/\tau) \cosh
(r/\tau)$\\[2pt]
$\displaystyle{ \Gamma_{\chi r}^\chi=\frac
1 {\tau{\tan (r/\tau)}} }$& 
 $ \Gamma_{\chi r}^\chi  =1/r $ &   
$\displaystyle{\Gamma_{\chi r}^\chi=\frac
1{\tau{\tanh (r/\tau)}} }$\\[8pt]
$\dd\area=\tau\sin (r/\tau)\,   \dd r\wedge\dd\chi$&
$\dd\area=  r\,   \dd r\wedge\dd\chi$&
$\dd\area=\tau\sinh (r/\tau)\,   \dd r\wedge\dd\chi$\\[6pt]
\hline
\end{tabular}\hfill}
\end{table}

\begin{table}[t]
{\footnotesize
 \noindent
\caption{{Generic form for the equations of  cycles,
(including as particular cases geodesics, equidistants and circles) in the
spaces $S^2_{[\k_1],\k_2}$  in the three geodesic coordinate systems. The 
circle equations can also be  read as expressions for the finite
(time-like) distance between two points, and the equidistant equations as
expressions for the finite (time-like) distance between a point and a
(space-like) line.}}
\label{table5}
\medskip
\noindent\hfill
\begin{tabular}{ll}
\hline
\\[-6pt]
&{Geodesic parallel I coordinates $(\aa,\yy)$}\\[6pt]
Cycles&$\displaystyle{\alpha_0\Ck_{\k_1}(\aa)+\alpha_1\Sk_{\k_1}(\aa)}=\frac
{\alpha}{\Ck_{\k_1\k_2}(\yy)} - \alpha_2
\Tk_{\k_1\k_2}(\yy)={\alpha}\,{\Ck_{-\k_1\k_2}(\yyl)} - \alpha_2
\Sk_{-\k_1\k_2}(\yyl)$\\[10pt]
&\   =
$\displaystyle{\left(\frac{\alpha\sqrt{\k_1\k_2}-\alpha_2}
{2\sqrt{\k_1\k_2}} \right)
\left(\frac{1+\sqrt{\k_1\k_2}\,\Tk_{\k_1\k_2}(\yy/2)}
{1-\sqrt{\k_1\k_2}\,\Tk_{\k_1\k_2}(\yy/2)} \right) +
\left(\frac{\alpha\sqrt{\k_1\k_2}+\alpha_2}{2\sqrt{\k_1\k_2}} \right)
\left(
\frac{1-\sqrt{\k_1\k_2}\,\Tk_{\k_1\k_2}(\yy/2)}
{1+\sqrt{\k_1\k_2}\,\Tk_{\k_1\k_2}(\yy/2)}    \right) } $ \\[12pt] 
Geodesics&$\Tk_{\k_1\k_2}(\yy)=
\beta_0\Ck_{\k_1}(\aa)+\beta_1\Sk_{\k_1}(\aa)$
\qquad  {and}\qquad  $\aa=\aa_0$\\[4pt]
Equidistants& 
$\displaystyle{
\Sk^2_{\k_1}(d)
=\k_2\,\frac{\left\{\Sk_{\k_1\k_2}(\yy) -\Ck_{\k_1\k_2}(\yy)(\beta_0
\Ck_{\k_1}(\aa)+\beta_1 \Sk_{\k_1}(\aa))
\right\}^2}{1+\k_2\beta_1^2+\k_1\k_2\beta_0^2}\qquad  (\k_2\ne
0)}$\\[10pt] 
Circles& 
 $\Ck_{\k_1}(\radio)=
\Ck_{\k_1\k_2}(\yy)
\Ck_{\k_1\k_2}(\yy_0)\Ck_{\k_1}(\aa-\aa_0)
+\k_1\k_2\Sk_{\k_1\k_2}(\yy)\Sk_{\k_1\k_2}(\yy_0)$
\\[2pt] 
 &$\Vk_{\k_1}(\radio)=\Ck_{\k_1\k_2}(\yy)
\Ck_{\k_1\k_2}(\yy_0)\Vk_{\k_1}(\aa-\aa_0)+
\k_2\Vk_{\k_1\k_2}(\yy-\yy_0)$\\[4pt]
\hline
\\[-6pt]
&{Geodesic parallel II coordinates $(\xx,\bb)$}\\[6pt]
Cycles&$
\displaystyle{\alpha_0\Ck_{\k_1\k_2}(\bb)+\alpha_2\Sk_{\k_1\k_2}(\bb)=\frac
{\alpha}{\Ck_{\k_1}(\xx)}-
\alpha_1\Tk_{\k_1}(\xx)={\alpha}\,{\Ck_{-\k_1}(\xxl)} - \alpha_1
\Sk_{-\k_1}(\xxl)
}$\\[10pt]
&\   =
$\displaystyle{\left(\frac{\alpha\sqrt{\k_1}-\alpha_1}{2\sqrt{\k_1}}\right) 
\left(
\frac{1+\sqrt{\k_1}\,\Tk_{\k_1}(\xx/2)}{1-\sqrt{\k_1}\,\Tk_{\k_1}(\xx/2)}   
\right) +
\left(\frac{\alpha\sqrt{\k_1}+\alpha_1}{2\sqrt{\k_1}} \right)
\left(
\frac{1-\sqrt{\k_1}\,\Tk_{\k_1}(\xx/2)}{1+\sqrt{\k_1}\,\Tk_{\k_1}(\xx/2)}   
 \right) }  $ \\ [12pt]
Geodesics&$\Tk_{\k_1}(\xx)=\beta_0\Ck_{\k_1\k_2}(\bb)
+\beta_2\Sk_{\k_1\k_2}(\bb)$
\qquad and \qquad $\bb=\bb_0$\\[6pt]
Equidistants& 
$\displaystyle{
\Sk^2_{\k_1}(d)
=\k_2\,\frac{\left\{\Sk_{\k_1}(\xx) -\Ck_{\k_1}(\xx)(\beta_0
\Ck_{\k_1\k_2}(\bb)+\beta_2 \Sk_{\k_1\k_2}(\bb))
\right\}^2}{\k_2+\beta_2^2+\k_1\k_2\beta_0^2}\qquad  (\k_2\ne
0)}$\\[10pt] 
Circles& 
 $\Ck_{\k_1}(\radio)=
\Ck_{\k_1}(\xx)
\Ck_{\k_1}(\xx_0)\Ck_{\k_1\k_2}(\bb-\bb_0)
+\k_1 \Sk_{\k_1 }(\xx)\Sk_{\k_1 }(\xx_0)$
\\[2pt] 
 &$\Vk_{\k_1}(\radio)=\k_2\Ck_{\k_1}(\xx)
\Ck_{\k_1}(\xx_0)\Vk_{\k_1\k_2}(\bb-\bb_0)
+\Vk_{\k_1}(\xx-\xx_0)$\\[4pt]
\hline
\\[-6pt]
&{Geodesic polar coordinates $(r,\te)$}\\[6pt]
Cycles&$\displaystyle{
\alpha_1\Ck_{\k_2}(\te)+\alpha_2\Sk_{\k_2}(\te)=\frac
{\alpha}{\Sk_{\k_1}(r)}-
 \frac  {\alpha_0}{\Tk_{\k_1}(r)}
=\k_1\left(\frac{\alpha+\alpha_0}{2}\right)\Tk_{\k_1}(r/2)
+\left(\frac{\alpha-\alpha_0}{2}\right)\frac
1{\Tk_{\k_1}(r/2)}}$\\[8pt]
Geodesics&$\displaystyle{ \frac  {1}{\Tk_{\k_1}(r)}
=\beta_1\Ck_{\k_2}(\te)+\beta_2\Sk_{\k_2}(\te)}$\qquad and
\qquad$\te=\te_0$\\[6pt]
Equidistants& 
$\displaystyle{
\Sk^2_{\k_1}(d)
=\k_2\,\frac{\left\{\Ck_{\k_1}(r) -\Sk_{\k_1}(r)(\beta_1
\Ck_{\k_2}(\phi)+\beta_2 \Sk_{\k_2}(\phi))
\right\}^2}{\beta_2^2+\k_2\beta_1^2+\k_1\k_2}\qquad  (\k_2\ne
0)}$\\[10pt] 
 Circles& 
 $\Ck_{\k_1}(\radio)=\Ck_{\k_1}(r)\Ck_{\k_1}(r_0)+\k_1\Sk_{\k_1}(r)
\Sk_{\k_1}(r_0)\Ck_{\k_2}(\te-\te_0)$ \\[2pt]
  &$\Vk_{\k_1}(\radio)=\Vk_{\k_1}(r-r_0)+ \k_2\Sk_{\k_1}(r)
\Sk_{\k_1}(r_0)\Vk_{\k_2}(\te-\te_0)$\\[4pt]
\hline
\end{tabular}\hfill}
\end{table}

\begin{table}[t]
{\footnotesize
 \noindent
\caption{{Particularized equations for geodesics (besides
$\aa=\aa_0$) and circles (constant geodesic `time-like' distance $\radio$
to a fixed center $(\aa_0,\yy_0)$) for the  nine spaces
$S^2_{[\k_1],\k_2}$ in geodesic parallel I coordinates and with the same
conventions as in table
\ref{table3}.  }}
\label{table6}
\medskip
\noindent\hfill
\begin{tabular}{lll}
\hline
\\[-6pt]
${\bf S}^2=S^2_{[+],+}$&${\bf
E}^2=S^2_{[0],+}$&
${\bf H}^2=S^2_{[-],+}$\\[6pt]
$\tan\yy=\beta_0 \cos\aa +\beta_1\sin\aa $&
$ \yy=\beta_0   +\beta_1 \aa $&
$\tanh\yy=\beta_0 \cosh\aa +\beta_1\sinh\aa $\\[4pt]
$\cos \radio=\cos\yy\cos\yy_0\cos(\aa-\aa_0)$&
$ \radio^2=(\aa-\aa_0)^2$&
$\cosh \radio=\cosh\yy\cosh\yy_0\cosh(\aa-\aa_0)$\\[2pt]
$\quad +\sin\yy\sin\yy_0$&
$\quad +(\yy-\yy_0)^2 $&
$\quad  -\sinh\yy\sinh\yy_0$\\[12pt]
 ${\bf NH}_+^{1+1}=S^2_{[+1/\tau^2],0}$&${\bf
G}^{1+1}=S^2_{[0],0}$&
${\bf NH}_-^{1+1}=S^2_{[-1/\tau^2],0}$\\[6pt] 
$ \yy= \beta_0 \cos (t/\tau)+\beta_1\tau  \sin (t/\tau) $&
$\yy= \beta_0 +\beta_1 t   $&
$\yy= \beta_0 \cosh (t/\tau)+\beta_1\tau  \sinh (t/\tau)  $\\[4pt]
$\cos (\radio/\tau)=\cos ((t-t_0)/\tau)$&
$ \radio^2=(t-t_0)^2$&
$\cosh (\radio/\tau)=\cosh ((t-t_0)/\tau)$\\[12pt]
 ${\bf AdS}^{1+1}=S^2_{[+1/\tau^2],-1/c^2}$&
${\bf M}^{1+1}=S^2_{[0],-1/c^2}$&${\bf
dS}^{1+1}=S^2_{[-1/\tau^2],-1/c^2}$\\ [6pt]
$c\tau \tanh(\yy/c\tau)= \beta_0 \cos (t/\tau)+\beta_1\tau  \sin (t/\tau)
$&$\yy= \beta_0 +\beta_1 t   $&
$c\tau \tan(\yy/c\tau)= \beta_0 \cosh (t/\tau)+\beta_1\tau  \sinh
(t/\tau)  $\\[4pt]
$\cos (\radio/\tau) = 
-\sinh (\yy/c\tau)\sinh(\yy_0/c\tau) $&
$ \radio^2=(t-t_0)^2$&
$\cosh (\radio/\tau) = 
\sin (\yy/c\tau)\sin(\yy_0/c\tau) $\\[2pt]
$\quad +\cosh(\yy/c\tau)\cosh(\yy_0/c\tau) \cos((t-t_0)/\tau) 
$&$\quad -\frac 1{c^2}(\yy-\yy_0)^2  $&$\quad +
\cos(\yy/c\tau)\cos(\yy_0/c\tau) \cosh((t-t_0)/\tau)$\\[6pt]
\hline
\end{tabular}\hfill}
\end{table}

\begin{table}[t]
{\footnotesize
 \noindent
\caption{{Parametrized differential realizations for the generators
$\{P_i,J_{12},D,L_i, G_i, R_i\}$ of conformal algebras
$\con_{\k_1,\k_2}$. Note the three pairs
$P_1, L_2; P_2, L_1; J, D$ with  a specially simple expression in each
of the three coordinate systems}.  }
\label{table7}
\medskip
\noindent\hfill
\begin{tabular}{l}
\hline
\\[-6pt]
Geodesic parallel I coordinates $(\aa,\yy)$\\[6pt]
$\displaystyle{P_1=-\partial_\aa\qquad 
P_2=-\k_1\k_2\Sk_{\k_1}(\aa)\Tk_{\k_1\k_2}(\yy)
\,\partial_\aa-\Ck_{\k_1}(\aa)\,\partial_\yy}$\\[4pt]
$\displaystyle{J_{12}=\k_2\Ck_{\k_1}(\aa)\Tk_{\k_1\k_2}(\yy)
\,\partial_\aa-\Sk_{\k_1}(\aa)\,\partial_\yy\qquad
D=-\frac{\Sk_{\k_1}(\aa)}{\Ck_{\k_1\k_2}(\yy)}\,
\partial_\aa-\Ck_{\k_1}(\aa)\Sk_{\k_1\k_2}(\yy)\,\partial_\yy}$\\[6pt]
$\displaystyle{L_1=-\frac{\Ck_{\k_1}(\aa)}{\Ck_{\k_1\k_2}(\yy)}
\,\partial_\aa +\k_1\Sk_{\k_1}(a)\Sk_{\k_1\k_2}(\yy)\,\partial_\yy
\qquad 
L_2=-\Ck_{\k_1\k_2}(\yy)\,\partial_\yy}$\\[10pt]
$\displaystyle{G_1=\frac 1{\Ck_{\k_1\k_2}(\yy)}\left(\Vk_{\k_1}(\aa)-\k_2
\Vk_{\k_1\k_2}(\yy)\right)\partial_\aa+
\Sk_{\k_1}(\aa)\Sk_{\k_1\k_2}(\yy)\,\partial_\yy}$\\[10pt]
$\displaystyle{G_2= \k_2\Sk_{\k_1}(\aa)\Tk_{\k_1\k_2}(\yy)\,\partial_\aa-
\left(\Vk_{\k_1}(\aa)-\k_2
\Vk_{\k_1\k_2}(\yy)\right)\partial_\yy}$\\[6pt]
$\displaystyle{
R_1=-\frac 12 \left(1+ \frac{\Ck_{\k_1}(\aa)}{\Ck_{\k_1\k_2}(\yy)}\right)
\partial_\aa+\frac
12 \k_1\Sk_{\k_1}(\aa)\Sk_{\k_1\k_2}(\yy)\,\partial_\yy}$\\[8pt]
$\displaystyle{R_2=-\frac 12 \k_1\k_2 
\Sk_{\k_1}(\aa)\Tk_{\k_1\k_2}(\yy)\,\partial_\aa
-\frac 12 \left(\Ck_{\k_1}(\aa)+
\Ck_{\k_1\k_2}(\yy)\right)\partial_\yy}$\\[8pt]
\hline
\\[-6pt]
Geodesic parallel II coordinates $(\xx,\bb)$\\[6pt]
$\displaystyle{P_1=-\Ck_{\k_1\k_2}(\bb)\,\partial_\xx-\k_1
\Tk_{\k_1}(\xx)\Sk_{\k_1\k_2}(\bb)
\,\partial_\bb\qquad  P_2=-\partial_\bb }$\\[4pt]
$\displaystyle{J_{12}=\k_2\Sk_{\k_1\k_2}(\bb)\,\partial_\xx-
\Tk_{\k_1}(\xx) \Ck_{\k_1\k_2}(\bb)
\,\partial_\bb\qquad
D=-\Sk_{\k_1 }(\xx)\Ck_{\k_1\k_2}(\bb)\,\partial_\xx
-\frac{\Sk_{\k_1\k_2}(\bb)}{\Ck_{\k_1 }(\xx)}\,
\partial_\bb }$\\[8pt]
$\displaystyle{L_1=-\Ck_{\k_1}(\xx)\,\partial_\xx\qquad 
L_2=\k_1\k_2\Sk_{\k_1}(\xx)\Sk_{\k_1\k_2}(\bb)\,\partial_\xx
-\frac{\Ck_{\k_1\k_2}(\bb)}{\Ck_{\k_1 }(\xx)}\,\partial_\bb }$ \\[10pt]
$\displaystyle{G_1= \left(\Vk_{\k_1}(\xx) -\k_2
\Vk_{\k_1\k_2}(\bb)\right)\partial_\xx+
\Tk_{\k_1 }(\xx) \Sk_{\k_1\k_2}(\bb)\,\partial_\bb }$\\[6pt]
$\displaystyle{G_2= \k_2\Sk_{\k_1}(\xx)\Sk_{\k_1\k_2}(\bb)\,\partial_\xx-
\frac 1{\Ck_{\k_1 }(\xx)}\left(\Vk_{\k_1}(\xx)-\k_2
\Vk_{\k_1\k_2}(\bb)\right)\partial_\bb}$\\[10pt]
$\displaystyle{
R_1=
-\frac 12 \left(\Ck_{\k_1}(\xx)+
\Ck_{\k_1\k_2}(\bb)\right)\partial_\xx
-\frac 12 \k_1
\Tk_{\k_1}(\xx)\Sk_{\k_1\k_2}(\bb)\,\partial_\bb}$\\[8pt]
$\displaystyle{
R_2=\frac
12 \k_1\k_2\Sk_{\k_1}(\xx)\Sk_{\k_1\k_2}(\bb)\,\partial_\xx
-\frac 12 \left(1+ \frac{\Ck_{\k_1\k_2}(\bb)}{\Ck_{\k_1}(\xx)}\right)
\partial_\bb}$\\[10pt]
\hline
\\[-6pt]
Geodesic polar coordinates $(r,\te)$\\[6pt]
$\displaystyle{ P_1=-\Ck_{\k_2}(\te)\,\partial_r
+\frac {\Sk_{\k_2}(\te)}{\Tk_{\k_1}(r)}\,\partial_\te
\qquad
 P_2=-\k_2 \Sk_{\k_2}(\te)\,\partial_r
-\frac {\Ck_{\k_2}(\te)}{\Tk_{\k_1}(r)}\,\partial_\te}$\\[8pt]
$\displaystyle{ J_{12}=-\partial_\te\qquad 
D=-\Sk_{\k_1}(r)\,
\partial_r}$ \\[6pt]
$\displaystyle{ L_1=-\Ck_{\k_2}(\te)\Ck_{\k_1}(r)\,\partial_r
+\frac {\Sk_{\k_2}(\te)}{\Sk_{\k_1}(r)}\,\partial_\te
\qquad 
L_2=-\k_2\Sk_{\k_2}(\te)\Ck_{\k_1}(r)\,\partial_r
-\frac {\Ck_{\k_2}(\te)}{\Sk_{\k_1}(r)}\,\partial_\te }$\\[10pt]
$\displaystyle{G_1=  \Ck_{\k_2}(\te)\Vk_{\k_1}(r)\,\partial_r
+\Sk_{\k_2}(\te)\frac {\Vk_{\k_1}(r)}{\Sk_{\k_1}(r)}\,\partial_\te
}$\\[10pt]
$\displaystyle{ G_2=\k_2 \Sk_{\k_2}(\te)\Vk_{\k_1}(r)\,\partial_r
-\Ck_{\k_2}(\te)\frac {\Vk_{\k_1}(r)}{\Sk_{\k_1}(r)}\,\partial_\te}$
\\[6pt]
$\displaystyle{
R_1=-\frac 1{2\,\Tk_{\k_1}(r/2)}
 \left( \Ck_{\k_2}(\te) \Sk_{\k_1}(r)\,\partial_r
-\Sk_{\k_2}(\te)\,\partial_\te\right) }$\\[10pt]
$\displaystyle{R_2=-\frac 1{2\,\Tk_{\k_1}(r/2)}
 \left( \k_2 \Sk_{\k_2}(\te)  \Sk_{\k_1}(r)\,\partial_r
+\Ck_{\k_2}(\te)\,\partial_\te\right) }$\\[10pt]
\hline
\end{tabular}\hfill}
\end{table}

\begin{table}[t]
{\footnotesize
 \noindent
\caption{{The Laplace--Beltrami operator $\cal C$ giving rise to
differential Laplace and wave-type  equations  ${\cal C}\Phi=0$ 
in geodesic parallel I $\Phi(a,y)\equiv \Phi(t,\yy)$ and polar
$\Phi(r,\te)\equiv \Phi(r,\chi)$ coordinates for the nine CK
spaces. This operator has the corresponding conformal algebra symmetry,
$\con_{\k_1,\k_2}$, which is independent of $\k_1$ and isomorphic to
either $so(3,1)$, $iso(2,1)$ or $so(2,2)$; this algebra appears as a
heading in each entry.}}
\label{table8}
\medskip
\noindent\hfill
\begin{tabular}{lll}
\hline
\\[-6pt]
$so(3,1):\ {\bf S}^2=S^2_{[+],+}$&$so(3,1):\ {\bf
E}^2=S^2_{[0],+}$&
$so(3,1):\ {\bf H}^2=S^2_{[-],+}$\\[4pt]
{$\displaystyle{   
\frac{1}{\cos^2\yy}\, \partial_\aa^2+\partial_\yy^2-
\tan\yy\, \partial_\yy}$}&
{$\displaystyle{ \partial_\aa^2+\partial_\yy^2}$}&
{$\displaystyle{ \frac{1}{\cosh^2\yy}\, \partial_\aa^2+\partial_\yy^2+
\tanh\yy\, \partial_\yy}$}\\[8pt]
{$\displaystyle{\frac{1}{\sin^2 r}\, \partial_\te^2+\partial_r^2+
\frac{1}{\tan r}\, \partial_r }$}&{$\displaystyle{\frac{1}{r^2}\,
\partial_\te^2+
\partial_r^2+ \frac{1}{r}\, \partial_r}$}&{$\displaystyle{
\frac{1}{\sinh^2 r}\, \partial_\te^2+
\partial_r^2+ \frac{1}{\tanh r}\, \partial_r}$}
\\[25pt]
$iso(2,1):\ {\bf NH}_+^{1+1}=S^2_{[+1/\tau^2],0}$&$iso(2,1):\ {\bf
G}^{1+1}=S^2_{[0],0}$&
$iso(2,1):\ {\bf NH}_-^{1+1}=S^2_{[-1/\tau^2],0}$\\[4pt] 
{$\displaystyle{\partial_\yy^2}$}&{$\displaystyle{
\partial_\yy^2}$}&{$\displaystyle{\partial_\yy^2}$}\\[2pt]
{$\displaystyle{  \frac{1}{\tau^2\sin^2(r/\tau)}\, \partial_\chi^2}$}&
{$\displaystyle{ \frac{1}{r^2} \,\partial_\chi^2}$}&
{$\displaystyle{\frac{1}{\tau^2\sinh^2 (r/\tau)} \,\partial_\chi^2}$}
\\[25pt]
$so(2,2):\ {\bf AdS}^{1+1}=S^2_{[+1/\tau^2],-1/c^2}$&
$so(2,2):\ {\bf M}^{1+1}=S^2_{[0],-1/c^2}$&$so(2,2):\ {\bf
dS}^{1+1}=S^2_{[-1/\tau^2],-1/c^2}$\\ [4pt]
{$\displaystyle{\frac{-1}{c^2\cosh^2(\yy/c\tau)}\,
\partial_t^2+\partial_\yy^2+\frac{\tanh(\yy/c\tau)}{c\tau}\,  \partial_\yy
}$}&{$\displaystyle{ -\frac{1}{c^2}\,\partial_t^2+\partial_\yy^2 }$}&
{$\displaystyle{\frac{-1}{c^2\cos^2(\yy/c\tau)}\,
\partial_t^2+\partial_\yy^2-
\frac{\tan(\yy/c\tau)}{c\tau}\, \partial_\yy}$}\\[8pt]
{$\displaystyle{
\frac{1}{\tau^2\sin^2 (r/\tau)}\,
\partial_\chi^2
-\frac{1}{c^2}\, \partial_r^2- \frac{1}{c^2\tau\tan
(r/\tau)}\,\partial_r }$}& {$\displaystyle{\frac{1}{r^2}\,
\partial_\chi^2-\frac{1}{c^2}\,\partial_r^2-
\frac{1}{c^2 r}\,\partial_r }$}&{$\displaystyle{ \frac{1}{\tau^2\sinh^2
(r/\tau)}\, \partial_\chi^2
-\frac{1}{c^2}\,\partial_r^2- \frac{1}{c^2\tau\tanh (r/\tau)}\,
\partial_r }$}
\\[10pt]
\hline
\end{tabular}\hfill}
\end{table}

\begin{table}[h]
{\footnotesize
 \noindent
\caption{{Parametrized expressions for the conformal embedding
$S^2_{[\k_1],\k_2}
\longmapsto
\cons^2_{[\k_1],\k_2}$ in three coordinate systems. Coordinates $(s^\cnu,
s^\cnd, s^1, s^2)$ parametrize the conformal ambient space, and hence
the cone $\Gamma_0$,
while $(\xxtilde^+,\xxtilde^-=-1,\xxtilde^1,\xxtilde^2)$ are a
parametrization of the conformal space itself $\cons^2_{[\k_1],\k_2}
\equiv \widetilde{S}^2_{[+],\k_2}$. We have introduced the shorthand
notation $\kkpp=\frac{{1+\k_1\ell^2}}{2\k_1
\ell^2}$ and
$\kkmm=\frac{{1-\k_1\ell^2}}{2\k_1 \ell^2}$.}}
\label{table9}
\medskip
\noindent\hfill
\begin{tabular}{lll}
\hline
\\[-6pt]
Parallel I coordinates&Parallel II coordinates
 &Polar coordinates\\[2pt]
 $(\aa,\yy)\in S^2_{[\k_1],\k_2}\to (A,Y)\in
\cons^2_{[\k_1],\k_2}$& 
$(\xx,\bb)\in S^2_{[\k_1],\k_2}\to (X,B)\in
\cons^2_{[\k_1],\k_2}$ & $(r,\te)\in S^2_{[\k_1],\k_2}\to (R,\Phi)\in
\cons^2_{[\k_1],\k_2}$\\[8pt]
$\displaystyle{ s^\pm  =\pm 1-  \kkppmm \left\{ 1- 
\Ck_{\k_1 }(\ssa )
\Ck_{\k_1\k_2 }(\ssb)    \right\} }$ &
$\displaystyle{ s^\pm  =\pm1-   \kkppmm \left\{ 1- 
\Ck_{\k_1 }(\xx )
\Ck_{\k_1\k_2 }(\bb)    \right\} }$&$\displaystyle{ s^\pm =\pm1- \kkppmm(1-
\Ck_{\k_1 }(\rra) ) }$\\
$\displaystyle{ s^1=   \Sk_{\k_1 }(\ssa)\Ck_{\k_1\k_2 }
(\ssb)/\ell}$ &
$\displaystyle{ s^1=   \Sk_{\k_1 }(\xx)/\ell}$&$\displaystyle{  s^1=  
\Sk_{\k_1 }(\rra)\Ck_{\k_2 }(\rrb) /\ell }$\\
$\displaystyle{s^2 =   \Sk_{\k_1 \k_2}(\ssb)/\ell}$ &
$\displaystyle{s^2 =  \Ck_{\k_1}(\xx) \Sk_{\k_1 \k_2}(\bb)/\ell
}$&$\displaystyle{s^2 =   \Sk_{\k_1 }(\rra)\Sk_{\k_2 }(\rrb)/\ell }$\\[8pt]
$\displaystyle{\xxtilde^\cnu=\frac{  1-  \kkpp \{ 1- 
\Ck_{\k_1 }(\ssa )\Ck_{\k_1\k_2 }(\ssb)\} }
{ 1+ \kkmm\left\{ 1- 
\Ck_{\k_1 }(\ssa )
\Ck_{\k_1\k_2 }(\ssb)    \right\} }}$ &
$\displaystyle{\xxtilde^\cnu=\frac{  1-  \kkpp \{ 1- 
\Ck_{\k_1 }(\xx )
\Ck_{\k_1\k_2 }(\bb)  \} }
{ 1+ \kkmm\left\{ 1- 
\Ck_{\k_1 }(\xx )
\Ck_{\k_1\k_2 }(\bb)  \right\} }}$ &$\displaystyle{\xxtilde^\cnu=
\frac{\ell^2-\Tk^2_{\k_1}(\rra/2)}{\ell^2+\Tk^2_{\k_1}(\rra/2)} }$\\[9pt]
$\displaystyle{\xxtilde^1=\frac{\Sk_{\k_1 }(\ssa)\Ck_{\k_1\k_2 }
(\ssb)/\ell }
{ 1+ \kkmm\left\{ 1- 
\Ck_{\k_1 }(\ssa )
\Ck_{\k_1\k_2 }(\ssb)    \right\} }}$ &
$\displaystyle{\xxtilde^1=\frac{\Sk_{\k_1 }(\xx)/\ell}
{ 1+ \kkmm\left\{ 1- 
\Ck_{\k_1 }(\xx )
\Ck_{\k_1\k_2 }(\bb)  \right\} }}$&$\displaystyle{ 
\xxtilde^1=\frac{2\ell\,\Tk_{\k_1}(\rra/2)\Ck_{\k_2 }(\rrb)}
{\ell^2+\Tk^2_{\k_1}(\rra/2)} }$\\[9pt]
$\displaystyle{\xxtilde^2=\frac{ \Sk_{\k_1\k_2 }
(\ssb)/\ell }
{ 1+ \kkmm\left\{ 1- 
\Ck_{\k_1 }(\ssa )
\Ck_{\k_1\k_2 }(\ssb)    \right\} }}$ &
$\displaystyle{\xxtilde^2=\frac{\Ck_{\k_1}(\xx) \Sk_{\k_1 \k_2}(\bb)/\ell}
{ 1+ \kkmm\left\{ 1- 
\Ck_{\k_1 }(\xx ) \Ck_{\k_1\k_2 }(\bb)  \right\}
}}$&$\displaystyle{\xxtilde^2=\frac{2\ell\,\Tk_{\k_1}(\rra/2)\Sk_{\k_2
}(\rrb)}{\ell^2+\Tk^2_{\k_1}(\rra/2)}}$\\[16pt]
$\displaystyle{\tan A=
\frac{\Tk_{\k_1 }(\ssa) }{\ell} 
 \frac{1-\k_1 \Tk^2_{\k_1}(\rra/2)}
{1-\frac{1}{\ell^2}\Tk^2_{\k_1}(\rra/2)} }$&
$\displaystyle{ \sin X=\frac{\Sk_{\k_1 }(\xx)}{\ell}
 \frac{1+\k_1 \Tk^2_{\k_1}(\rra/2)}
{1+\frac{1}{\ell^2}\Tk^2_{\k_1}(\rra/2)}}$&
$\displaystyle{ \tan^2(R/2) =\frac{1}{\ell^2} \Tk^2_{\k_1}(\rra/2)}$\\[8pt]
$\displaystyle{\Sk_{\k_2 }(Y)=\frac{ \Sk_{\k_1\k_2 }
(\ssb)}{\ell }
 \frac{1+\k_1 \Tk^2_{\k_1}(\rra/2)}
{1+\frac{1}{\ell^2}\Tk^2_{\k_1}(\rra/2)}    }$&
$\displaystyle{\Tk_{\k_2 }(B)= \frac{\Tk_{\k_1\k_2 }(\bb) }{\ell} 
 \frac{1-\k_1 \Tk^2_{\k_1}(\rra/2)}
{1-\frac{1}{\ell^2}\Tk^2_{\k_1}(\rra/2)} }$&
$\displaystyle{ \Phi=\rrb}$\\[12pt]
\hline
\end{tabular}\hfill}
\end{table}

\begin{table}[t]
{\footnotesize
 \noindent
\caption{{The conformal embedding 
in parallel I coordinates for the nine CK spaces and for their
compactifications: $(\aa,\yy)\equiv (t,\yy)\in S^2_{[\k_1],\k_2}
\longmapsto (A,Y)\equiv (T,Y)\in \cons^2_{[\k_1],\k_2} \equiv
\widetilde{S}^2_{[+],\k_2}$. The length $\ell$ is chosen in such a manner
that $\k_1\ell^2\in\{1,0,-1\}$ and $\k_2\in\{1,0,-1/c^2\}$.}}
\label{table10}
\medskip
\noindent\hfill
\begin{tabular}{lll}
\hline
\\[-6pt]
${\bf S}^2\longmapsto {\bf S}^2$&${\bf
E}^2\longmapsto {\bf S}^2$&
${\bf H}^2\longmapsto {\bf S}^2$\\[2pt]
$\k_1\ell^2=1,\ \k_2=1$&$\k_1\ell^2=0,\ \k_2=1$&
$\k_1\ell^2=-1,\ \k_2=1$\\[6pt]
$\xxtilde^\cnu =\cos (\aa/\ell)  \cos (\yy/\ell)$ 
&$\displaystyle{\xxtilde^\cnu   =\frac{4 -  \left\{ (\ssa/\ell)^2 +  
(\ssb/\ell)^2\right\}}{4 +  \left\{ (\ssa/\ell)^2 +(\ssb/\ell)^2\right\}
}}$ &$\displaystyle{ \xxtilde^\cnu = \frac{1}{\cosh (\aa/\ell)
\cosh (\yy/\ell)}  } $\\[8pt]
$\xxtilde^1 =\sin (\aa/\ell)   \cos (\yy/\ell)$  &$\displaystyle{
\xxtilde^1 = \frac{ 4\, \ssa/\ell }{4 + \left\{ (\ssa/\ell)^2
+(\ssb/\ell)^2\right\}}  }$ &$   \xxtilde^1=\tanh (\aa/\ell) $\\[8pt]
$\xxtilde^2 =\sin (\yy/\ell) $  & $\displaystyle{ \xxtilde^2 =
\frac{4\, \ssb/\ell }{4+ \left\{ (\ssa/\ell)^2 +(\ssb/\ell)^2\right\} }
}$&  
$\displaystyle{  \xxtilde^2 =\frac{\tanh (\yy/\ell)}{\cosh (\aa/\ell)}
}$\\[10pt]
$\tan A =\tan (\aa/\ell) $  & $\displaystyle{ \tan A=
\frac{4\, \aa/\ell }{4- \left\{ (\ssa/\ell)^2 +(\ssb/\ell)^2\right\} } }$&  
$\displaystyle{  \tan A = \sinh (\aa/\ell)  \cosh (\yy/\ell)  }$\\[8pt]
$\sin Y =\sin  (\yy/\ell) $  & $\displaystyle{ \sin Y=
\frac{4\, \yy/\ell }{4+ \left\{ (\ssa/\ell)^2 +(\ssb/\ell)^2\right\} } }$&  
$\displaystyle{  \sin Y = \frac{\tanh (\yy/\ell)}{ \cosh (\aa/\ell)} 
}$\\[18pt]
 ${\bf NH}_+^{1+1}\longmapsto \widetilde{{\bf NH}}_+^{1+1}$&${\bf
G}^{1+1}\longmapsto \widetilde{{\bf NH}}_+^{1+1}$&
${\bf NH}_-^{1+1}\longmapsto \widetilde{{\bf NH}}_+^{1+1}$\\[2pt] 
$\k_1\ell^2=1,\ \k_2=0$&$\k_1\ell^2=0,\ \k_2=0$&
$\k_1\ell^2=-1,\ \k_2=0$\\[6pt] 
$\xxtilde^\cnu=\cos (t/\ell)$  &$\displaystyle{\xxtilde^\cnu  
=\frac{4 -  (t/\ell)^2}{4 +  (t/\ell)^2}}$&$\displaystyle{ \xxtilde^\cnu = 
\frac{1}{\cosh (t/\ell) }  } $\\[8pt]
$\xxtilde^1= \sin (t/\ell)$  &$\displaystyle{ \xxtilde^1 = \frac{
4\,t/\ell }{4 +  (t/\ell)^2 }  }$&$  
\xxtilde^1=\tanh (t/\ell) $\\[8pt] 
$\xxtilde^2 =  y/\ell $  & $\displaystyle{ \xxtilde^2 =
\frac{4\,\ssb/\ell }{4+ (t/\ell)^2}}$&$\displaystyle{  \xxtilde^2 =\frac{ 
 \yy/\ell }{\cosh (t/\ell)} }$\\[10pt]
$\tan T =\tan (t/\ell) $  & $\displaystyle{ \tan T=
\frac{4\, t/\ell }{4-   (t/\ell)^2   } }$&  
$\displaystyle{  \tan T = \sinh (t/\ell)   }$\\[8pt]
$  Y = \yy/\ell  $  & $\displaystyle{  Y=
\frac{4\,\yy/\ell }{4+   (t/\ell)^2  } }$&  
$\displaystyle{  Y = \frac{  \yy/\ell }{ \cosh (t/\ell)}  }$\\[18pt]
${\bf AdS}^{1+1}\longmapsto \widetilde{{\bf AdS}}^{1+1}$&${\bf
M}^{1+1}\longmapsto \widetilde{{\bf AdS}}^{1+1}$&
${\bf dS}^{1+1}\longmapsto \widetilde{{\bf AdS}}^{1+1}$\\[2pt] 
$\k_1\ell^2=1,\ \k_2=-1/c^2$&$\k_1\ell^2=0,\ \k_2=-1/c^2$&
$\k_1\ell^2=-1,\ \k_2=-1/c^2$\\[6pt] 
$\xxtilde^\cnu =\cos (t/\ell)  \cosh (\yy/c\ell)$ 
&$\displaystyle{\xxtilde^\cnu   =\frac{4 -  \left\{ (t/\ell)^2 -  
(\ssb/c\ell)^2\right\}}{4 +  \left\{ (t/\ell)^2 -(\ssb/c\ell)^2\right\}
}}$ &$\displaystyle{ \xxtilde^\cnu = \frac{1}{\cosh (t/\ell)
\cos (\yy/c\ell)}  } $\\[8pt]
$\xxtilde^1 =\sin (t/\ell)   \cosh (\yy/c\ell)$  &$\displaystyle{
\xxtilde^1 =
\frac{ 4\,t/\ell }{4 + \left\{ (t/\ell)^2 -(\ssb/c\ell)^2\right\}}  }$ &$  
\xxtilde^1=\tanh (t/\ell) $\\[8pt]
$\xxtilde^2 =c  \sinh (\yy/c\ell) $  & $\displaystyle{ \xxtilde^2 =
\frac{4\,\ssb/\ell }{4+ \left\{ (t/\ell)^2 -(\ssb/c\ell)^2\right\} } }$&  
$\displaystyle{  \xxtilde^2 =\frac{c \tan (\yy/c\ell)}{\cosh (t/\ell)}
}$\\[10pt]
$\tan T =\tan (t/\ell) $  & $\displaystyle{ \tan T=
\frac{4\, t/\ell }{4- \left\{ (t/\ell)^2 -(\ssb/c\ell)^2\right\} } }$&  
$\displaystyle{  \tan T = \sinh (t/\ell)  \cos (\yy/c\ell)  }$\\[8pt]
$\sinh (Y/c) =\sinh  (\yy/c\ell) $  & $\displaystyle{ \sinh( Y/c)=
\frac{4 (\yy/c\ell) }{4+ \left\{ (t/\ell)^2 -(\ssb/c\ell)^2\right\} } }$&  
$\displaystyle{  \sinh (Y/c) = \frac{\tan (\yy/c\ell)}{ \cosh (t/\ell)} 
}$\\[12pt]
\hline
\end{tabular}\hfill}
\end{table}

\end{document}